\documentclass[manuscript, screen]{acmart}

\AtBeginDocument{%
  }

\setcopyright{acmlicensed}
\copyrightyear{2018}
\acmYear{2018}
\acmDOI{XXXXXXX.XXXXXXX}

\acmJournal{JACM}
\acmVolume{37}
\acmNumber{4}
\acmArticle{111}
\acmMonth{8}

\newtheorem{myTheo}{Example}[section]
\usepackage{array}
\usepackage{booktabs}
\usepackage{algorithm} 
\usepackage{algorithmic}

\begin{document}

% \title{Large Language Model for ID-based Recommendation}
\title{Enhancing ID-based Recommendation with Large Language Models}

% \affiliation{%

\author{Lei Chen} 
\email{chenlei_hfut@mail.tsinghua.edu.cn} 
\affiliation{%
  \institution{Tsinghua University} 
    \city{Beijing}
  \postcode{100084} 
  \country{China} 
}

\author{Chen Gao} 
\email{chgao96@gmail.com}
\affiliation{%
  \institution{Tsinghua University} 
    \city{Beijing}
  \postcode{100084} 
  \country{China} 
}

\author{Xiaoyi Du} 
\email{duxiaoyi@meituan.com}
\affiliation{%
  \institution{Meituan} 
    \city{Beijing}
  \postcode{100084} 
  \country{China} 
}

\author{Hengliang Luo} 
\email{luohengliang@meituan.com}
\affiliation{%
  \institution{Meituan} 
    \city{Beijing}
  \postcode{100084} 
  \country{China} 
}

\author{Depeng Jin}  
\email{jindp@tsinghua.edu.cn}
\affiliation{%
  \institution{Tsinghua University} 
    \city{Beijing}
  \postcode{100084} 
  \country{China} 
}

\author{Yong Li} 
\email{liyong07@tsinghua.edu.cn}
\affiliation{%
  \institution{Tsinghua University} 
  \city{Beijing}
  \postcode{100084} 
  \country{China} 
}

\author{Meng Wang} 
\email{eric.mengwang@gmail.com	}
\affiliation{%
  \institution{Hefei University of Technology	} 
  \city{Hefei}
  \postcode{230601}
  \country{China} 
}

\begin{abstract}
Large Language Models (LLMs) have recently garnered significant attention in various domains, including recommendation systems. Recent research leverages the capabilities of LLMs to improve the performance and user modeling aspects of recommender systems. These studies primarily focus on utilizing LLMs to interpret textual data in recommendation tasks. However, it's worth noting that in ID-based recommendations, textual data is absent, and only ID data is available. The untapped potential of LLMs for ID data within the ID-based recommendation paradigm remains relatively unexplored. To this end, we introduce a pioneering approach called "LLM for ID-based Recommendation" (LLM4IDRec). This innovative approach integrates the capabilities of LLMs while exclusively relying on ID data, thus diverging from the previous reliance on textual data. The basic idea of LLM4IDRec is that by employing LLM to augment ID data, if augmented ID data can improve recommendation performance, it demonstrates the ability of LLM to interpret ID data effectively, exploring an innovative way for the integration of LLM in ID-based recommendation. Specifically, we first define a prompt template to enhance LLM's ability to comprehend ID data and the ID-based recommendation task. Next, during the process of generating training data using this prompt template, we develop two efficient methods to capture both the local and global structure of ID data. We feed this generated training data into the LLM and employ LoRA for fine-tuning LLM. Following the fine-tuning phase, we utilize the fine-tuned LLM to generate ID data that aligns with users' preferences. We design two filtering strategies to eliminate invalid generated data. Thirdly, we can merge the original ID data with the generated ID data, creating augmented data. Finally, we input this augmented data into the existing ID-based recommendation models without any modifications to the recommendation model itself. We evaluate the effectiveness of our LLM4IDRec approach using three widely-used datasets. Our results demonstrate a notable improvement in recommendation performance, with our approach consistently outperforming existing methods in ID-based recommendation by solely augmenting input data.

\end{abstract}
% Our work is motivated by 
 % ID-associated text segments (such
% as user name and item title) a


\begin{CCSXML}
<ccs2012>
   <concept>
       <concept_id>10002951.10003317.10003338.10003341</concept_id>
       <concept_desc>Information systems~Language models</concept_desc>
       <concept_significance>500</concept_significance>
       </concept>
   <concept>
       <concept_id>10002951.10003317.10003347.10003350</concept_id>
       <concept_desc>Information systems~Recommender systems</concept_desc>
       <concept_significance>500</concept_significance>
       </concept>
 </ccs2012>
\end{CCSXML}

\ccsdesc[500]{Information systems~Language models}
\ccsdesc[500]{Information systems~Recommender systems}

\keywords{Large Language Model, ID-based recommendation, Data augmentation}

\received{20 February 2007}
\received[revised]{12 March 2009}
\received[accepted]{5 June 2009}

\maketitle

\section{Introduction}
%%RS很重要，探索了很多深度学习模型去建模用户偏好，LLM在多项任务表现出推理饿能力，那么一个自然的想法就是用LLM去做RS
% 大模型现有的方法分为LLM as RS和LLM + RS,其中LLM as RS只能在zero-shot和few-shot上表现的更好，比并且严重依赖与text等描述性语言。LLM +RS的模型多依赖于text等描述性语言，而很少去探索ID的结果.
%我们的方法
Recommender systems play a central role and have emerged as indispensable tools in online services~\cite{covington2016deep,guo2017deepfm,dai2021adversarial,yu2023self,chen2023bias}. They serve the critical function of offering personalized recommendations in the face of information overload, effectively aligning with user preferences across a range of tasks~\cite{zhou2018deep,chen2020try}. While various recommendation tasks exist~\cite{covington2016deep,lin2021graph,xi2023bird}, including top-N recommendation, next-item recommendation, and rating prediction, the common approach involves learning user representations to model their preferences and intentions. These learned representations are subsequently employed to generate decisions regarding recommended items for users. In recent years, Large Language Models (LLMs) have exhibited remarkable proficiency in approximating human intentions and excelling in a wide array of tasks, including reasoning and decision-making~\cite{huang2022towards,brown2020language,zhao2023survey}.
{Inspired by the great success of Large Language Models (LLMs) \cite{gpt3, llama, vicuna}, exploring the potential of LLMs in recommendation is attracting attention~\cite{RecFormer, TALLRec, P5, M6-Rec, uncovering_ChatGPT4Rec,agent4rec,ToolRec}, especially driven by innate reasoning capabilities and approximating human intentions of LLMs.
% With exceptional learning and modeling ability, the remarkable effectiveness of LLMs has motivated the exploration of LLM for recommendation systems~\cite{gao2023chat,hou2023large,liu2023chatgpt}.
With the exploration of LLM-based recommendation~\cite{gao2023chat,hou2023large,liu2023chatgpt}, this direction has emerged as a promising approach for the next-generation recommendation systems~\cite{zeng2021knowledge,liu2023pre,wang2023generative}. %For example, ** model. 
}

\begin{figure}[t]
    \centering
    \includegraphics[width=0.95\textwidth]{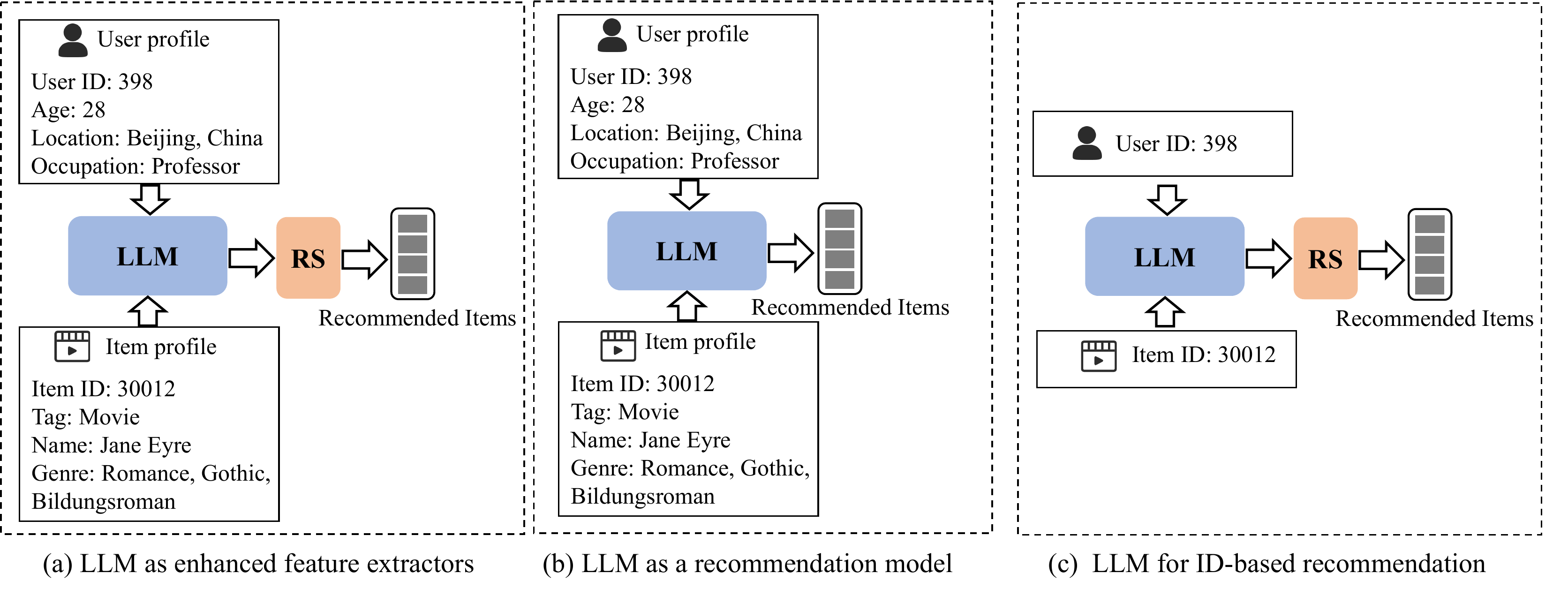} 
    \caption{An example of comparing different structures of LLM in Recommender Systems~(RS). Figure 1(a) and (b) depict current approaches for incorporating LLM into recommendation models. Both methods utilize textual data as input, such as user and item profiles. Nevertheless, in the context of ID-based recommendations, solely ID data is available, devoid of any textual information. When dealing with recommendation scenarios that solely rely on ID data, we investigated the application of LLM for ID-based recommendation, as shown in Figure 1(c).}
    \label{fig:intro}
\end{figure}

%图a和图b展示了现有方法将LLM引入到推荐模型中。这2类方法输入都依赖文本数据。然而在ID推荐中只有ID数据而没有文本数据。针对只有ID数据的推荐场景，我们探索了LLM用于ID推荐。

% \textcolor{blue}{
In the context of recommender systems, accurately capturing patterns in user interaction data (ID data) is crucial for generating relevant recommendations. The core goal of recommendation algorithms is to predict the items a user is likely to engage with based on their interaction history. Over the years, various methods have been developed to improve recommendation performance, including Markov Chains~\cite{he2016fusing,rendle2010factorizing}, RNN/CNN models~\cite{he2016fusing,rendle2010factorizing}, self-attentive models~\cite{kang2018self,li2020time,zhao2021variational}, and more recently, graph convolutional network (GCN)-based models~\cite{wang2019neural,chen2020revisiting,yu2022graph}. These advancements stem from increasingly sophisticated architectures designed to better capture the complex relationships in interaction data. GCN-based models, for example, excel at capturing intricate relationships between users and items compared to traditional CNN-based models.
However, the emergence of LLMs introduces a new paradigm with even greater capacity for modeling and reasoning, surpassing existing structures like GCNs~\cite{minaee2024large,wang2024llms,ren2024survey,wu2024survey}. Given LLMs' ability to model and interpret complex data, they present a promising direction for processing user interaction data (ID data). Various studies have demonstrated that LLMs can successfully address graph-related tasks, which rely heavily on understanding and reasoning about relationships within ID data~\cite{zhang2024llm4dyg,guo2023gpt4graph,ye2024language,geng2022recommendation,sun2024determlr}. For instance, LLM4DyG~\cite{zhang2024llm4dyg} constructs prompts using node and edge IDs, which are then processed by LLMs for inference, effectively tackling graph-based tasks. This suggests that employing LLMs to model and interpret ID data is not only feasible but also a research direction worth exploring further.

For the studies of LLMs in recommendations, the current approaches can be broadly categorized into two groups. As shown in Figure~\ref{fig:intro}(a), the first category involves leveraging LLMs as enhanced feature extractors to improve the performance of recommendation models in terms of accuracy or interpretability~\cite{li2023exploring,li2023ctrl,liu2023first,wei2023llmrec}. %can add some paper details
For example, LLMRec~\cite{wei2023llmrec} uses LLMs to explicitly model with three types of features, such as user profile.
In this category, LLMs serve as tools to extract valuable features from textual data, which contribute to improving recommendation models.
As shown in Figure~\ref{fig:intro}(b), the second category employs LLMs as a powerful recommendation system to directly generate recommendation results, such as a list of recommended items~\cite{hou2023large,sun2023chatgpt,dai2023uncovering}. LLM-driven recommendation systems leverage the inherent capabilities of LLMs to generate recommendations based on the textual data they process.
It is worth noting that both of these approaches primarily rely on textual data, such as book/movie titles and knowledge labels. The textual data forms the foundation for constructing recommendation paradigms, capitalizing on the generalization capabilities of LLMs to enhance recommendation performance. However, in ID-based recommendation, textual data is absent, containing only core ID information—namely, user IDs and item IDs—without corresponding textual descriptions for users and items~\cite{yuan2023go,rendle2012bpr,yuan2022tenrec}. Therefore, instead of focusing on designing an LLM-based recommendation system utilizing textual data, we explore LLM utilizing pure ID data in ID-based recommendation, as depicted in Figure~\ref{fig:intro}(c). Specifically, we investigate the following two questions:

\textbf{Q1: Can LLM feed ID data to generate results that meet recommendation requirements?}
Many studies have shown that the representations learned by LLMs are applicable to various tasks enriched with textual side information~\cite{borisov2023language,carranza2023privacy,mysore2023large}. Nevertheless, in existing recommendation systems, countless models have been developed that rely solely on user/item IDs to generate high-quality recommendations, eschewing the use of any textual or content-based information~\cite{rendle2009bpr,he2020lightgcn,chen2020revisiting,yu2022graph,sun2019bert4rec}. In this study, we leverage ID-based recommendation as a domain task to investigate the applicability of LLM when confronted with ID data.

%现在有一些工作将LM用于推荐系统，因此
%在大模型之前，已经有很多的预训练语言模型并且
%这些模型的优势在于2个方面
%https://arxiv.org/pdf/2007.15356.pdf BERT-know
Before the emergence of large language models, there were already many pre-trained language models, such as BERT~\cite{devlin2018bert} and T5~\cite{raffel2020exploring}. These pre-trained language models have been successfully applied to various recommendation tasks~\cite{qu2019bert,sakata2019faq,yang2019end,yang2019simple,hou2022towards}. For example, {UniSRec\cite{hou2022towards} improving recommendation models using pre-trained language modelsr~(e.g. BERT). P5~\cite{geng2022recommendation} pretrains on an encoder–decoder Transformer model to effectively model user behavior sequences (i.e., ID sequences). } The success of using pre-trained language models in recommendation tasks can be attributed to their inherent linguistic capabilities. These linguistic capabilities encompass various facets, including the storage of factual knowledge and model sequential data. Meanwhile, LLMs have witnessed superior outperformance compared to traditional pre-trained language models, so utilization of LLM for recommendation tasks is an appealing avenue, particularly in the context of countless ID-based recommendations.

 % way machines interact with human language.
% This particular advantage empowers pre-trained language models to leverage linguistic capabilities, including the storage of factual knowledge, for recommendation tasks while incurring minimal computational and migration costs.   
% Another advantage of these pre-trained language models is store factual knowledge in their parameters by modeling human language. 
%上述2方面的优势使得预训练的模型能够以极小的计算和迁移成本将预训练的能力应用到IR任务上，例如，。

\textbf{Q2: How can we effectively leverage LLM to enhance ID-based recommendation?}
ID-based recommendation contain several essential components, such as interaction data, network structure, and ranking loss functions~\cite{rendle2009bpr,chen2020revisiting,yu2022graph}. LLMs can assume different roles within the ID-based recommendation pipeline. A straightforward question arises about where to integrate LLM in an ID-based recommendation model. Consequently, depending on the specific role, we must design different paradigms for LLMs, such as fine-tuning and prompting. The primary focus is on how to adapt LLM for an ID-based recommendation model effectively.

% The core ID data utilized within ID-based recommendation model consists of interaction data, representing the most abundant and easily accessible form of data. When LLM exhibits competence in the processing and understanding of ID data, it holds the promise of mitigating challenges encountered by recommendation systems, such as data sparsity, ultimately enhancing the precision of recommendations. Therefore, we integrate LLM as a fundamental component to augment interactive data within the ID-based recommendation model.

{The utilization of core ID data in an ID-based recommendation model primarily relies on interaction data, which is abundant and readily accessible. If a large language model (LLM) demonstrates proficiency in processing and understanding ID data, it presents an opportunity to integrate LLM as an element of recommendation. Specifically, by augmenting interactive data within the ID-based recommendation model using LLM capabilities, we can enhance the precision of recommendations.}

% The utilization of core ID data in an ID-based recommendation model primarily relies on interaction data, which is abundant and readily accessible. When a large language model (LLM) demonstrates proficiency in processing and understanding ID data, it presents an opportunity to address challenges faced by recommendation systems, such as data sparsity. By integrating LLM as a foundational element, we aim to enhance the precision of recommendations by augmenting interactive data within the ID-based recommendation model.

In this paper, we seek to present innovative solutions for integrating LLM into ID-based recommender systems. The basic idea is to feed pure ID data from recommendation tasks into a pre-trained LLM. Suppose the output of the LLM is meaningful and can improve recommendation performance. In that case, it can reflect that the LLM can understand and process ID data within the context of recommendation tasks. Conversely, suppose the output of the LLM is random, inaccurate, or lacks meaningfulness, resulting in a deterioration in recommendation performance. In that case, the LLM still lacks the ability to understand ID data and recommendation tasks intuitively.
Based on the above idea, we propose a novel framework, namely the Large Language Model for ID-based Recommendation (LLM4IDRec).
We designed a prompt template to process ID data and ensure that the output meets the recommended requirements (\textbf{Q1}). Meanwhile, LLM4IDRec utilizes LLM to augment the ID data and then enhance the ID-based recommendation model  (\textbf{Q2}).
Specifically, the ID data is initially transformed into training data through a designed prompt template that is understandable to language models. These generative training data are fed into the pre-trained LLM and adopt the fine-tuning strategy. After training, the fine-tuned LLM can generate ID data, effectively augmenting the original ID data.
LLM4IDRec represents a pioneering approach to enhancing the capabilities of ID-based recommendation models through the power of LLM.

Our main contributions are summarized as follows:
\begin{itemize}
  \item We intend to explore the potential of LLM and investigate a key question: Can the LLM adapt to the ID-based recommendation? This is different from existing work centred around textual data. {While prior studies predominantly utilize textual data, such as attribute or title, to represent items and users in prompts, ID-based recommendations lack such textual information, relying solely on ID data. This unexplored potential of LLMs in handling ID data within recommendation paradigms is a driving force behind their integration. We propose that prompting LLMs mainly with textual data overlooks their full potential in recommendation tasks. Instead, delving deeper into ID-based data patterns inherent in interaction data can unlock novel insights and improve recommendation precision, thus justifying the use of LLMs.}

  \item  We propose a novel approach, LLM4IDRec, which leverages LLM to augment ID data and subsequently utilizes this augmented data to improve the performance of ID-based recommendations.{Specifically, we design a tailored prompt template specifically for ID-based recommendation tasks. This template, combined with two methods for generating training data to capture both local and global ID data structures. Additionally, combining fine-tuning strategy and two filter strategies guide LLMs in capturing ID-based data and generating interaction data aligned with user preferences. These modules significantly enhance the effectiveness of LLM4IDRec.}

  \item  In experiments conducted using three publicly available datasets, LLM4IDRec consistently outperforms existing ID-based recommendation methods by only augmenting the input ID data. These results demonstrate the rationality of LLM4IDRec and the feasibility of LLM in the ID-based recommendation.

\end{itemize}

  % We first propose a novel training framework, which is capable of aligning signals from ID data and language models, introducing semantic knowledge into the collaborative models.
% 构建出 developing a vast amount of textual data from diverse sources,  is an ambitious goal

% What are the key challenges and potential solutions in applying LLMs to ID data?

\section{Related Work}
Our work is closely related to several other lines of research. First, we discuss ID-based recommendation models, and then we explore the integration of LLM with the recommendation model.

% Several lines of work are closely related to ours. We first
% discuss ID-based recommendation models, followed by integration of LLM with the recommended model.

\subsection{ID-based Recommendation}
% In the existing recommendation literature, there are countless models built entirely on user/item ID, from early item-to-item collaborative filtering [35], shallow factorization models [28, 48], to deep neural models [21, 22].

% The modern recommendation models usually use unique identities (ID) to represent users and items, which are subsequently converted to embedding vectors as learnable parameters. These ID-based recommendation models (IDRec) have been well-established and dominated the RS field for over a decade until now [28, 49, 77].

In recommendation scenarios, the prevalent practice is to employ unique identifiers (IDs) to represent users and items, and then the user behavior (such as click and purchase behavior) is represented as an interaction matrix between user IDs and item IDs. Thus, modern recommendation models~\cite{koren2009matrix,rendle2012bpr,yuan2022tenrec} have placed a significant emphasis on ID data and have, in fact, constructed numerous models tailored specifically to the handling of user and item IDs. The developmental trajectory of recommendation models has evolved from the traditional collaborative filtering~\cite{linden2003amazon}, shallow factorization models~\cite{koren2009matrix,rendle2010factorization} to the more deep neural network models~\cite{he2017neural,hidasi2015session}.

%对于ID推荐的常见的可以分为2种：一种是用户交互行为之间没有时间关系，
% The interaction matrix between user IDs and item IDs is often too sparse. 
%用户ID和项目ID之间的交互行为构建出交互矩阵。根据行为之间先后关系，我们可以将数据分为2种：一种是用户交互行为之间没有时间关系，另外是序列推荐，行为之间有明确的先后关系。针对第一种数据，现有的方法基于GCN的推荐模型已经成为主流。例如NCL，随着对比学习的发展最近一些推荐模型将对比学习融合到基于GCN的推荐模型，并提出SimGCL和XsimGCL模型。针对序列推荐数据，现有的推荐模型是基于对序列有建模能力的结构去建模序列行为，例如SASRec和BERT4REC模型均是利用attention结构去更好利用序列行为数据。随着对比学习的发展最近一些推荐模型也将对比学习融合到序列推荐中，提出了CL4SRec。

The interaction behavior between user IDs and item IDs constructs the interaction matrix. This matrix can be categorized into two types based on the temporal information associated with these behaviors: one involves user interaction behavior without temporal relationships, and the other pertains to the sequential recommendation, where a clear sequential relationship between behaviors exists.
For the first interaction data type, existing research methods mainly rely on recommendation models based on  Graph Convolutional Networks {(GCNs)~\cite{chen2020revisiting,he2020lightgcn}} which have become mainstream methods. For example, {NCL~\cite{lin2022improving}} and other models have consistently achieved significant results. More recently, as contrastive learning has gained prominence, some recommendation models have integrated it into {GCN-based models~~\cite{yu2022graph,yu2023xsimgcl}}. Examples include {SimGCL~\cite{yu2022graph} and XsimGCL~\cite{yu2023xsimgcl}}, which excel at capturing the correlations within ID data.
For sequential recommendation data, existing recommendation models focus on modeling the structure of sequential behaviors to better understand the temporal order between {them~\cite{kang2018self,xie2022contrastive}}. Models like {SASRec~\cite{kang2018self}} and {BERT4REC~\cite{sun2019bert4rec}} leverage attention mechanisms to effectively utilize sequence behavior data. It's worth mentioning that recent studies have introduced contrastive learning into the sequential recommendation, leading to models like {CL4SRec~\cite{xie2022contrastive}}. These models enhance the performance of sequence recommendation by incorporating contrastive learning and improving the understanding of user behavior sequences.
In summary, various ID-based recommendation models have been developed in the research field for the interaction data with only user IDs and item IDs. Based on these models, we use LLM for data augmentation to improve the performance of existing ID-based recommendation models, thereby providing feasible solutions for exploring ID-based recommendation tasks in the current development of LLM.

% Choosing the appropriate model based on different data types can better model user behavior and provide personalized recommendation services. This constantly evolving research trend is expected to bring more innovation and improvement to the field of recommendation systems, improving user experience and recommendation accuracy.

% The interaction matrix between user IDs and item IDs is often too sparse. Effectively solving this sparsity issue usually addresses it from two aspects: one is from the perspective of supplementing data, and the other is from the perspective of better-utilizing data.
% In terms of loss functions, they can be broadly categorized into three types: pairwise-based loss functions, setwise-based loss functions, and listwise-based loss functions. Among these, pairwise-based loss is the most widely adopted setting due to its simplicity and computational efficiency. However, it is noted that while the pairwise-based approach offers expedience, it may not fully exploit the underlying information as effectively as setwise-based loss and listwise-based loss. 
% This, in turn, can result in potential performance disparities that favor the latter two methodologies in specific contexts.

%这样反过来setwise和listwise的效果更好。
%用户行为（如点击、购买行为）表示成用户ID和产品ID的交互矩阵。
%由于用户和产品数量众多，往往用户和产品的交互行为过于稀疏。
%如何利用好稀疏交互数据方面，有从模型角度来利用，有从损失函数角度来利用。
%然后举例来说bpr，gcn这类模型。
%采用的loss函数通常分为3类，pairwise，setwise，listwise，其中pairwise使用最多。因为其简单高效，但是对信息的利用不如setwise和listwise，所以效果上通常不如pairwise。

\subsection{LLM with the recommended model}

For the adaption of language models in recommendations, specifically concerning the modeling paradigm, existing research can be broadly categorized into the following two main groups:\textbf{LLM as RS} and \textbf{LLM + RS}.

\subsubsection{LLM as RS}
LLM as RS is to leverage LLM as a powerful recommendation model~\cite{hou2023large,sun2023chatgpt,dai2023uncovering}. Typically, the input sequence consists of textual descriptions, user behavior prompts, and instructions for the recommendation task. Then the "LLM as RS" generates the desired output, which may include predictions of target item scores and the ranking of the candidate item list. Some recent works explore the utilization of LLMs as RS.
% Uncovering chatgpt’s capabilities in recommender systems.  RecSys2023 
Dai et al.~\cite{dai2023uncovering} proposes to enhance ChatGPT’s recommendation capabilities by aligning it with ranking capabilities. This approach leverages item titles as descriptors within the prompt, followed by conducting a preliminary evaluation using generated prompts.
% use the titles of the items as descriptions in the prompt and conduct a preliminary evaluation with generated prompts.
% Is chatgpt good at search? investigating large language models as reranking agent.  EMNLP 2023 
Sun et al.~\cite{sun2023chatgpt} explore the two instructional strategies for LLMs in ranking tasks, they evaluate the capabilities of LLMs on three passage re-ranking datasets, including the multilingual passage retrieval dataset named Mr.TyDi.
% Personalized Prompt Learning for Explainable Recommendation  ACM TOIS
The PEPLER method~\cite{li2023personalized} employs two training strategies for turning LLMs and generates language explanations for recommendations. To ensure the explanation quality, the input data for PEPLER contains an explanation and, crucially, a minimum of one item feature.
%Unitrec: A unified text-to-text transformer and joint contrastive learning framework for textbased recommendation ACL-2023 Short Paper
UniTRec~\cite{mao2023unitrec} leverages pre-trained language models to assess the alignment of user candidate text and enhance text-based recommendations.
{TransRec~\cite{lin2023multi} introduces multi-faceted identifiers that include ID, title, and attribute to propose a new recommendation model.}

%no-truning
% Chat-rec: Towards interactive and explainable llms-augmented recommender system. 
% Generative recommendation: Towards next-generation recommender paradigm.
% Large language models are zero-shot rankers for recommender systems.
%  Large language models are effective text rankers with pairwise ranking prompting. 
%Is ChatGPT a Good Recommender? A Preliminary Study
% Large Language Models (LLMs) have demonstrated a remarkable ability to generalize zeroshot to various language-related tasks.

%turning
% Recommendation as Instruction Following: A Large Language Model Empowered Recommendation Approach
%Do LLMs Understand User Preferences? Evaluating LLMs On User Rating Prediction
% Exploring Large Language Model for Graph Data Understanding in Online Job Recommendations

\subsubsection{LLM + RS}
LLM + RS is to leverage LLM as a feature extractor~\cite{li2023exploring,li2023ctrl,liu2023first}. In this approach, the input sequence consists of item-specific and user-specific features, and then the output is to obtain embeddings that correspond to these features. The obtained embeddings by LLM can be integrated into the modeling process of a recommendation system, such as capturing potential preferences. For example,
% Towards unified conversational recommender systems via knowledge-enhanced prompt learning. Accepted by KDD 2022
UniCRS~\cite{wang2022towards} adopts knowledge-enhanced prompts and bases on knowledge-enhanced prompt learning to effectively address various conversational subtasks within a unified framework.
% ONCE: Boosting Content-based Recommendation with Both Open- and Closed-source Large Language Models
Liu et al.~\cite{liu2013once} employ open-source LLMs to augment content at the embedding level and utilize close-source LLMs to enrich the training data at the token level. It is noteworthy that both open- and close-source LLMs also can enhance content-based recommendation systems.

% Towards open-world recommendation with knowledge augmentation from large language models
% When Large Language Model based Agent Meets User Behavior Analysis: A Novel User Simulation Paradigm
% Exploring the upper limits of text-based collaborative filtering using large language models: Discoveries and insights. 
% Ctrl: Connect tabular and language model for ctr prediction

%现有方法依赖于文本数据,而没有探索大模型在纯ID数据上可行性。对于LLM在纯ID数据上的还值得探索，还可以证明LLM的能力和适用范围

However, existing methods rely on textual data, which neglects the potential of LLMs on pure ID data. In our work, we aim to enhance ID-based recommender paradigms by incorporating LLMs. Our approach presents a pioneering paradigm for next-generation recommender systems, and investigates the capacity and suitability of LLMs in this domain.

\section{Preliminary}
In this section, we present the ID-based recommendation model and reveal the large language model in modeling ID information. The core data of a recommendation model is to represent users and items and their interaction data. Denote~$U$ (of number~$|U|$) and~$V$~(of number~$|V|$) as the set of users and items, respectively. For each user~$u \in U$, we can represent it by its unique ID. Similarly, for each item~$v \in V$, we can represent it by its unique ID. Users and items are only represented by ID and do not contain modality information. The interaction data are represented by a binary matrix~$R={r_{uv}}$, where~$r_{uv} \in \{0,1\}$ indicates whether user~$u$ has interacted with item~$v$.

In the ID-based recommendation model, users and items are represented by an embedding matrix~$\mathbf{E} \in \mathbb{R}^{(|U|+|V|) \times d}$. According to the interaction data~$R$, the ID-based recommendation model retrieves the embedding of users/items and then feeds it to the recommendation network. During training, the ID-based recommendation model typically designs and optimizes a loss function~$L$, where~$L$ can be a pairwise BPR loss or other widely-used loss. In a word, various ID-based recommendation models are input as interaction data~$R$, and these models are optimized through a loss function to make the predicted candidate items close to the real interaction data~$R$.

\textbf{LLM and ID-based recommendation.}
Within the domain of language models, specifically LLM, the objective usually is to project the probability distribution of the probability word in a given textual context, denoted as:
\begin{equation}
p(word|context).
\end{equation}
{
In LLM, input data is typically represented as tokens. Therefore, the above formula is essentially predicting the next most likely token given a series of tokens, denoted as:
\begin{equation}
p(next~token|tokens).
\end{equation}
When dealing with a user denoted as $u$ and their list of interacted items represented as $R_u$, our approach involves treating users and items as tokens and modeling interaction data as textual sequences for fine-tuning LLMs. This transformation enables LLMs to predict the next most likely tokens (the most likely items) based on given tokens (the interaction data), effectively allowing LLMs to model interaction data. The objective of LLM is denoted as:
\begin{equation}
p(next~token=\{v_i\}|tokens=\{u,R_u\}). 
\end{equation}
Whether the data comprises purely textual information or includes interaction data, LLMs aim to estimate the probability distribution of the next token. In the context of fine-tuning LLMs, the goal is to empower the pretrained LLM to acquire domain-specific knowledge, particularly implicit feedback data, by learning the distribution relationship between tokens (the interaction data). Hence, it is possible for LLM to generate items that align with user preferences within the specified context~$tokens=\{u, R_u\}$. These generated items by LLM can then be used to augment the existing sparse interaction data $R$, ultimately enhancing the overall performance of the recommendation system.
}

\begin{figure}[t]
    \centering
    \includegraphics[width=0.98\textwidth]{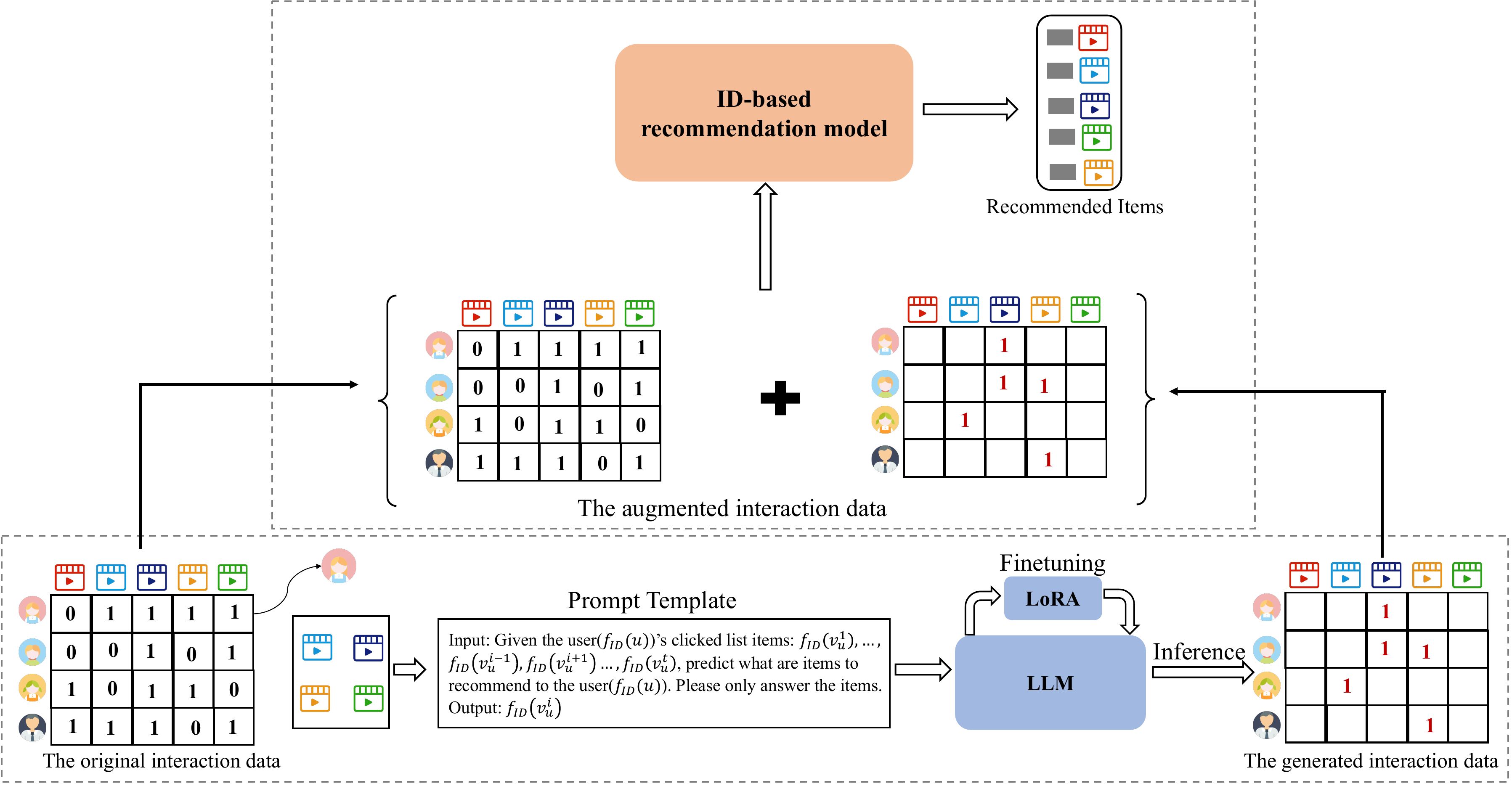} 
    \caption{The architecture of our proposed LLM4IDRec.}
    \label{fig:model}
\end{figure}
 % . For the sake of brevity, we only illustrate the process of transferring information from the domain D  to promote performance in the domain D

\section{Method}
As shown in Figure~\ref{fig:model}, the LLM4IDRec framework aims to enhance the alignment of LLMs with ID-based recommendation systems. We employ LLMs for data augmentation within ID-based models, leveraging the advantages of LLMs in modeling and inference capabilities. This collaboration with ID-based models allows us to effectively capture collaborative information, thereby improving the overall performance of recommendations. The proposed LLM4IDRec adopts a two-stage paradigm. The first stage is the LLM-based Data Generation. The LLM-based Data Generation feeds the historical interaction data into an LLM and then infers potential interaction data. Combining the generated potential interaction data with the original historical interaction data creates augmented interaction data. The augmented interaction data can be naturally plugged into existing ID-based recommendation models. The second stage is the ID-based recommendation model, which is a common ID-based recommendation model.

% As depicted in Figure~\ref{fig:model}, the proposed LLM4IDRec framework aims to facilitate the alignment of LLMs with ID-based recommendation. We use LLM for data augmentation in ID-based models, leveraging the advantages of LLM in modeling and inference capabilities and collaborating with ID-based models to model collaborative information, thereby improving recommendation performance.

%我们通过大模型来处理ID数据并用于数据增强，发挥了大模型在建模和推理能力上的优势并协同ID-based模型对协同信息的建模，从提升推荐性能。

% we introduce the TALLRec framework, which aims to facilitate the effective and efficient alignment of LLMs with recommendation tasks, particularly in low GPU memory consumption settings. Specifically, we first present two TALLRec tuning
% stages with lightweight implementation, followed by the backbone selection. As shown in Figure 2, TALLRec comprises two tuning stages: alpaca tuning and rec-tuning. The former stage is the common training process of LLM that enhances LLM’s generalization ability, while the latter stage emulates the pattern of instruction tuning and tunes LLMs for the recommendation task.

%将生存的交互数据和原始的历史交互数据结合生成数据增强的交互数据，

% Prompt engineering for extract local/global structures
% Fine-tune for data adaption and alignment
% Data augumention-based small-large model cooperation 

\subsection{The LLM-based Data Generation}
In the interaction data~$R$, given a user~$u$ with interacted item sequence~$S_u=\{v^1_u,v^2_u,...,v^{t}_u,\}$, we want to predict~$n$ items~$\hat{S}_u^n=\{v^{t+1}_u,...,v^{t+n}_u\}$ that user~$u$ may like. For almost large language model recommender systems~(LLMRec), each item~$i$ is associated with a textual description~$d(i)$, such as a book title. These LLMRec models initially process each item's description~$d(i)$ by tokenizing the text and formulate a prompt~$f_{prompt}(*)$ which serves to convert the interacted item sequence $S_u$ into a contextual sequence $f(d(S_u)) = f_{prompt}({d(v^1_u), d(v^2_u), \ldots, d(v^{t}_u)})$. Subsequently, LLMRec feeds this contextual sequence, $f_{prompt}(d(S_u))$, into a comprehensive language model to generate recommendations tailored to the preferences of user~$u$.

It's important to underscore that existing LLMRec models rely exclusively on textual descriptions and are primarily concerned with text-based data. However, the challenge that arises is that in ID-based recommendation tasks, there is no textual description associated with the items. Consequently, our proposed approach grapples with two key questions: firstly, how can ID-based data be appropriately represented within the LMRec framework? secondly, can a large language model effectively capture and model ID data?

% These models first map each item description~$d(i)$ to its tokenized text and design a prompt~$f(*)$ to convert the interacted item sequence~$S_u$ into the context sequence~$f(d(S_u))=f(\{d(v^1_u),d(v^2_u),...,d(v^{t}_u)\})$. Then LLMRec inputs the context sequence~$f(d(S_u))$ into a large language model and outputs items that user~$u$ may like. Note that the existing LLMRec relies on text description and focuses only on the text itself. However, there is no text description in the ID-based recommendation task. The key challenge of our proposed method is two questions: can a large language model be used to model ID-based data, and how to model ID-based data.

\subsection{Prompt Template}
We first present the process of constructing prompt templates. These templates serve the purpose of converting ID-based data into textual data for each instance. A proper prompt should contain the interaction data about the user and the item and also provide a clear task description. The following steps are undertaken to create a proper prompt: 

1) Establishing context and defining goal: Begin by providing context regarding the user and the recommendation task. It necessitates the clear articulation of the task's objectives, explicitly stating that the LLM model is tasked with offering recommended items.

2) List the interacted items: A fundamental component of the prompt template is the enumeration of items that the user has previously interacted with or clicked. This enumeration ensures that the LLM model is equipped with the historical data requisite for generating relevant recommendations. 

3) In the process of fine-tuning the LLM model, an additional requirement is introduced wherein labeled data with user interactions and desired recommendations is provided. This labeled data can guide the optimization of the LLM model and improve performance based on real-world data.

% There are follow these steps: 1) Set the context and goal: Begin by providing context about the user and the recommendation task. Clearly state the goal of the recommendation task. Explain that the LLM model should recommend items. 2) List the interacted items: Enumerate the items that the user has already clicked. This provides the LLM model with the historical data it needs to generate relevant recommendations. When fine-tuning the LLM model, we also provide labeled data with user interactions and desired recommendations, allowing the LLM model to learn from this data.

For this purpose, we define the prompt template.% design the following template to construct the prompts:
\begin{definition}
\label{def:prompt}
For user~$u$ and interaction item sequence~$S_u=\{v^1_u,v^2_u,...,v^{t}_u,\}$, We first use the ID mapping function~$f_{ID}(*)$ to obtain the unique ID corresponding to the user~$u$ and item sequence~$S_u$. Then we design the following template to construct the prompts:

\begin{itemize}
\item Input: Given the user($f_{ID}(u)$)'s clicked list items:$f_{ID}(v^1_u),...,f_{ID}(v^{i-1}_u),f_{ID}(v^{i+1}_u),...,f_{ID}(v^{t}_u)$, predict what are items to recommend to the user($f_{ID}(u)$). Please only answer the items.

\item Output:$f_{ID}(v^i_u)$
\end{itemize}
Please note that the item~$v^i_u \in S_u$ in the output is any one item in item sequence~$S_u$. {In this prompt template, the prompts are tailored to what the pretrained LLM can understand, aiming to trigger its reasoning ability based on encoded knowledge. For instance, consider the relational phrase "the user($f_{ID}(u)$)'s clicked list items $f_{ID}(v^1_u),...,f_{ID}(v^{i-1}u),f{ID}(v^{i+1}u),$ $...,f{ID}(v^{t}_u)$" in the prompt, which guides the LLM to recognize the newly-introduced token is a user subject~$f_{ID}(u)$ and the tokens~$f_{ID}(v^1_u),...,f_{ID}(v^{i-1}_u),f_{ID}(v^{i+1}_u),...,f_{ID}(v^{t}_u)$ in the prompt are the objects of the interacted item sequneces. Moreover, the contexts "predict what items to recommend to the user($f_{ID}(u)$)" help the content LLM better understand the goal. The specific formulation of the prompt plays a crucial role in semantically aiding LLMs to comprehend interactive data, minimizing semantic and goal differences during fine-tuning.}

\end{definition}

\begin{figure}[t]
    \centering
    \includegraphics[width=0.98\textwidth]{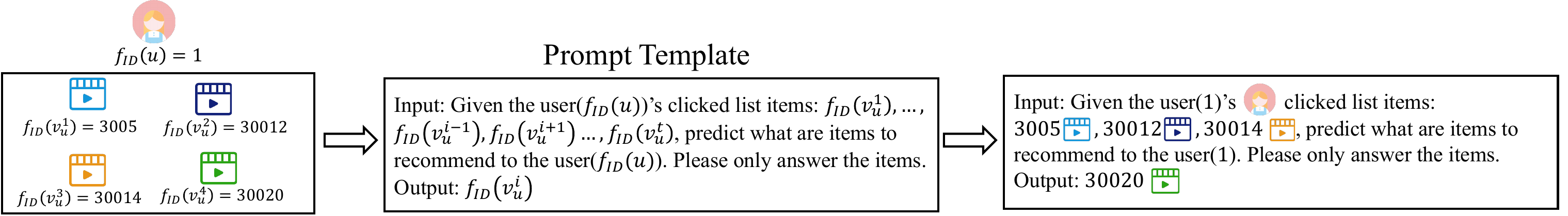} 
    \caption{An example of generating data based on user sequence~$S_u$ and prompt template.}
    \label{fig:example}
\end{figure}

In the prompt template, the input sentence describes the interacted data, imparts contextual information, and clarifies the goal of the recommendation task. The output sentence describes the desired target outcome. In the practical implementation, The "user ID" and "Item ID" are replaced by actual user and item IDs. To offer a more illustrative illustration, consider the following example: User~$u$ ($f_{ID}(u)=1)$) has clicked items $S_u$, $S_u$ includes item~$v_u^1$ ($f_{ID}(v_u^1)=3005$), item~$v_u^2$ ($f_{ID}(v_u^2)=30012$), item~$v_u^3$ ($f_{ID}(v_u^3)=30014$), and item~$v_u^4$ ($f_{ID}(v_1^4)=30020$).
% User (ID: 1) has clicked items $S_1$, $S_1$ includes the items (ID: 3005), (ID: 30012), (ID: 30014), and (ID: 30020). T
the process of generating data according to the prompt template definition~\ref{def:prompt} is shown in Figure~\ref{fig:example}, and the generating data is structured as follows:
% The data by prompt template definition~\ref{def:prompt} is structured as follows:
\begin{myTheo}
\label{exampl1}
~

\begin{itemize}
\item Input: Given the user(1)'s clicked list items:3005, 30012,30014, predict what are items to recommend to the user(1). Please only answer the items.

\item Output: 30020
\end{itemize}

\end{myTheo}
% 例子中可以看到构造的提示数据仅仅包括ID和推荐任务的描述，而不包含任何对用户或产品的文本描述等其他额外信息。
In Example~\ref{exampl1}, the generating prompt data only contains the ID data and description of the recommendation task, without any additional information such as textual descriptions of the user or item.

\subsection{Generating Training Data by Prompt Template} 
%根据提示模版，还需要结合用户交互数据，才能构造出训练数据用于微调大模型。在提示模版中，对于输入部分的产品序列是需要填入具体的产品ID的。由于不同用户的交互产品数量不同，那么填入的产品ID数量也不一致。针对不同长度产品序列ID，我们提出了2种实现方式来更好利用不同长度的用户序列。
% 当用户交互的产品太多，那么将所有交互的产品序列都填入到提示模版中。我们提出了2种实现方式，来更好的填入产品ID。
%一种直接的方式就是将R_u中的所有产品ID都填入到提示模版中，称之为：full generating Training Data ，另外一种就是将R_u中随机采样出固定长度的产品ID填入到提示模版中，称之为：random generating Training Data。

According to the prompt template, the construction of training data for finetuning the LLM requires the combination of user interaction data. Within the prompt template, specific item IDs must be inserted into the input section, designated as "clicked list items: {Item ID, Item ID, ..., Item ID}". It is worth noting that the number of interactive items may vary among different users, resulting in different quantities of item IDs to be filled in. We propose two implementation methods to better utilize clicked list items, $S_u$, with varying lengths.

% One direct way is to fill in full item IDs in $R_u$ into the prompt template, called full generating training data. Another way is to randomly sample fixed-length item IDs from $R_u$ into the prompt template, called random generating training data.
We designed two methods to integrate item IDs from $R_u$ into the prompt template:
1. \textbf{Fully generating training data}: In this approach, we insert the complete item IDs from $R_u$ directly into the prompt template. The training data generated using this method emphasizes the user's overall behavior, enabling LLM to infer which items the user might prefer based on their comprehensive behavior data.
2. \textbf{Randomly generating training data}: Alternatively, we employ a random sampling technique to select fixed-length item IDs from $S_u$ and incorporate them into the prompt template. The training data generated through this method concentrates on the user's local behavior, enabling the LLM to analyze the user's specific, localized behavior data to make inferences about the items that might interest the user.
These two methods offer different ways to incorporate item IDs into the template, allowing for flexibility in generating training data.

% 这两种方法是将项目ID通过提示模版生成训练数据中的有效方法。现在让我们给出例子来说明2种方法的差异和具体实现。

These two methods serve as effective ways to generate training data for incorporating item IDs through prompt templates. Now let's provide intuitive examples to illustrate the differences between the two methods. We use the same example in Section 4.1.1: User~$u$ ($f_{ID}(u)=1)$) has clicked items $S_u$, $S_u$ includes item~$v_u^1$ ($f_{ID}(v_u^1)=3005$), item~$v_u^2$ ($f_{ID}(v_u^2)=30012$), item~$v_u^3$ ($f_{ID}(v_u^3)=30014$), and item~$v_u^4$ ($f_{ID}(v_1^4)=30020$). 

We use \emph{fully generating training data} to generate training data as follows:
% The training data by \emph{full generating training data} is structured as follows:

\begin{myTheo}
\label{exampl2}
~

\begin{itemize}
\item Input: Given the user(1)'s clicked list items:3005, 30012,30014, predict what are items to recommend to the user(1). Please only answer the items.

\item Output: 30020
\end{itemize}
\end{myTheo}

% We use \emph{randomly generated training data} and set the fixed length to 1 to generate training data as follows.
We generate training data using \emph{randomly generated training data} with a fixed length set to 2 as follows:
\begin{myTheo}
\label{exampl3}
~

\begin{itemize}
\item Input: Given the user(1)'s clicked list items:3005,30012 predict what are items to recommend to the user(1). Please only answer the items.

\item Output: 30020
\end{itemize}

\begin{itemize}
\item Input: Given the user(1)'s clicked list items:30012,30014 predict what are items to recommend to the user(1). Please only answer the items.

\item Output: 30020
\end{itemize}

\begin{itemize}
\item Input: Given the user(1)'s clicked list items:30014,3005 predict what are items to recommend to the user(1). Please only answer the items.

\item Output: 30020
\end{itemize}

\end{myTheo}

The output of all examples is item~$v_u^4$~($f_{ID}(v_1^4)=30020$), which can be replaced with any item in item sequence~$S_u$. For the sake of simplicity, we do not provide examples of other items as outputs here.

% From the above example~\ref{exampl2} and~\ref{exampl3}, it can be seen that the two methods we designed generated different training data. Example~\ref{exampl2} captures users' general preferences from their overall behavior, while Example~\ref{exampl3} focuses more on fine-grained information and the relationships between some user behaviors, which can better capture the diversity of user preferences. Consequently, \emph{fully generating training data} and \emph{randomly generated training data} utilize user behavior data from both global and local perspectives, providing sufficient training data for following finetuning of LLM.

From the above Example~\ref{exampl2} and~\ref{exampl3}, it can be seen that the two methods we designed generated different training data. Example~\ref{exampl2} captures users' general preferences from their overall behavior, while Example~\ref{exampl3} focuses more on fine-grained information and the relationships between some user behaviors, which can better capture the diversity of user preferences. Consequently, \emph{fully generating training data} and \emph{randomly generated training data} utilize user behavior data from both global and local perspectives, providing sufficient training data for the following fine-tuning of LLM.

\subsection{Fine-tuning for Data Adaption and Inference for Data Generation} 
Utilizing the provided prompt template and interaction data~$R$, we have the capability to construct ID-based textual data. This constructed data can be effectively employed for training an LLM to model ID-based information. It is worth noting that directly fine-tuning the LLM is a computationally intensive and time-consuming process. To address these challenges, we propose the adoption of an efficient fine-tuning strategy aimed at significantly reducing the number of trainable parameters. One such approach is LoRA, which excels in parameter efficiency. LoRA not only reduces the storage requirements for adapting LLMs to specific tasks but also demonstrates superior performance compared to several other methods.
% Therefore, we employ LoRA to efficiently model ID-based information while maintaining the original parameters in a frozen strategy. Combining the logical reasoning ability of LLM with ID data, that is, understanding and predicting ID data in language models not collaborate filtering.
Consequently, we employ LoRA as an efficient means to model ID-based information while maintaining the original LLM parameters. This approach allows us to leverage the logical reasoning ability of LLM when dealing with ID data, facilitating their understanding and prediction within the language model. This enables us to use the potential of LLM in handling ID-based information, which may not be achieved through collaborative filtering alone. In total, we use the objective of LoRA as follows:

\begin{equation}
\max_{ \Theta } \sum_{u \in U} \sum_{(S_u,\hat{S}_u^n)} \sum_{1}^n \log(P_{\Theta_{LLM}+\Theta_{LoRA}}({S}_u^{sub}|\{S_u-{S}_u^{sub})\}),
\end{equation}
where~$\Theta_{LLM}$ is the original LLM parameters and~$\Theta_{LoRA}$ is the LoRA parameters. The~$S_u$ is the user interaction data, and ${S}_u^{sub}$ is a subset of user interaction data~$S_u$.~$\{S_u-{S}_u^{sub})\}$ represents the data after removing subset~${S}_u^{sub}$ from the original data~$S_u$. For instance, given the~$S_1=\{3005,30012,30014,20020\}$ and the~${S}_1^{sub}=\{30020\}$,then~$\{S_u-{S}_u^{sub})\}=\{3005,30012,30014\}$.

When deployed for inference, we can set~$S_u$ as the full user interaction data and generate corresponding prompts by the prompt template. The resulting output from fine-tuning LLM consists of potential items that a user might click. Due to the inherent diversity and randomness of the output, we employ two filter strategies. The first strategy involves ensuring that the output's item IDs belong to the item set~$V$, thus confirming the validity of the item IDs, rather than containing irrelevant or useless data. In practice, it has been observed that outputs containing minimal ID data, such as just one item ID, often represent random guesses. Therefore, the second strategy is designed to identify outputs that contain multiple item IDs, as these are more likely to be meaningful recommendations. By employing these strategies, we are able to leverage the valid outputs for recommendation tasks, while disregarding the invalid ones.

% \subsection{The LLM-based Generating Loss}
% In the ID-based recommendation model, various formats of loss functions exist, such as pairwise loss, setwise loss, and listwise loss. Distinguishing and Ranking lie at the core of the loss function, as they establish an ordering of target items based on the user preferences. Typically, this component is formulated as the task of identifying the boundary between positive and negative items and depends on this boundary to rank the target items. Such a goal could be achieved by various kinds of loss functions. For example, BPR loss assigns a higher rank to positive items and a lower rank to negative items. Rather than relying on manually designed loss functions, LLMs offer a promising direction to explore in ranking tasks. We use LLM to translate the items slated for ranking into the corresponding loss function.

%就设计了prompts,去直接排序，直接排序很难。我们对pairwise方法，设计了magin的方式。

% mall-scale language models in combination with collaborative knowledge injection?

% the fine-tuned LLM in combination with collaborative information injection

% Injecting the fine-tuned LLM into the ID-based recommendation model.

\subsection{Injecting the Fine-tuned LLM into the ID-based Recommendation Model}
Let's denote the set of interactions generated through the fine-tuned LLM as $R_{LLM}$. Then, we combine the original interaction data $R$ and the generated interaction data $R_{LLM}$ to create the augmented interaction data $R_{aug} = \{R, R_{LLM}\}$. The augmented data $R_{aug}$ relies on fine-tuned LLM to understand ID-based data and generate interaction data, which means that the quality of the augmented data directly reflects the generalization performance of LLM on recommendation tasks and its ability to understand ID-based data. 

To measure the quality of the augmented data~$R_{aug}$, we employ a straightforward approach. We input the augmented data~$R_{aug}$ into an ID-based recommendation model without making any alterations to the recommendation model itself. If the augmented data $R_{aug}$ allows us to train a more accurate ID-based recommendation model, it demonstrates the effectiveness of our data augmentation process. This suggests that finetuning LLM can understand the ID-based data within the recommendation task and output results that align with the task's requirements.

\subsection{Model Discussion}

\subsubsection{Model Analysis.}
%为何有效的分析。
%为了给出更多的见解，我们分析模型为何有效，模型又是在优化什么。一个直接的问题就是，ID对于LLM是没有意义的，或者说仅仅只是个数字。对于一个纯数字，LLM是无法理解到ID的实际意义。

%我们为何模型在优化什么，以至于有效。一个直接的问题就是，ID对于LLM是没有意义的，或者说仅仅只是个数字。对于一个纯数字，LLM是无法理解到ID的实际意义。
%LLM是预测给定context的情况下，nextword的概率，并输出最大概率的word。在LLM中对于每个word或context都转化为了token进行处理。换句话说，LLM是在给定的tokens下，预测并输出最大概率的token。在我们模型中微调大模型也是让大模型在给定的tokens下，预测并输出最大概率的token。由于ID对应的token，
%

%为了提供更多的见解，我们z进行了一项分析，旨在描述模型的有效性和优化目标。将ID数据输入到LLM模型会出现一个主要问题，即，LLM输入的 “ID”是纯数值但是LLM直观上是无法理解数字“ID”在推荐场景中的语义意义。将ID数据输入到LLM直观上是难以相信会有好结果的.如同在其他将LLM用于垂直领域的工作上，一个

%将LLM用于垂直领域都是面临的一个难题，即如何让LLM理解特定领域的知识和概念。例如**2个概念在常识中和医学上的意义并不相同。同样的，我们将LLM用于推荐任务，一个值得关心的问题是如何让LLM理解ID数据在推荐场景的意义，而不是当成一个个的数字去理解。

The primary concern of using LLM in specific domains is how to enable LLM to understand specific domain knowledge and concepts~\cite{gou2023tora,rana2023sayplan,schick2023toolformer,li2023sailer}. For instance, LLMs lack access to external knowledge sources, thereby constraining their efficacy in applications within chemistry~\cite{bran2023augmenting}. Similarly, when we employ LLM for ID-based recommendation, a pertinent concern is how to make LLM understand the meaning of ID data in recommendation scenarios, rather than interpreting them as mere individual numbers. 

Our proposed LLM4Rec is designed to facilitate the organization of ID data, interaction data, and variables in a manner that aligns with user preferences, with the ultimate aim of fine-turning LLM for an enhanced understanding of ID data. Specifically, we are designing different structures for ID relationships in interaction data. For an item list~$R_u$ of user~$u$, we constrain that the expected output ID of LLM is also in the item list~$R_u$ given any subset of the item list~$R_u$. This structure models the correlation between IDs. In addition, it is emphasized in the prompt that it is specific to the current user, ensuring that even if different users input the same subset of the item list into LLM, the ID output from MLM will not be the same. This structure ensures that the correlation between IDs is personalized. Through the above structures, we can ensure that the output ID of LLM is highly correlated with the current input interaction data and the input user after fine-tuning. Therefore, it can be said that LLM can understand ID data in recommendation tasks.

%ID这些数据在LLM是有对应的token的，但是ID直接的关系，在LLM的token之间的关系不成立。
%LLM是预测给定context的情况下，预测next word的概率，并输出最大概率的word。在LLM中对于每个word或context都转化为了token进行处理。换句话说，LLM是在给定的tokens下，预测并输出最大概率的token。在我们模型中微调大模型也是让大模型在给定的tokens下，预测并输出最大概率的token。只不过这里token是对应到ID数据。

% 化学：ChemCrow: Augmenting large-language models with chemistry tools. 
% 数学：ToRA: A Tool-Integrated Reasoning Agent for Mathematical Problem Solving
% 机器人：SayPlan: Grounding Large Language Models using 3D Scene Graphs for Scalable Robot Task Planning
% API的使用：Toolformer: Language Models Can Teach Themselves to Use Tools

% a pre-trained LLM cannot intuitively understand the "ID.
% To offer deeper insights into our research, we undertook an analysis with a specific focus on elucidating the efficacy and optimization objectives inherent in our proposed model. A fundamental issue emerges when one attempts to input identification (ID) data into a pre-trained large language model (LLM), such as Llama. This issue pertains to the inherently numerical nature of the "ID" input within the context of a pre-trained LLM. These models, despite their remarkable capabilities, lack the inherent capacity to discern the semantic significance of the "ID" when embedded within a recommendation dataset. In essence, a pre-trained LLM does not possess the innate ability to intuitively comprehend the meaning and relevance of the "ID" in this context.

\subsubsection{Connections with Previous Works.}
%2种区别，一个是和bert用于rec的区别。另外一种是现在的LLM+RS的区别
%bert用于rec的区别。
%联系都是用预训练的语言模型来适应到推荐任务中。
%区别是LLM对语言理解和序列数据的建模能力更强。

%现在的LLM+RS的区别
%联系是都利用了大模型对推荐任务提出新的next-方案
%他们还是依赖于文本数据，而我们探索的是纯ID数据。从不同角度给人启发。

% For graph collaborative filtering, theconstruction of neighborhood is more important than other collaborative filtering methods [36], since it is based on the graph structure. 

Here, we discuss the correlation between our proposed model and previous research works within the same domain.

While pre-trained language models preceding the emergence of LLMs have been employed in recommendation systems and achieved performance improvements, there are several relations between our proposed approach and previous works: 
1) The fusion of pre-trained language models and recommendation models has previously demonstrated the capacity of language models to effectively model user interaction behavior~(ID data). Building upon this foundation, we delve into the application of LLM to the task of modeling user interaction behavior (ID data). 
2) The most critical difference lies in the different abilities of the language models employed. In contrast to previous pre-trained models like BERT, LLMs constitute a larger and more potent class of models, characterized by their heightened proficiency in modeling sequences and comprehending human language. These stronger modeling and understanding capabilities enable LLMs to capture users' historical interaction data, resulting in better results.

Pure ID data has remained relatively underexplored in LLM-based recommendation systems, despite its fundamental importance in modern recommendation models, such as graph collaborative filtering. In this study, we employ prompting and fine-tuning techniques to effectively incorporate ID data—a departure from existing LLM-based recommendation methodologies.
Previous LLM-based recommendation approaches primarily focus on extracting features or modeling user preferences from textual data. In contrast, our objective is to comprehensively understand ID data and generate useful ID data through fine-tuning LLM. This source data distinguishes our work from prior LLM-based recommendation research.
Moreover, our model deviates from existing methods, where the integration of LLM occurs directly within the recommendation model itself, or the recommendation results. Instead, we introduce prototypes as a data augmentation mechanism and seamlessly integrate them with the recommendation inputs. This novel approach enhances the sparse ID data and provides more data, thereby elevating the overall performance and effectiveness of our recommendation system.

\section{Experiments}
In this section, we conduct experiments to evaluate our proposed method to answer the following research questions:
\begin{enumerate}

  \item [\textbf{Q1}:] Can our proposed LLM4IDRec improve the performance of existing ID-based recommendation models, such as the collaborative filtering model and the sequential recommendation model? If only data augmentation by LLM4IDRec is done, it can improve the performance of existing models~(the collaborative filtering model and the sequential recommendation model), which indicates the effectiveness of LLM4IDRec. 
  % Can our proposed LLM4IDRec improve the performance of existing ID-based recommendation models?  
  
  \item [\textbf{Q2}:] What is the composition of the data generated by our proposed LLM4IDRec? Does LLM4IDRec generate data similar to the test/real data, or does LLM4IDRec generate data that matches user preferences but is different from the test/real data? The latter situation better illustrates that LLM4IDRec provides more information to assist in ID-based recommendation, providing a new solution for applying LLM to ID-based recommendation.

  % What is the composition of the data generated by our proposed LLM4IDRec? 
  % How does our proposed LLM-based generating data (generating training data with LLM) work for the recommendation task?
  % \item [\textbf{Q3}:] How does our proposed LLM-based generating loss (generating loss function with LLM) work for the recommendation task?
  
  \item [\textbf{Q3}:] Has the utilization of LLMs consistently shown advantageous for all users? For example, does LLM4IDRec only improve the performance of a certain user group, or does LLM4IDRec improve the performance of all user groups? Obviously, the latter situation better illustrates that our proposed LLM4IDRec has a more comprehensive understanding of ID data and recommendation tasks.

 % \item [\textbf{Q4}:]

\end{enumerate}

%我们提出的方法是针对ID数据做数据增强，针对ID数据，我们挑选了2个典型的推荐场景，一个是ID行为数据之间没有先后关系，一个是ID行为数据之间有先后关系。为此我们选择了2个没有时间信息的数据集和1个有时间信息的数据集。此外，一个值得注意的事情是，我们的模型只是做数据增强，而不改变现有的模型，因此我们提出的模型是可以适用用任意推荐模型，是和模型无关的方案

Our proposed LLM4IDRec aims to augment ID data, and we validate it in two typical recommendation scenarios: one with no sequential relationship between ID behavior data and the other with a sequential relationship. To verify the generality and effectiveness of LLM4IDRec, we selected two datasets without sequence information and one dataset with sequence information for the experiment.
It's essential to emphasize that our model is only used for data augmentation and does not make any modifications to existing recommendation models. Therefore, our proposed method is model-agnostic and can integrate with any ID-based recommendation model without requiring modifications to the original model.

%增加数据对结果的影响。多个基础模型，2-3个数据集，这样就2-3个表了
%增加的数据和test数据的重合程度，来说明，LLM是生成了新的数据，也就提供了更多的信息。也能说明这些信息是有用的，提升了性能。1个表。
%生成的loss函数和之前的loss优化曲线对比一下，说明我们方法能够有效快速的降低训练时间。？？这点还不太确定。要不就是在不同的loss情况下，1个图。
%消融实验。去掉数据，去掉loss。以及去掉2个部分。1个表。
%一些失败的case study，来说明大模型的不稳定性，对ID数据的处理能力还值得探索。
%人群的对比，不同user group
%

\subsection{Experimental Settings}
\textbf{Datasets.} In our study, we conducted experiments utilizing three publicly available datasets, namely Yelp, Amazon-kindle, and Amazon-beauty. The characteristics of these datasets are outlined in Table~\ref{tab:datasets}. Yelp dataset has been sourced from the 2018 edition of the Yelp challenge. Additionally, the Amazon-kindle and Amazon-beauty datasets have been sourced from Amazon reviews, with a focus on selecting Kindle and beauty from the available collection. During the data preprocessing phase, a common step involves the exclusion of users and items that possess fewer than 10 interaction records. There is no sequential relationship between the ID data in the Yelp and Amazon-kindle datasets. Amazon-beauty is used for sequential recommendation, and there is a sequential relationship between the ID data. {In Amazon-beauty, we adopted a leave-one-out strategy, splitting the data by the last two interactions to create validation and test sets.}

\begin{table}[htb]
    \centering
    % \vspace{-0.3cm}
    \linespread{2}
\setlength\tabcolsep{13pt}{
    \caption{Statistics of the two datasets.} \label{tab:datasets} 
    \begin{tabular}{c|c|c|c|c}
    \hline
        Datasets & Users & Items & Interaction &  Density \\  \hline 
        Yelp & 31,668 & 38,048 & 1,561,406  & 0.130\%  \\  \hline 
        Amazon-kindle & 138,333 & 98,572 & 1,909,965  & 0.014\% \\  \hline 
        Amazon-beauty &22,363  &12,092  &198,502   &0.07\%  \\  \hline 
    \end{tabular} 
   }
\end{table}

\textbf{Evaluation Metrics.}
In our study, our emphasis is on recommending personalized item lists for individual users. To evaluate the efficacy of our recommendation system, we employ two widely ranking metrics, namely, Recall@K and NDCG@K. For each user in the test data, we treat all items with which the user has not interacted as negative items. Then, each recommendation method outputs preference scores for all items, excluding those already identified as positive items in the training data.

\textbf{Baselines.}
To show the effectiveness of our proposed LLM4IDRec, we conducted a series of experiments to compare our method with some representative baselines.

% BPR: It is one of the representative collaborative filtering baselines. It utilizes matrix factorization to model user embeddings and item embeddings with pairwise loss. 
%该模型结构简单，其性能能够直观的反应出数据集质量的高低。因为当数据稀疏的情况下，BPR建模用户偏好就不准确，提升数据质量之后，建模用户偏好就准确些。说明我们的模型在数据增强上的效果。

\begin{itemize}
  \item \textbf{BPR~\cite{rendle2009bpr}:} In collaborative filtering, BPR stands as a representative baseline model. This approach leverages matrix factorization techniques to construct user embeddings and item embeddings while optimizing through pairwise loss functions. The simplicity of the BPR framework offers an intuitive metric for reflecting dataset quality. For example, when compared to modeling user preferences on sparse/low-quality data, BPR models user preferences more accurately and performs better on denser/high-quality datasets.

% SimGCL
% XSimGCL(TKDE'23)
% NCL(WWW'22)

  \item \textbf{SimGCL~\cite{yu2022graph}:} It employs a graph neural network architecture and utilizes contrastive learning to model user-item interactions. This approach proposes a straightforward contrastive learning method that eliminates graph augmentations and, instead, introduces uniform noise into the embedding space to generate contrastive views.

   % NCL(WWW'22)
   \item \textbf{NCL~\cite{lin2022improving}:} It explicitly incorporates potential neighbors into contrastive pairs. It achieves this by introducing the neighbors of a user or an item from both the graph structure and the semantic space. The NCL model effectively captures the neighboring relations among users or items and maximizes the benefits of contrastive learning for the recommendation.

% XSimGCL: It employs a graph neural network architecture to model user-item interactions instead of the inner product used by matrix factorization. We put forward a novel CL-based recommendation model XSimGCL that surpasses its redecessor SimGCL in terms of effectiveness and efficiency. Additionally, we provide theoretical analysis explaining the superiority of XSimGCL through the lens of graph spectrum.

 \item \textbf{XSimGCL~\cite{yu2023xsimgcl}:} XSimGCL extends SimGCL, enhancing its noise-based augmentation approach and streamlining the computation process by unifying the recommendation and contrastive tasks through a shared single pass. XSimGCL outperforms its predecessor, SimGCL, in both effectiveness and efficiency.

% SASRec(ICDM'18 )
  \item \textbf{SASRec~\cite{kang2018self}:} It is a self-attention-based sequential model that combines the benefits of capturing long-term semantics while making predictions based on relatively few actions. Then SASRec can discern the items deemed 'relevant' within a user's action history and utilize this information to predict the subsequent item.

% BERT4Rec(CIKM'19)
\item \textbf{BERT4Rec~\cite{sun2019bert4rec}:} It is a sequential recommendation model that incorporates a bidirectional self-attention network to capture user behavior sequences through the Cloze task. In the context of sequential recommendation, the Cloze objective entails predicting randomly masked items within the sequence while taking into account both the preceding and following context.

% CL4SRec(ICDE'22.)
\item \textbf{CL4SRec~\cite{xie2022contrastive}:} It combines the traditional sequential recommendation task with contrastive learning to model self-supervision signals from the user interaction sequences. This approach allows for the extraction of more meaningful user patterns and enhances the effectiveness of encoding user representations.

% CL4SRec(ICDE'22.)
\item {
\textbf{P5~\cite{geng2022recommendation}:} It is an unified  paradigm which integrates various recommendation related tasks into a
shared conditional language generation framework.}
% combines the traditional sequential recommendation task with contrastive learning to model self-supervision signals from the user interaction sequences. This approach allows for the extraction of more meaningful user patterns and enhances the effectiveness of encoding user representations. 

\item {
\textbf{CID+IID~\cite{hua2023index}:} It is an item indexing method for LLM-based recommendation. We implement the indexing method~(CID+IID) under P5, same as the setting in~\cite{hua2023index}.}

\end{itemize}

Please note that the above baselines are categorized into two groups: one is for modeling Yelp and Kindle datasets without sequence information, including BPR, SimGCL, NCL, XSimGCL and P5. The other group focuses on modeling the Amazon-beauty dataset with sequence information, including SASRec, BERT4Rec, CL4SRec, P5 and CID+IID.

Our proposed LLM4IDRec is model-agnostic and can be applied to ID-based recommendation models. For the sake of convenience, we use \textbf{"?+LLM4IDRec"} to represent different base ID-based recommendation models of our proposed LLM4IDRec. For example, BPR+LLM4IDRec means that we first use LLM4IDRec for data augmentation and then input the augmented data into the BPR model to obtain the final recommendation result.

% The paper suggests that merging original ID data with generated ID data creates augmented data that improves recommendation performance. However, there raises a valid point regarding the potential for similar or even better performance improvements through simple merging strategies involving existing  ID-based recommenders like BERT4Rec and BPR. It is crucial for the authors to address this concern by providing experimental evidence LLM4IDRec is superior to these simpler merging between these small ID-based recommenders.

\textbf{Implementation Details.}
In our proposed LLM4IDRec, we chose the open-source Large Language Model LLama2-7B and used it for fine-tuning. To maintain consistency in the sequence length of the generated training data, we set the maximum length to 2048, and sequences exceeding this length will be truncated. To implement the ID-based recommendation model, we use open-source code without any modifications or customization. { The total number of training tokens used amounts to approximately one billion for Yelp, five billion for Amazon-kindle, and nine hundred million for Amazon-beauty, respectively. Our fine-tuning process involved a maximum of 400 steps with a learning rate of 1e-3. ID data is represented in prompts using the LLama-7B tokenizer, which employs a token-based representation for IDs. In the inference process of LLM4IDRec, we utilized top-p sampling with a set value of top\_p=0.9. Our approach do not involve dividing LLM into separate user groups for individual fine-tuning. Instead, we adopt a unified approach where fine-tuning and data generation were carried out without distinguishing between user groups. This unified fine-tuning process allowed us to evaluate the framework's effectiveness across varying levels of interaction sparsity without bias towards specific user groups.} To ensure a fair performance comparison, we uniformly set the embedding length of all ID-based recommendation models to 64. To obtain a more stable performance evaluation, we ran different random seeds on all methods five times and finally reported the average of the results of these five runs.

\begin{table}[]
\caption{Performance comparisons on Yelp data. LLM4IDRec augments the Yelp dataset by an additional 0.88\% data through data augmentation. Taking the BPR model as an example, we obtain the results for BPR by training it on the original data~$R$. On the other hand, BPR+LLM4IDRec is trained using a dataset~$R_{aug}$ that incorporates a 0.88\% augmentation.}
\label{tab:yelp_overall}
% \linespread{2}
% \setlength\tabcolsep{13pt}{
\begin{tabular}{lcccccc}

\hline
Model                                  & Recall@10 & NDCG@10 & Recall@20 & NDCG@20               & Recall@50 & NDCG@50 \\  \specialrule{0.1em}{2pt}{2pt}  
\multicolumn{1}{l}{BPR}               & 0.02778   & 0.03106 & 0.04914 & 0.03946  &  0.09772   & 0.05788   \\ \specialrule{0.em}{2pt}{2pt}
\multicolumn{1}{l}{BPR+LLM4IDRec}     & 0.02879  & 0.03215 & 0.05111          & 0.04064 & 0.10113           &  0.05927        \\ 
\multicolumn{1}{l}{Improv.}     & 3.64\%   & 3.51\% &  4.01\%         & 2.99\% & 3.49\%  & 2.40\%     \\ \specialrule{0.1em}{3pt}{3pt}

\multicolumn{1}{l}{SimGCL}            & 0.04218   & 0.04839 & 0.07137  & 0.05892 & 0.13569   &  0.08261   \\ \specialrule{0.em}{2pt}{2pt}
\multicolumn{1}{l}{SimGCL+LLM4IDRec}  & 0.04393    & 0.05019 & 0.07371   & 0.06089 &0.13991            & 0.08533         \\ %\specialrule{0.1em}{2pt}{2pt}  
\multicolumn{1}{l}{Improv.}     & 4.15\%   & 3.72\% & 3.28\%           & 3.34\% & 3.11\%           &   3.29\%       \\ \specialrule{0.1em}{3pt}{3pt}

\multicolumn{1}{l}{NCL} &0.03946	&0.04516	&0.06693	&0.05513 &0.12726	&0.0782          \\ \specialrule{0.em}{2pt}{2pt} 
\multicolumn{1}{l}{NCL+LLM4IDRec}    &0.04098	&0.04694	&0.06921	&0.06921 &0.13112	&0.08014         \\ %\specialrule{0.1em}{2pt}{2pt}  
\multicolumn{1}{l}{Improv.}     & 3.85\%   & 3.94\% &3.41\%   &2.55\% &3.03\% &2.48\%          \\ \specialrule{0.1em}{3pt}{3pt}

\multicolumn{1}{l}{XSimGCL}           & 0.04249   & 0.04859 & 0.07234          & 0.05943 &  0.13864          &0.08389          \\ \specialrule{0.em}{3pt}{3pt}
\multicolumn{1}{l}{XSimGCL+LLM4IDRec} & 0.04411   & 0.05036 & 0.07394          & 0.06104 &  0.14075    &  0.08581  \\ %\specialrule{0.1em}{2pt}{2pt} 
\multicolumn{1}{l}{Improv.}     & 3.81\%   & 3.64\% &2.21\%           & 2.71\% &1.52\%            & 2.29\%  \\ \specialrule{0.1em}{3pt}{3pt}

\multicolumn{1}{l}{P5}           & 0.04286  & 0.04951 & 0.07307  & 0.06050 &  0.13974  &0.08558         \\ \specialrule{0.em}{3pt}{3pt}
\multicolumn{1}{l}{P5+LLM4IDRec} & 0.04428   & 0.05110 & 0.07556   & 0.06220 &  0.14395    &  0.08804  \\ %\specialrule{0.1em}{2pt}{2pt} 
\multicolumn{1}{l}{Improv.}     & 3.32\%   & 3.21\% &3.41\%    & 2.81\% &3.01\%  & 2.88\%  \\ \specialrule{0.1em}{3pt}{3pt}

\end{tabular} 
% }
\end{table}

\begin{table}[]
\caption{Performance comparisons on Amazon-kindle data. LLM4IDRec augments the Amazon-kindle dataset by an additional 0.48\% data through data augmentation. Taking the BPR model as an example, we obtain the results for BPR by training it on the original data~$R$. On the other hand, BPR+LLM4IDRec is trained using a dataset~$R_{aug}$ that incorporates a 0.48\% augmentation.}
\label{tab:kindle_overall}
% \linespread{2}
% \setlength\tabcolsep{13pt}{
\begin{tabular}{lcccccc}

\hline
Model                                  & Recall@10 & NDCG@10 & Recall@20 & NDCG@20               & Recall@50 & NDCG@50 \\  \specialrule{0.1em}{2pt}{2pt}  
\multicolumn{1}{l}{BPR}    &0.12292	&0.09161	&0.16725	&0.10502 &0.24249	&0.12452   \\ \specialrule{0.em}{2pt}{2pt}
\multicolumn{1}{l}{BPR+LLM4IDRec}    & 0.13138	& 0.09786	& 0.17447	& 0.11098 & 0.24404	& 0.12912        \\ 
\multicolumn{1}{l}{Improv.}     & 6.88\%   & 6.82\% &  4.32\%  &5.68\% & 0.64\%  & 3.69\%     \\ \specialrule{0.1em}{3pt}{3pt}

\multicolumn{1}{l}{SimGCL}          &0.13876	&0.10146	&0.19028	&0.11691 &0.26872	&0.13728   \\ \specialrule{0.em}{2pt}{2pt}
\multicolumn{1}{l}{SimGCL+LLM4IDRec}  &0.14511	&0.10813	&0.19652	&0.12344 &0.27350	&0.14350        \\ %\specialrule{0.1em}{2pt}{2pt}  
\multicolumn{1}{l}{Improv.}     & 4.58\%   & 6.57\% & 3.28\%   & 5.59\% & 1.78\% & 4.53\%       \\ \specialrule{0.1em}{3pt}{3pt}

\multicolumn{1}{l}{NCL} &0.13676	 & 0.09876	 & 0.18582	 & 0.11364  & 0.26234	 & 0.13368       \\ \specialrule{0.em}{2pt}{2pt} 
\multicolumn{1}{l}{NCL+LLM4IDRec}   &0.14079	& 0.10299	& 0.19014	& 0.11803 & 0.26658	& 0.13801        \\ %\specialrule{0.1em}{2pt}{2pt}  
\multicolumn{1}{l}{Improv.}     & 2.95\%   & 4.28\% &2.32\%   &3.86\% &1.62\% &3.24\%          \\ \specialrule{0.1em}{3pt}{3pt}

\multicolumn{1}{l}{XSimGCL}          &0.14392	&0.10612	&0.19514	&0.12143 &0.27379	&0.14185         \\ \specialrule{0.em}{3pt}{3pt}
\multicolumn{1}{l}{XSimGCL+LLM4IDRec} &0.14976	&0.11235	&0.20147	&0.12776 &0.27956	&0.14810 \\ %\specialrule{0.1em}{2pt}{2pt} 
\multicolumn{1}{l}{Improv.}     & 4.06\%   &5.87\% &3.24\%  &5.21\% &2.11\% &4.41\%  \\ \specialrule{0.1em}{3pt}{3pt}

\multicolumn{1}{l}{P5}    &0.14536	&0.12361	&0.19809	&0.14244 &0.27554	&0.16123        \\ \specialrule{0.em}{3pt}{3pt}
\multicolumn{1}{l}{P5+LLM4IDRec} & 0.15212   & 0.13111 & 0.20488  & 0.15016 & 0.27849   &  0.16673 \\ %\specialrule{0.1em}{2pt}{2pt} 
\multicolumn{1}{l}{Improv.}     & 4.65\%   & 6.07\% &3.43\%    & 5.42\% &1.07\%  & 3.41\%  \\ \specialrule{0.1em}{3pt}{3pt}

\end{tabular} 
% }
\end{table}

\begin{table}[]
\caption{Performance comparisons on Amazon-beauty data. LLM4IDRec augments the Amazon-beauty dataset by an additional 12.70\% data through data augmentation. Taking the BPR model as an example, we obtain the results for BPR by training it on the original data~$R$. On the other hand, BPR+LLM4IDRec is trained using a dataset~$R_{aug}$ that incorporates a 12.70\% augmentation.}
\label{tab:beauty_overall}
% \linespread{2}
% \setlength\tabcolsep{13pt}{
\begin{tabular}{lcccccc}

\hline
Model                                  & Recall@10 & NDCG@10 & Recall@20 & NDCG@20               & Recall@50 & NDCG@50 \\  \specialrule{0.1em}{2pt}{2pt}  
\multicolumn{1}{l}{SASRec}  &0.04991 &0.03284	&0.06771 &0.03834 &0.10637	&0.04926  \\ \specialrule{0.em}{2pt}{2pt}
\multicolumn{1}{l}{SASRec+LLM4IDRec}    &0.05446	&0.03798	&0.07271	&0.04257 &0.11439	&0.0508        \\ 
\multicolumn{1}{l}{Improv.}     & 9.12\%   & 15.65\% &  7.38\% & 11.03\% & 7.54\%  & 3.13\%     \\ \specialrule{0.1em}{3pt}{3pt}

\multicolumn{1}{l}{BERT4Rec}    &0.0351	&0.01461	&0.06055	&0.02099   & 0.10897	   & 0.03055             \\ \specialrule{0.em}{2pt}{2pt} 
\multicolumn{1}{l}{BERT4Rec+LLM4IDRec}     	&0.03839		&0.01705		&0.06544		&0.02385 	&0.12008	&0.03464   \\ %\specialrule{0.1em}{2pt}{2pt}  
\multicolumn{1}{l}{Improv.}   & 9.37\% & 16.70\% &8.08\%  & 13.63\% &10.19\% &13.39\%          \\ \specialrule{0.1em}{3pt}{3pt}

\multicolumn{1}{l}{CL4SRec}  &0.05224	&0.03222	&0.07268	&0.03756 &0.11326	&0.04524  \\ \specialrule{0.em}{2pt}{2pt}
\multicolumn{1}{l}{CL4SRec+LLM4IDRec}  &0.05455	&0.03736	&0.07642	&0.04283 &0.1236	&0.05214    \\ %\specialrule{0.1em}{2pt}{2pt}  
\multicolumn{1}{l}{Improv.}     & 4.42\%   & 15.95\% & 5.15\%  & 14.03\% & 9.13\%           &  15.25\%       \\ \specialrule{0.1em}{3pt}{3pt}

\multicolumn{1}{l}{P5}  &0.05011 &0.03324	&0.06861 &0.03944 &0.11042	&0.05031  \\ \specialrule{0.em}{2pt}{2pt}
\multicolumn{1}{l}{P5+LLM4IDRec}    &0.05417	&0.03734	&0.07359	&0.03984 &0.12073	&0.05613        \\ 
\multicolumn{1}{l}{Improv.}     & 8.11\%   & 12.35\% &  7.26\% & 10.17\% & 9.34\%  & 11.56\%     \\ \specialrule{0.1em}{3pt}{3pt}

\multicolumn{1}{l}{CID+IID}  &0.05010 &0.03295	&0.06881 &0.03966 &0.10976	&0.04988  \\ \specialrule{0.em}{2pt}{2pt}
\multicolumn{1}{l}{CID+IID+LLM4IDRec}    &0.05334	&0.03645	&0.07251	&0.04355 &0.11771	&0.05549        \\ 
\multicolumn{1}{l}{Improv.}     & 6.47\%   & 10.62\% &  5.38\% & 9.81\% & 7.24\%  & 11.25\%     \\ \specialrule{0.1em}{3pt}{3pt}

\end{tabular} 
% }
\end{table}

\subsection{Performance Comparison~(Q1)}
Table~\ref{tab:yelp_overall},~\ref{tab:beauty_overall} and~\ref{tab:kindle_overall} present the overall performance comparisons. According to the results, we have the following observations:

\begin{itemize}
    \item \textbf{LLM4IDRec can generally improve performance across all metrics and baselines.} Our proposed LLM4IDRec improves the performance of all ID-based recommendation models by leveraging the power of LLM to augment input data. To illustrate this, let's consider the BPR model as an example. When considering the BPR model, its input is represented as $R$, while the input of the BPR+LLM4IDRec model is represented as $R_{aug}$. The only difference between BPR and BPR+LLM4IDRec lies in the input data. Compared with the BPR model, the BPR+LLM4IDRec model achieved better performance with an average performance improvement of 3\%. LLM4IDRec also achieved better performance on all baselines. These observations indicate that LLM is capable of generating and satisfying data for recommendation tasks, thereby having a positive impact on overall recommendation results. It is essential for LLM to possess an understanding of ID-based data and recommendation tasks to generate data that aligns with the recommendation task's goals and effectively improves recommendation performance. Otherwise, either LLM cannot generate data that is consistent with the recommendation task, or LLM can generate data that meets the requirements of the task but fails to improve recommendation performance. In light of the above findings, we can answer the two questions raised in the introduction section. We designed the LLM4IDRec model and verified through experiments that LLM can generate data that meets the requirements of recommendation tasks and demonstrates its potential to enhance ID-based recommendation models.
    
    % Our proposed LLM4IDRec significantly outperforms all ID-based recommendation models by enhancing the input data with the power of LLM. Especially in sequence recommendation tasks, our proposed model shows a higher performance improvement. When considering the BPR model, its input is represented as~$R$, while the input for the BPR+LLM4IDRec model is denoted as $R_{aug}$. the only difference between BPR and BPR+LLM4IDRec is the input. The augmented data $R_{aug}$ generated through LLM, directly enhances the recommendation performance. This observation shows LLM's ability to generate data that aligns with the recommendation task, thereby positively influencing the overall recommendation results. This phenomenon illustrates the capacity of LLM to comprehend both ID-based data and recommendation tasks. Otherwise, where LLM fails to generate data in alignment with the recommendation task or generates data that meets the task's requirements but fails to improve recommendation performance, the limitations of LLM become apparent.

    \item  \textbf{LLM4IDRec achieves significant improvement in the sequential recommendation.} The Yelp and Amazon-kindle datasets are general ID-based recommendation data without sequential interaction behavior relationships. In contrast, the Amazon-beauty dataset is ID-based sequential data with clear sequential interaction patterns. As shown in Tables~\ref{tab:yelp_overall} and~\ref{tab:beauty_overall}, our proposed LLM4IDRec consistently improves performance across the Yelp and Beauty datasets, achieving an average improvement of approximately 3\% over all baselines. Notably, on the Amazon-beauty dataset, LLM4IDRec demonstrates significant performance gains averaging around 9\% across all baselines.
    Our proposed LLM4IDRec model achieves the most significant enhancement on the Amazon-beauty dataset. This is primarily attributed to the inherent strengths of the LLM itself as a model designed for processing language sequences. With its pre-training on a vast corpus of language data, LLM exhibits a superior capacity for understanding and modeling sequential data. Consequently, our proposed LLM-based approach, LLM4IDRec, performs better in the sequential recommendation.

    \item \textbf{Comparing the NDCG and Recall metrics to show the LLM4IDRec's ability.}Table~\ref{tab:beauty_overall} and~\ref{tab:kindle_overall} show that the average improvement of our proposed LLM4IDRec on the NDCG metric is higher than the average improvement on the Recall metric. For example, LLM4IDRec on the Amazon-beauty dataset shows an average improvement of 13\% on the NDCG metric and 7.8\% on the Recall metric. These results indicate that LLM4IDRec not only increases the recall of relevant items but also elevates their ranking, resulting in more accurate rankings and, consequently, a higher NDCG improvement. Achieving more accurate ranking results hinges on precise modeling of user preferences. Remarkably, we incorporated only a minimal amount of generated data via LLM4IDRec, building upon the original dataset and thereby enhancing the baseline's ability to capture user preferences accurately. This situation confirms that the limited data generated by LLM4IDRec aligns with user preferences and aids in refining user preference modeling, thus validating the rationality and effectiveness of our proposed LLM4IDRec.

    \item \textbf{Comparing different top K values on Recall@K and NDCG@K metrics.} From Tables~\ref{tab:yelp_overall} and~\ref{tab:kindle_overall}, it can be observed that the performance of our proposed LLM4IDRec mostly shows a smaller improvement as the K value increases. That is to say, LLM4IDRec achieves the highest performance improvement on the NDCG@10 and Recall@10. For example, in Tables~\ref{tab:yelp_overall}, LLM4IDRec shows an average improvement of 3.70\% on the NDCG@10, an average improvement of 2.89\% on the NDCG@20, and an average improvement of 2.61\% on the NDCG@50. The results indicate that LLM4IDRec is more accurate in ranking the top items. In addition, LLM4IDRec still achieves stable improvement on all K values. These experiments highlight that our method enhances all baseline accuracy even with limited data, resulting in an effective and stable improvement in recommendation performance.

    \item  \textbf{Discussion of some outlier results.} In Table~\ref{tab:yelp_overall}, when examining the Recall@50 for XSimGCL+LLM4IDRec, we observe that the increase is smaller compared to improvements seen in other metrics. A similar pattern emerges in Table~\ref{tab:beauty_overall}, where SASRec+LLM4IDRec shows a smaller improvement on NDCG@50 compared to other metrics. This trend also appears in Table~\ref{tab:kindle_overall}, where, for BPR+LLM4IDRec, the increase on Recall@50 is less than the improvement in other metrics.
    One potential explanation for these outlier results lies in the inherent instability and randomness associated with LLM. The outcomes generated by LLM exhibit randomness, which can introduce noise into the final results. Despite our efforts to design filtering strategies, it remains challenging to completely eliminate noise from the generated data~$R_{LLM}$. Moreover, the length~$|R_{LLM}|$ that has been added to the augmented data~$R_{aug}=R+R_{LLM}$ is relatively small, resulting in a correspondingly limited amount of noise. The impact of this minimal noise on the results primarily becomes noticeable when K=50. Nevertheless, it's worth noting that these outliers are infrequent, and overall, LLM4IDRec achieves superior performance across all metrics.

\end{itemize}

%Recall和NDCG的对比，
%不同任务的对比，
%不同数据集的对比，to do
%不同K的对比
%不同基础模型的对比,to do
%Recall和NDCG的对比
%讨论一下异常值。

\begin{figure}[t]
    \centering
    \includegraphics[width=0.95\textwidth]{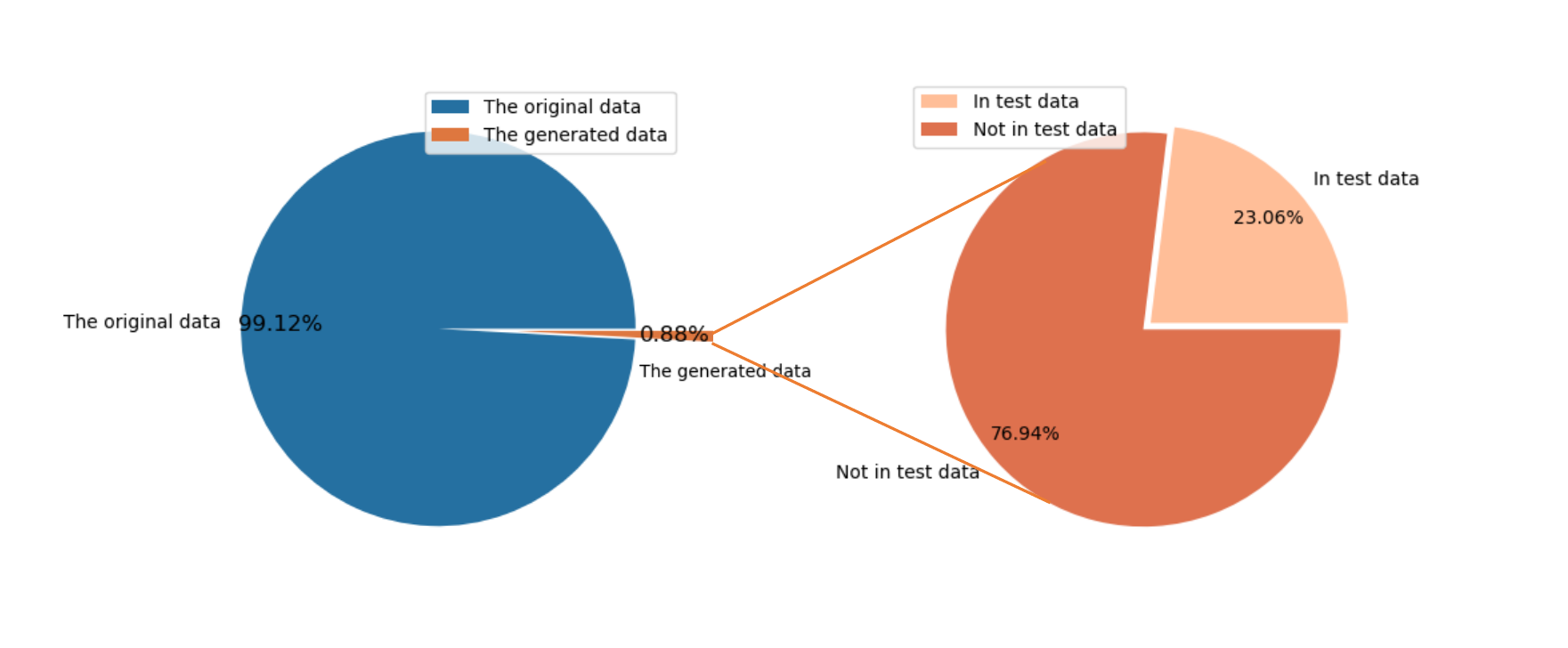} 
    \caption{Analysis of augmented data~$R_{aug}$ composition on Yelp data. Augmented data~$R_{aug}$ consists of the original data~$R$ and the generated data~$R_{LLM}$ by LLM4IDRec. The generated data~$R_{LLM}$ only accounts for 0.88\% of the total data~$R_{aug}$. Only 23.06\% of the generated data~$R_{LLM}$ coincides with the test data on Yelp data.}
    \label{fig:yelp_data}
\end{figure}

\begin{figure}[t]
    \centering
    \includegraphics[width=0.95\textwidth]{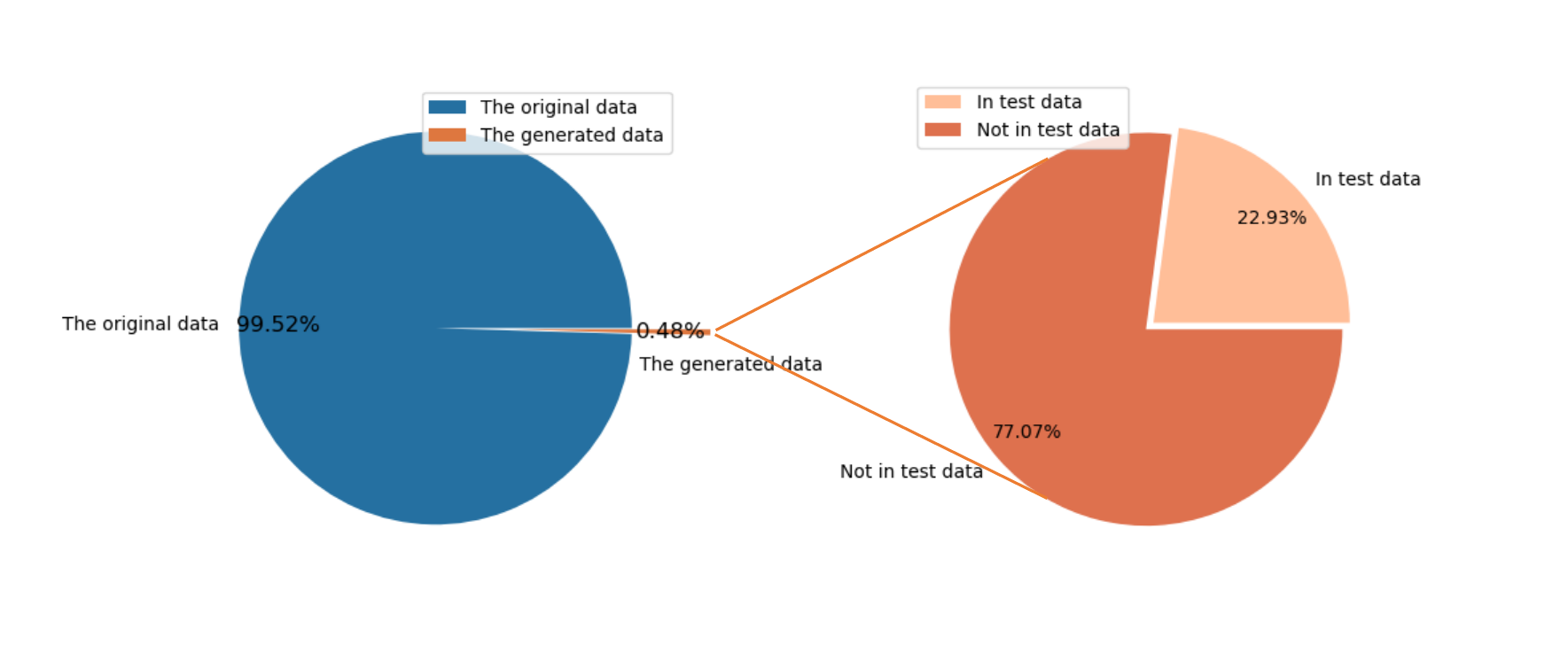} 
    \caption{Analysis of augmented data~$R_{aug}$ composition on Amazon-kindle data. Augmented data~$R_{aug}$ consists of the original data~$R$ and the generated data~$R_{LLM}$ by LLM4IDRec. The generated data~$R_{LLM}$ only accounts for 0.48\% of the total data~$R_{aug}$. Only 22.93\% of the generated data~$R_{LLM}$ coincides with the test data on Amazon-kindle data.}
    \label{fig:kindle_add}
\end{figure}

\begin{figure}[t]
    \centering
    \includegraphics[width=0.95\textwidth]{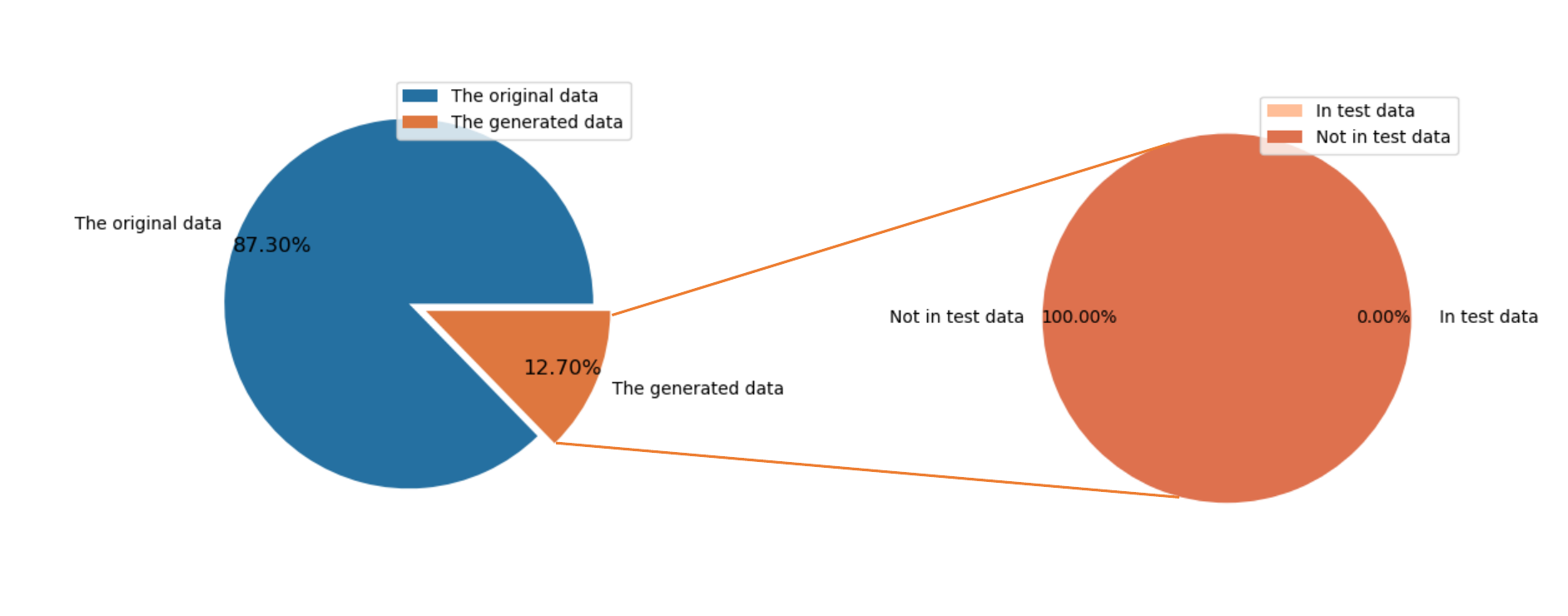} 
    \caption{Analysis of augmented data~$R_{aug}$ composition on Amazon-beauty data. Augmented data~$R_{aug}$ consists of the original data~$R$ and the generated data~$R_{LLM}$ by LLM4IDRec. The generated data~$R_{LLM}$ accounts for 12.70\% of the total data~$R_{aug}$. There is no overlap between the generated data and the test data on Amazon-beauty data.}
    \label{fig:beauty_add}
\end{figure}

\subsection{Analysis of Augmented Data~$R_{aug}$ Composition~(Q2)}
Our proposed LLM4IDRec generates additional data~$R_{LLM}$ to augment the original dataset~$R$. These additional data~$R_{LLM}$ are generated by finetuning the LLM based on user interaction behavior and the description of recommendation tasks. We not only compare the overall recommendation performance to evaluate the effectiveness of generated data and the performance of LLM4IDRec, but we also need to do an in-depth analysis of the generated data, as this is crucial for evaluating the performance of LLM4IDRec. In this section, we investigate in detail the data generated on different datasets to further understand their composition.

\begin{itemize}
    \item From Figures~\ref{fig:yelp_data} and~\ref{fig:kindle_add}, it can be observed that the proportion of generated data $R_{LLM}$ to augmented data $R_{aug}$ is 0.88\% on the Yelp dataset and 0.48\% on the Amazon-kindle dataset, respectively. The purpose of generated data is to improve recommendation performance. By combining the results from Table~\ref{tab:yelp_overall} and Table~\ref{tab:kindle_overall}, it can be illustrated that there is a significant performance improvement when the data volume is added by less than 1\%. This phenomenon shows that the data generated by our model on the Yelp and Amazon-kindle datasets is closely related to actual user preferences and the recommendation task, ensuring that the generated data is valuable for modeling user preferences and improving performance.

    \item  %fig4.5生成的数据中只有20的数据和test数据重合，而80数据都是无关的。尽管生成的数据大多是都是无关的，但是依旧能提升推荐性能，也就是说明了生成数据的是符合用户兴趣的。这样也就说明了我们生成数据的合理性。
    The generated data in Figures 4 and 5 only have approximately 20\% overlap with the test data, while 80\% of the generated data is independent of the test data. This phenomenon can be analyzed and explained in depth from two perspectives. Firstly, considering the low overlap between the generated data and the test data, there is still 20\% of the data that matches. This indicates that LLM4IDRec can learn some patterns and features from the user's historical data (training data~$R$), thereby generating content consistent with the test data. This further confirms that LLM4IDRec has a certain ability to understand user historical behavior and can generate relevant data according to the requirements of recommendation tasks. This can be seen as evidence that LLM4IDRec accurately infers user interests to a certain extent. Secondly, although most of the generated data is unrelated to the test data, it is still important. These generated data are used to synthesize augmented data~$R_{aug}$, significantly improving recommendation performance on datasets such as Yelp and Amazon-kindle datasets. This indicates that although the generated data have a low correlation with the test data, they are still related to the user's interests and historical behavior, thus helping to improve the accuracy of recommendation results. This also shows the rationality of generating data, as they reflect, to some extent, the interests and behavioral patterns of users. Overall, by analyzing the composition of the generated data~$R_{LLM}$, we can conclude that the LLM4IDRec model can understand ID-based data and infer recommendation results that match the user's current behavior. Although the overlap between generated data and test data is low, they still have value and can be used to improve recommendation performance, reflecting the model's understanding of user interests. This further confirms the effectiveness and potential of the model.
    
    \item  %fig6生成的数据相比另外2个数据集多一点，但是总体来说并不多。特别是生成的数据和test数据一点关系都没有。
    From Figure~\ref{fig:beauty_add}, it can be seen that the generated data~$R_{LLM}$ only accounts for 12.7\% of the augmented data~$R_{aug}$. Although the amount of data generated for the Amazon-beauty dataset is significantly larger than that for the Yelp and Amazon-kindle datasets, the amount of generated data~$R_{LLM}$ is relatively small. Here, we will further analyze why more data needs to be generated for the beauty Amazon-dataset and provide more reasons. Firstly, the Amazon-beauty dataset is for the sequential recommendation task that predicts the next item. In sequential recommendation, the diversity of user interests is usually high. This means that users may be interested in various types of items, and these interests may change over time. Therefore, in order to better model user interests, we need more data to cover different situations and user behavior patterns. Generating more data can help us gain a more comprehensive understanding of user interests and behaviors, thereby improving the performance of recommendation systems. Secondly, the Amazon-beauty dataset is smaller than the Yelp and Amazon-kindle datasets, making it more difficult to model user interests. Generating more data can provide more information, making it easier for the model to capture potential user interests and item correlations. In the case of limited data, the model is prone to overfitting the training data, resulting in a decrease in performance on new data. By increasing the amount of data, over-fitting problems can be alleviated, making the model more generalizable and better adaptable to unseen users and products. In summary, generating more data is crucial for improving recommendation performance on low-volume Amazon-beauty datasets. It can better capture the diversity of user interests, improve model performance, alleviate overfitting problems, and provide users with more accurate and personalized recommendation services. Moreover, Table~\ref{tab:beauty_overall} shows the results on Amazon-beauty dataset, it can be seen that adding more data has a more significant improvement in recommendation performance compared to the Yelp and Amazon-kindle datasets. This also reflects the rationality of generating more data for the Amazon-beauty dataset.

    \item %%fig6生成的数据相比另外2个数据集多一点，但是总体来说并不多。特别是生成的数据和test数据一点关系都没有。
    %%fig6生成的数据和test数据无重合，不同于在yelp和kindle数据集上生成的数据和test存在重合。
    % fig6生成的数据和test数据无重合，一方面是由于beauty数据集是序列推荐任务，test数据集仅包括next-item，所以数量很少。这样也就导致了生成的数据和test数据无重合的现象。另外一方面，尽管生成的数据和test数据集没有重合但是对推荐性能的提升依旧很明显，从table2可以看到。这也说明了生成数据的合理性。
    In Figure~\ref{fig:beauty_add}, the phenomenon of no overlap between the generated data and the test data can be attributed to multiple reasons. Firstly, this is because the Amazon-beauty dataset is for the sequential recommendation, while the test dataset only contains the next item in the user behavior sequence, so the test data only contains a very limited amount of information. The test dataset on Amazon-beauty is relatively small, and the generated data may contain other items that users may like, which can make it more difficult to overlap between the generated data and the test data. However, although there is no overlap between the generated data and the test data, it can be seen from the results in Table~\ref{tab:beauty_overall} that the generated data still significantly improves recommendation performance. This indicates that the generated data is effective in improving the performance of recommendation systems. This may be because the generated data can capture more user behavior patterns and preferences, providing more information for the ID-based recommendation model. In addition, the generated data can also be used to expand the size of the training dataset, improve the model's generalization ability, and further enhance recommendation performance. In summary, although there is no overlap between the generated data and the test data, this does not affect the improvement of recommendation performance by the generated data. The rationality of generating data can be verified through its positive impact on model performance, which provides strong support for using LLM to improve ID-based recommendation models.

\end{itemize}

% 综合来看，描述了一个研究的方法，旨在通过生成额外的数据来改进推荐系统的性能。强调了对生成数据的深入分析的重要性，以更好地理解其对模型性能的影响。这个分析可以帮助研究人员优化生成数据的生成过程，以更好地满足推荐任务的需求。

\begin{figure}[ht]
\centering
\includegraphics[width=\textwidth]{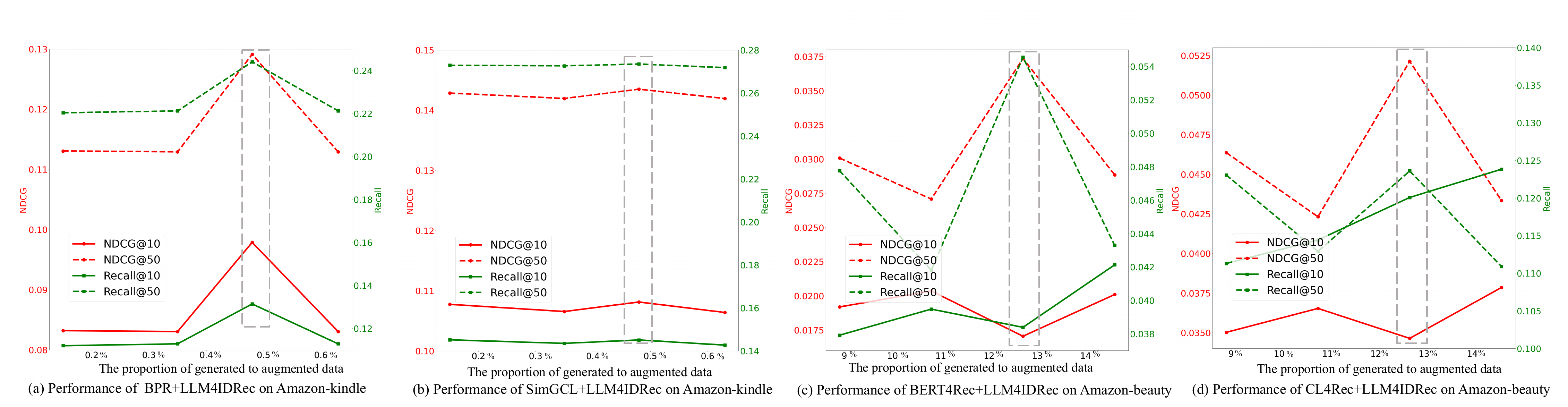}
\caption{The performance of LLM4IDRec across the different sizes of generated training data. The results reported in the previous version are framed by gray dashed lines.} 
\label{fig:ratio}
\end{figure}

{
\textbf{Influence of Augmented Data~$R_{aug}$ Size}
We conduct comparisons across different sizes of generated training data to provide more insights, as shown in Figure~\ref{fig:ratio}. For simplicity, we select several models (BPR+LLM4IDRec, SimGCL+LLM4IDRec, BERT4Rec+LLM4IDRec, CL4Rec+LLM4IDRec) to present their results. It's worth noting that other models (such as NCL+LLM4IDRec, XSimGCL+LLM4IDRec, SASRec+LLM4IDRec) exhibit a similar trend in results. From the results in Fig.~\ref{fig:ratio}, it can be seen that the amount of generated data we have presented in the previous version almost achieves optimal performance. This is because our model currently does not constrain the amount of generated data. If the amount of generated data is forcibly increased or decreased, the naturally generated noise data will increase, leading to performance degradation. This observation has inspired us to enhance the proposed LLM4IDRec to generate more effective data.
}

\begin{figure}[ht]
    \centering
    \includegraphics[width=0.9\textwidth]{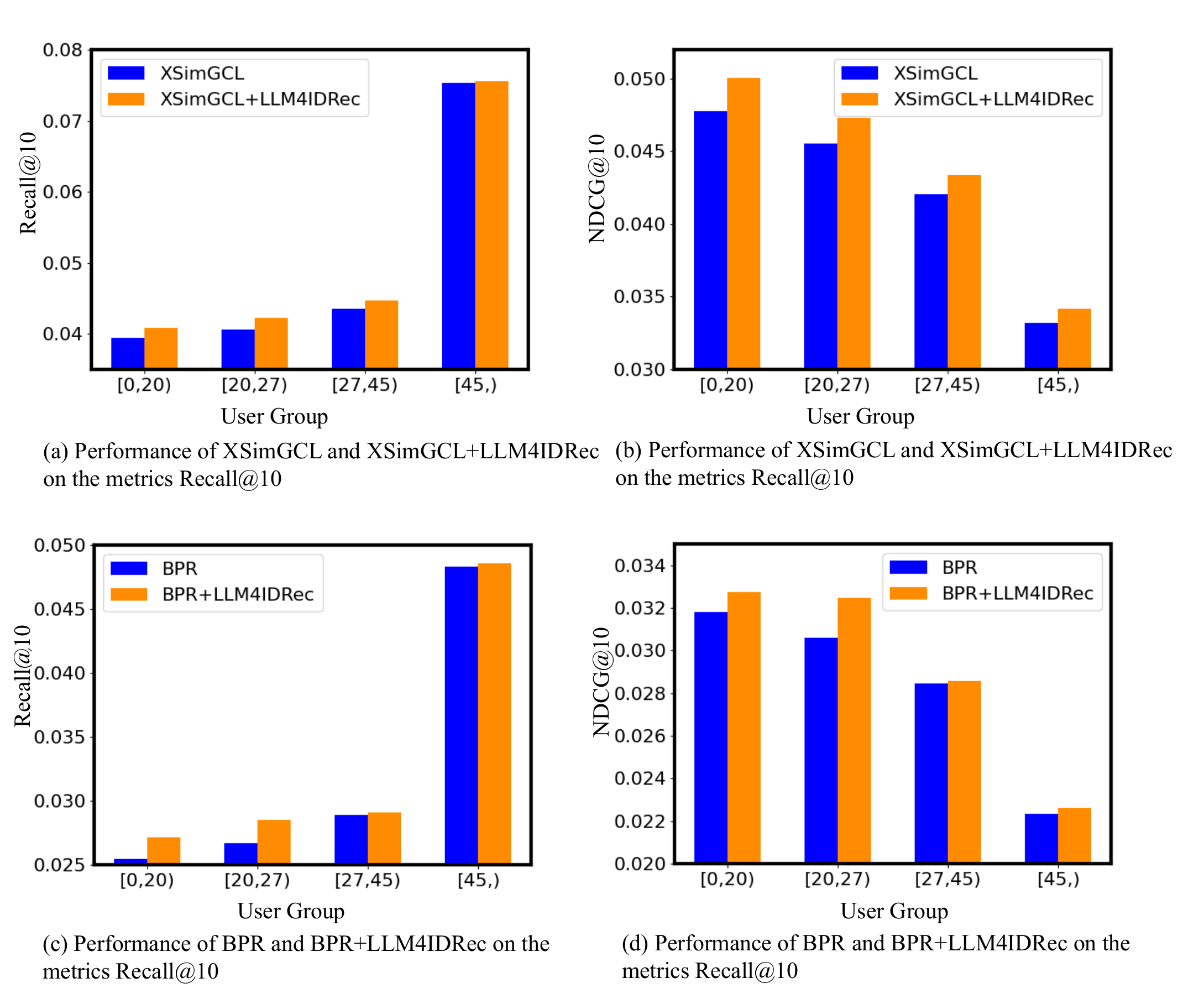} 
    \caption{Performance comparison across user groups on Yelp Dataset.}
    % Performance comparison under different sparsity of user groups on Yelp dataset.}
    \label{fig:yelp_top10_group}
\end{figure}

\begin{figure}[ht]
    \centering
    \includegraphics[width=0.9\textwidth]{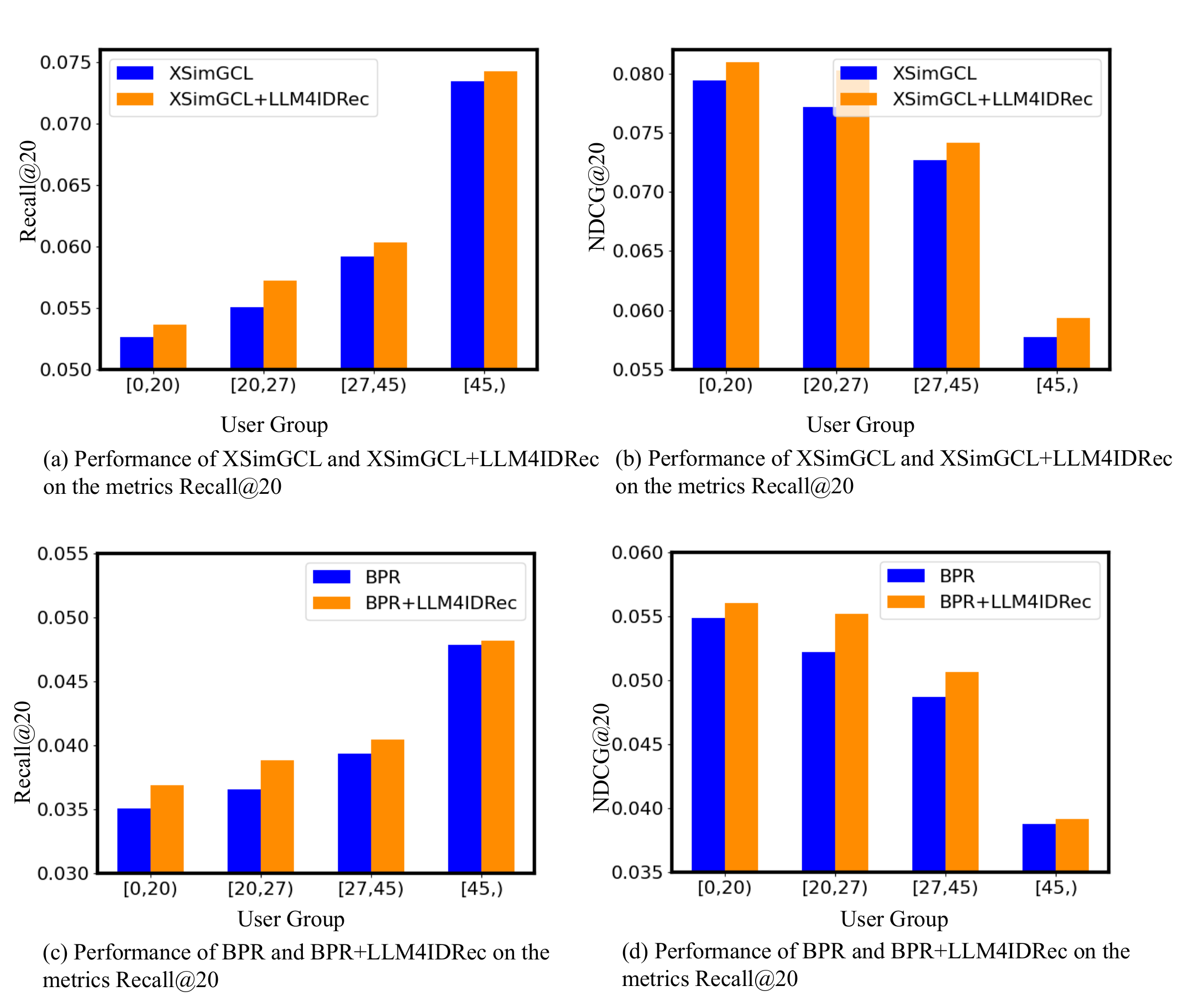} 
    \caption{Performance comparison across user groups on Yelp Dataset.}
    \label{fig:yelp_top20_group}
\end{figure}

\begin{figure}[ht]
    \centering
    \includegraphics[width=0.9\textwidth]{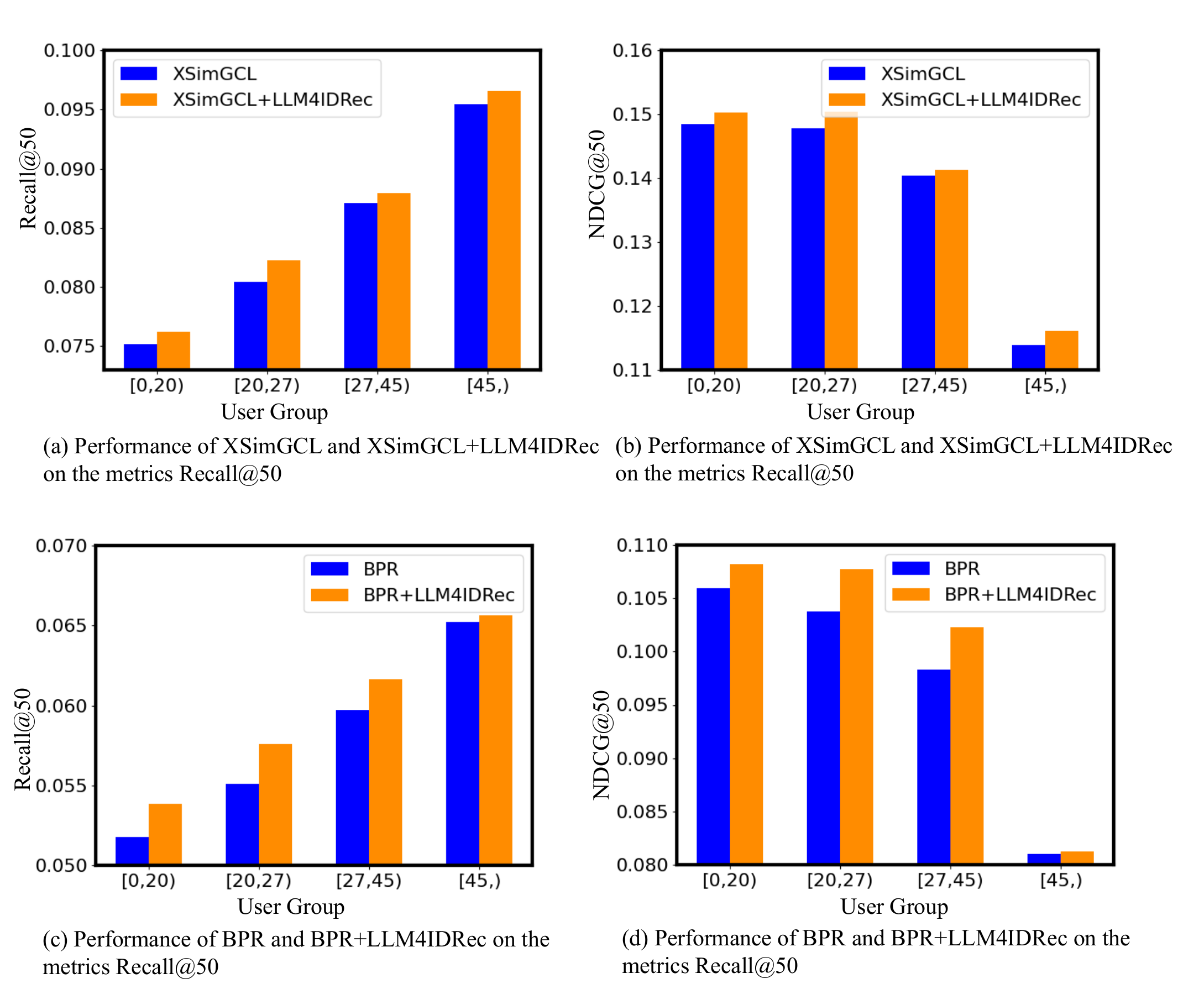} 
    \caption{Performance comparison across user groups on Yelp Dataset.}
    \label{fig:yelp_top50_group}
\end{figure}

\subsection{Comparing User Groups of Different Sparsity Levels~(Q3)}

The data sparsity issue has always been a significant concern in recommendation systems. In practical applications, most users tend to interact with only a few items, while a large portion of items receive minimal user engagement. Consequently, this results in sparse user-item interaction data~$R$, making it challenging to model user preferences accurately. Our proposed solution, LLM4IDRec, can address the sparsity issue through the utilization of LLM-based data augmentation.

To provide a clearer explanation, we categorize user groups based on their sparsity levels. This categorization allows us to gain the performance of different user groups. Sparsity levels are determined by the number of interactions per user. Based on these sparsity levels, we divided the users in the training set into four groups, each containing an equal number of users. Using the Yelp dataset as an example, these user groups exhibited interaction frequencies of less than 20, 27, 45, and more than 45, respectively. It's worth noting that while we have chosen to present the results of LLM4IDRec using the Yelp dataset to reinforce our points, we achieved similar outcomes with other datasets as well. Due to similar trends, we omit the details related to those datasets.

In Figures~\ref{fig:yelp_top10_group},~\ref{fig:yelp_top20_group}, and~\ref{fig:yelp_top50_group}, we show the results for different user groups across various K values (K=10, 20, 50). These results reveal several critical observations:

\begin{itemize}
    \item \textbf{Performance improvement across all user groups}: Regardless of the user group's sparsity level, all of them demonstrated improved performance. This underscores the fact that the augmented data~$R_{aug}$ obtained through LLM4IDRec effectively mitigates the sparsity issue. This result emphasizes the universality of LLM4IDRec, which can generate recommended items that match the preferences of various types of users, thus illustrating the rationality and feasibility of LLM4IDRec.

    \item \textbf{Consistency across different backbone models}: We noticed consistent trends across different backbone models, such as BPR and XSimGCL. This indicates that the augmented data~$R_{aug}$ obtained through LLM4IDRec has a similar positive impact on various ID-based recommendation models. Regardless of the specific ID-based model used, LLM4IDRec assists in more accurately modeling user preferences. This further supports the notion that the data generated by LLM4IDRec holds meaningful value for users, as it results in recommendations that align with their preferences.
\end{itemize}

In summary, these experiments provide robust results supporting the feasibility and effectiveness of employing LLM to alleviate sparsity issues in ID-based recommendation models.
% LLM4IDRec is anticipated to deliver more precise recommendations for inactive users and sparse data, ultimately enhancing the performance of personalized recommendation systems.

\begin{table}[]
\caption{Performance comparisons on Amazon-beauty data. The "?+SASRec" represents utilizing SASRec as data augmentation, integrating different ID-based recommendation models. Taking the BERT4Rec+SASRec model as an example, we use SASRec to augment the interaction data, then train BERT4Rec on the augmented data and generate recommended items. Please note that this process is the same as BERT4Rec+LLM4IDRec.}
\label{tab:beauty_sasrec}
% \linespread{2}
% \setlength\tabcolsep{13pt}{
\begin{tabular}{lcccccc}

\hline
Model                                  & Recall@10 & NDCG@10 & Recall@20 & NDCG@20               & Recall@50 & NDCG@50 \\  \specialrule{0.1em}{2pt}{2pt}  
\multicolumn{1}{l}{SASRec}  &0.04991 &0.03284	&0.06771 &0.03834 &0.10637	&0.04926  \\ \specialrule{0.em}{2pt}{2pt}
\multicolumn{1}{l}{SASRec+SASRec}     & 0.02978   & 0.01189 & 0.05572 & 0.01844 & 0.10723  & 0.02859 \\ 
\multicolumn{1}{l}{SASRec+LLM4IDRec}    &0.05446	&0.03798	&0.07271	&0.04257 &0.11439	&0.0508        \\ 
\specialrule{0.1em}{3pt}{3pt}

\multicolumn{1}{l}{BERT4Rec}    &0.0351	&0.01461	&0.06055	&0.02099   & 0.10897	   & 0.03055             \\ \specialrule{0.em}{2pt}{2pt} 
\multicolumn{1}{l}{BERT4Rec+SRSRec}   & 0.02603 & 0.0114 &0.047 & 0.01668 &0.09113 &0.0254         \\ 
\multicolumn{1}{l}{BERT4Rec+LLM4IDRec}     	&0.03839		&0.01705		&0.06544		&0.02385 	&0.12008	&0.03464   \\ %\specialrule{0.1em}{2pt}{2pt}  
\specialrule{0.1em}{3pt}{3pt}

\multicolumn{1}{l}{CL4SRec}  &0.05224	&0.03222	&0.07268	&0.03756 &0.11326	&0.04524  \\ \specialrule{0.em}{2pt}{2pt}
\multicolumn{1}{l}{CL4SRec+SASRec}     & 0.02938   & 0.01149 & 0.05442  & 0.01777 & 0.10401  &  0.02753      \\ 
\multicolumn{1}{l}{CL4SRec+LLM4IDRec}  &0.05455	&0.03736	&0.07642	&0.04283 &0.1236	&0.05214    \\ %\specialrule{0.1em}{2pt}{2pt}  
\specialrule{0.1em}{3pt}{3pt}

\end{tabular} 
% }
\end{table}

\begin{table}[]
\caption{Performance comparisons on Yelp data. The "?+BPR" represents utilizing BPR as data augmentation, integrating different ID-based recommendation models. Taking the SimGCL+BPR model as an example, we use BPR to augment the interaction data, then train SimGCL on the augmented data and generate recommended items. Please note that this process is the same as SimGCL+LLM4IDRec.}
\label{tab:yelp_bpr}
% \linespread{2}
% \setlength\tabcolsep{13pt}{
\begin{tabular}{lcccccc}

\hline
Model                                  & Recall@10 & NDCG@10 & Recall@20 & NDCG@20               & Recall@50 & NDCG@50 \\  \specialrule{0.1em}{2pt}{2pt}  
\multicolumn{1}{l}{BPR}               & 0.02778   & 0.03106 & 0.04914 & 0.03946  &  0.09772   & 0.05788   \\ \specialrule{0.em}{2pt}{2pt}
\multicolumn{1}{l}{BPR+BPR}     & 0.02709  & 0.03036 &  0.04850   & 0.03985 & 0.09636  & 0.05647     \\ 
\multicolumn{1}{l}{BPR+LLM4IDRec}     & 0.02879  & 0.03215 & 0.05111          & 0.04064 & 0.10113           &  0.05927        \\ 
\specialrule{0.1em}{3pt}{3pt}

\multicolumn{1}{l}{SimGCL}            & 0.04218   & 0.04839 & 0.07137  & 0.05892 & 0.13569   &  0.08261   \\ \specialrule{0.em}{2pt}{2pt}
\multicolumn{1}{l}{SimGCL+BPR}  & 0.04047    &0.04884 & 0.06724   & 0.05827 &0.12872  &0.08109         \\
\multicolumn{1}{l}{SimGCL+LLM4IDRec}  & 0.04393    & 0.05019 & 0.07371   & 0.06089 &0.13991            & 0.08533         \\ %\specialrule{0.1em}{2pt}{2pt}  
\specialrule{0.1em}{3pt}{3pt}

\multicolumn{1}{l}{NCL} &0.03946	&0.04516	&0.06693	&0.05513 &0.12726	&0.07820          \\ \specialrule{0.em}{2pt}{2pt} 
\multicolumn{1}{l}{NCL+BPR}     & 0.03846   & 0.0456 &0.06384   &0.05467 &0.12298 &0.07667        \\ 
\multicolumn{1}{l}{NCL+LLM4IDRec}    &0.04098	&0.04694	&0.06921	&0.06921 &0.13112	&0.08014         \\ %\specialrule{0.1em}{2pt}{2pt}  
\specialrule{0.1em}{3pt}{3pt}

\multicolumn{1}{l}{XSimGCL}           & 0.04249   & 0.04859 & 0.07234          & 0.05943 &  0.13864          &0.08389          \\ \specialrule{0.em}{3pt}{3pt}
\multicolumn{1}{l}{XSimGCL+BPR}     & 0.04068   & 0.04902 &0.06742           & 0.05842 &0.1298   & 0.08155 \\ 
\multicolumn{1}{l}{XSimGCL+LLM4IDRec} & 0.04411   & 0.05036 & 0.07394          & 0.06104 &  0.14075    &  0.08581  \\ %\specialrule{0.1em}{2pt}{2pt} 
\specialrule{0.1em}{3pt}{3pt}

% \multicolumn{1}{l}{P5}           & 0.04286  & 0.04951 & 0.07307  & 0.06050 &  0.13974  &0.08558         \\ \specialrule{0.em}{3pt}{3pt}
% \multicolumn{1}{l}{P5+LLM4IDRec} & 0.04428   & 0.05110 & 0.07556   & 0.06220 &  0.14395    &  0.08804  \\ %\specialrule{0.1em}{2pt}{2pt} 
% \multicolumn{1}{l}{Improv.}     & 3.32\%   & 3.21\% &3.41\%    & 2.81\% &3.01\%  & 2.88\%  \\ \specialrule{0.1em}{3pt}{3pt}

\end{tabular} 
% }
\end{table}

\begin{table}[]
\caption{Performance comparisons on Amazon-kindle data. The "?+BPR" represents utilizing BPR as data augmentation, integrating different ID-based recommendation models. Taking the SimGCL+BPR model as an example, we use BPR to augment the interaction data, then train SimGCL on the augmented data and generate recommended items. Please note that this process is the same as SimGCL+LLM4IDRec.}
\label{tab:kindle_bpr}
% \linespread{2}
% \setlength\tabcolsep{13pt}{
\begin{tabular}{lcccccc}

\hline
Model                                  & Recall@10 & NDCG@10 & Recall@20 & NDCG@20               & Recall@50 & NDCG@50 \\  \specialrule{0.1em}{2pt}{2pt} 

\multicolumn{1}{l}{BPR}    &0.12292	&0.09161	&0.16725	&0.10502 &0.24249	&0.12452   \\ \specialrule{0.em}{2pt}{2pt}
\multicolumn{1}{l}{BPR+BPR}     & 0.11305   & 0.08362 & 0.1538 &0.09586 &0.22189 & 0.11351    \\ 
\multicolumn{1}{l}{BPR+LLM4IDRec}    & 0.13138	& 0.09786	& 0.17447	& 0.11098 & 0.24404	& 0.12912        \\ 
\specialrule{0.1em}{3pt}{3pt}

\multicolumn{1}{l}{SimGCL}          &0.13876	&0.10146	&0.19028	&0.11691 &0.26872	&0.13728   \\ \specialrule{0.em}{2pt}{2pt}
\multicolumn{1}{l}{SimGCL+BPR}     & 0.13461  & 0.09738 & 0.18583   &0.11274& 0.26357 & 0.13304      \\
\multicolumn{1}{l}{SimGCL+LLM4IDRec}  &0.14511	&0.10813	&0.19652	&0.12344 &0.27350	&0.14350        \\ %\specialrule{0.1em}{2pt}{2pt}  
 \specialrule{0.1em}{3pt}{3pt}

\multicolumn{1}{l}{NCL} &0.13676	 & 0.09876	 & 0.18582	 & 0.11364  & 0.26234	 & 0.13368       \\ \specialrule{0.em}{2pt}{2pt} 
\multicolumn{1}{l}{NCL+BPR}     & 0.13845  & 0.10023 &0.187   &0.11503 &0.26331 &0.13497         \\
\multicolumn{1}{l}{NCL+LLM4IDRec}   &0.14079	& 0.10299	& 0.19014	& 0.11803 & 0.26658	& 0.13801        \\ %\specialrule{0.1em}{2pt}{2pt}  
 \specialrule{0.1em}{3pt}{3pt}

\multicolumn{1}{l}{XSimGCL}          &0.14392	&0.10612	&0.19514	&0.12143 &0.27379	&0.14185         \\ \specialrule{0.em}{3pt}{3pt}
\multicolumn{1}{l}{XSimGCL+BPR}     &0.13836  &0.10110 &0.18946 &0.11640 &0.26834 &0.13690 \\
\multicolumn{1}{l}{XSimGCL+LLM4IDRec} &0.14976	&0.11235	&0.20147	&0.12776 &0.27956	&0.14810 \\ %\specialrule{0.1em}{2pt}{2pt} 
 \specialrule{0.1em}{3pt}{3pt}

\end{tabular} 
% }
\end{table}

{
\subsection{Replacing LLMs with existing ID-based recommendation models }
To better illustrate the effectiveness of our proposed LLM4IDRec in augmented data, we also compared it with simple merging strategies involving existing ID-based recommendation models like SASRec and BPR. The \textbf{"?+SASRec"} represents utilizing SASRec as data augmentation, integrating different ID-based recommendation models. Taking the BERT4Rec+SASRec model as an example, we use SASRec to augment the interaction data, then train BERT4Rec on the augmented data and generate recommended items. Please note that this process is the same as BERT4Rec+LLM4IDRec. %Similarly, the \textbf{"?+BPR"} represents utilizing BPR as data augmentation, integrating different ID-based recommendation models.
}

In Table~\ref{tab:beauty_sasrec},~\ref{tab:yelp_bpr} and~\ref{tab:kindle_bpr}, the experimental results demonstrated that when used for data augmentation, LLM4IDRec outperformed SASRec and BPR, resulting in a better performance improvement. Conversely, using SASRec or BPR for data augmentation led to a decrease in recommendation performance. This discrepancy can be attributed to the different approaches employed by SASRec/BPR and LLM4IDRec in handling feature space expansion. Large language models like LLM4IDRec can effectively explore a broad feature space during data augmentation, enabling the ID-based model to better learn user preferences from enhanced data. In contrast, SASRec's or BPR's data augmentation method, constrained by a smaller model, may overly memorize specific patterns in the training data, resulting in reduced generalization ability on the test set and a decline in performance for the ID-based model after incorporating this data. 
% \textcolor{blue}{
When ID-based recommenders are used in augmentation, they tend to emphasize similar patterns and data points, which could reinforce existing biases in the recommendation process. This can lead to the amplification of incorrect recommendations, as the model continues to rely on the same data patterns. For example, if an ID-based recommender mistakenly suggests an item that a user dislikes, and this item becomes part of the augmented dataset, the likelihood of recommending the similar incorrect item increases through similar ID-based models, ultimately degrading overall performance. On the other hand, LLM4IDRec, with its advanced reasoning and modeling capabilities, introduces a more diverse set of patterns and relationships that ID-based models might miss. Specifically, when comparing the overlap between the data generated by LLM4IDRec and that produced by ID-based models, we find that the overlap is less than 1\% in both cases. Under the same data generation conditions, the overlap rate between the BPR model and LLM4IDRec on the Yelp dataset is only 0.99\%. Similarly, on the Amazon-Beauty dataset, the overlap rate between the SASRec model and LLM4IDRec is just 0.01\%. These results strongly indicate that LLM4IDRec is capable of uncovering new data patterns and previously unseen interactive relationships that have not been captured by ID-based models. By identifying and utilizing these previously untapped connections, LLM4IDRec provides a broader, more balanced perspective, introducing a new dimension of insights to the recommendations. This ability to capture novel patterns ultimately enhances performance when combined with traditional ID-based models.
% }

\begin{figure}[htp]
    \centering
    \includegraphics[width=0.95\textwidth]{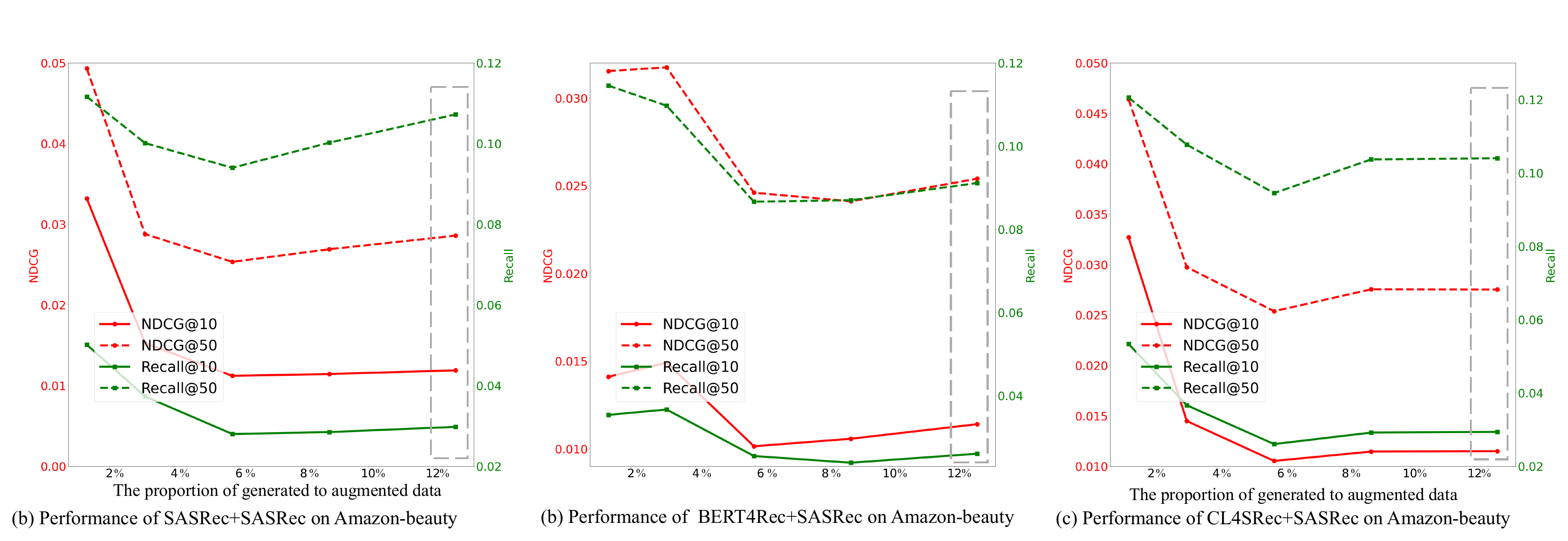} 
    \caption{The performance of SASRec as augmentation method across the different sizes of generated training data. The results reported in the previous version are framed by gray dashed lines.}
    \label{fig:diff_size_sasrec}
\end{figure}

% \textcolor{blue}{
We explored how varying ratios of synthetic data could offer more nuanced insights into the impact of ID-based model augmentation on recommendation performance. To achieve this, we conduct additional experiments using SASRec to generate different proportions of synthetic data on the Amazon-Beauty dataset. The results, shown in Figure~\ref{fig:diff_size_sasrec}, indicate that performance remains stable or even improves slightly when a small amount of synthetic data (e.g., 1\%) is introduced. However, once the proportion exceeds a small threshold (starting from 3\%), the model's performance declines significantly. These experiments confirm that while a small amount of synthetic data can be beneficial, increasing the proportion leads to rapid performance deterioration. This suggests that while data augmentation with SASRec may offer short-term gains, it falls short of the performance achieved by LLM4IDRec. Nonetheless, these results highlight that the proportion of synthetic data is a key factor in influencing recommendation performance.
% Nevertheless, the different proportions of synthesized data can affect the recommendation performance. What is more important is the quality of the synthesized data. Our proposed LLM4IDRec achieved better performance compare with ID-based model, which also indicates that LLM4IDRec generates better data quality.
% we acknowledge the significance of how 
% different proportions of synthetic data impact model performance.
% This investigation offer a more profound understanding of the relationship between synthetic data and the quality of recommendations.
% }

% \textcolor{blue}{
% 1. The key difference between merging existing ID-based recommenders and integrating LLM4IDRec lies in how each approach expands the feature space. Traditional ID-based recommenders, such as SASRec and BPR, tend to operate within a relatively limited feature space due to their smaller model structures and the nature of ID-based data. In contrast, LLM4IDRec, driven by a large language model (LLM), can explore a much broader and more complex feature space. The LLM's ability to process vast amounts of contextual and relational information introduces new perspectives and patterns that were previously untapped in ID-based models. This enhanced understanding allows for the discovery of novel insights, which is likely why combining LLM4IDRec with existing ID-based recommenders leads to performance improvements. The LLM essentially enriches the dataset by bringing in fresh information that was previously overlooked, thus enhancing the overall recommendation quality.
% }

\section{Conclusion}
In this paper, we have introduced an innovative approach called LLM4IDRec that leverages LLM in recommendation systems, particularly in scenarios where only ID data is available. Our objective was to explore the potential of LLMs in ID-based recommendations, diverging from the reliance on textual data that has been the focus of previous studies. The key idea behind LLM4IDRec was to employ LLMs to augment ID data and assess whether this augmentation could enhance recommendation performance. Our experimental results, conducted on three widely-used datasets, demonstrate the effectiveness of the LLM4IDRec approach. We consistently observed a notable improvement in recommendation performance compared to existing methods, highlighting the potential of LLMs in ID-based recommendations solely through data augmentation.  In future work, we plan to combine LLM4IDRec with traditional ID-based models to generate more effective data, as the overlap between data generated by LLM4IDRec and that produced by conventional models is less than 1\% across all datasets, allowing for mutual reinforcement.

% In future work, we will consider combining LLM4IDRec with traditional ID-based models to generate more effective data, as the overlap between data generated by LLM4IDRec and that produced by conventional models is less than 1\% across all datasets, allowing for mutual reinforcement.
% Revise grammar and make it more fluent

% In the future, we will consider improving the proposed model to generate more effective data.

% \begin{acks}
% \end{acks}

\newpage
\bibliographystyle{ACM-Reference-Format}
\bibliography{paper_list}


\begin{thebibliography}{86}

%%% the Web address.

\ifx \showCODEN    \undefined \def \showCODEN     #1{\unskip}     \fi
\ifx \showDOI      \undefined \def \showDOI       #1{#1}\fi
\ifx \showISBNx    \undefined \def \showISBNx     #1{\unskip}     \fi
\ifx \showISBNxiii \undefined \def \showISBNxiii  #1{\unskip}     \fi
\ifx \showISSN     \undefined \def \showISSN      #1{\unskip}     \fi
\ifx \showLCCN     \undefined \def \showLCCN      #1{\unskip}     \fi
\ifx \shownote     \undefined \def \shownote      #1{#1}          \fi
\ifx \showarticletitle \undefined \def \showarticletitle #1{#1}   \fi
\ifx \showURL      \undefined \def \showURL       {\relax}        \fi
% invisible to TeX
\providecommand\bibfield[2]{#2}
\providecommand\bibinfo[2]{#2}
\providecommand\natexlab[1]{#1}
\providecommand\showeprint[2][]{arXiv:#2}

\bibitem[Bao et~al\mbox{.}(2023)]%
        {TALLRec}
\bibfield{author}{\bibinfo{person}{Keqin Bao}, \bibinfo{person}{Jizhi Zhang}, \bibinfo{person}{Yang Zhang}, \bibinfo{person}{Wenjie Wang}, \bibinfo{person}{Fuli Feng}, {and} \bibinfo{person}{Xiangnan He}.} \bibinfo{year}{2023}\natexlab{}.
\newblock \showarticletitle{TALLRec: An Effective and Efficient Tuning Framework to Align Large Language Model with Recommendation}. In \bibinfo{booktitle}{\emph{RecSys}}. \bibinfo{publisher}{{ACM}}, \bibinfo{pages}{1007--1014}.
\newblock


\bibitem[Borisov et~al\mbox{.}(2023)]%
        {borisov2023language}
\bibfield{author}{\bibinfo{person}{Vadim Borisov}, \bibinfo{person}{Kathrin Sessler}, \bibinfo{person}{Tobias Leemann}, \bibinfo{person}{Martin Pawelczyk}, {and} \bibinfo{person}{Gjergji Kasneci}.} \bibinfo{year}{2023}\natexlab{}.
\newblock \showarticletitle{Language Models are Realistic Tabular Data Generators}. In \bibinfo{booktitle}{\emph{The International Conference on Learning Representations}}.
\newblock
\urldef\tempurl%
\url{https://openreview.net/forum?id=cEygmQNOeI}
\showURL{%
\tempurl}


\bibitem[Bran et~al\mbox{.}(2023)]%
        {bran2023augmenting}
\bibfield{author}{\bibinfo{person}{Andres~M Bran}, \bibinfo{person}{Sam Cox}, \bibinfo{person}{Oliver Schilter}, \bibinfo{person}{Carlo Baldassari}, \bibinfo{person}{Andrew White}, {and} \bibinfo{person}{Philippe Schwaller}.} \bibinfo{year}{2023}\natexlab{}.
\newblock \showarticletitle{Augmenting large language models with chemistry tools}. In \bibinfo{booktitle}{\emph{Conference on Neural Information Processing Systems (AI for Science Workshop)}}.
\newblock


\bibitem[Brown et~al\mbox{.}(2020b)]%
        {brown2020language}
\bibfield{author}{\bibinfo{person}{Tom Brown}, \bibinfo{person}{Benjamin Mann}, \bibinfo{person}{Nick Ryder}, \bibinfo{person}{Melanie Subbiah}, \bibinfo{person}{Jared~D Kaplan}, \bibinfo{person}{Prafulla Dhariwal}, \bibinfo{person}{Arvind Neelakantan}, \bibinfo{person}{Pranav Shyam}, \bibinfo{person}{Girish Sastry}, \bibinfo{person}{Amanda Askell}, {et~al\mbox{.}}} \bibinfo{year}{2020}\natexlab{b}.
\newblock \showarticletitle{Language models are few-shot learners}.
\newblock \bibinfo{journal}{\emph{Advances in Neural Information Processing Systems}}  \bibinfo{volume}{33} (\bibinfo{year}{2020}), \bibinfo{pages}{1877--1901}.
\newblock


\bibitem[Brown et~al\mbox{.}(2020a)]%
        {gpt3}
\bibfield{author}{\bibinfo{person}{Tom~B. Brown}, \bibinfo{person}{Benjamin Mann}, \bibinfo{person}{Nick Ryder}, \bibinfo{person}{Melanie Subbiah}, \bibinfo{person}{Jared Kaplan}, \bibinfo{person}{Prafulla Dhariwal}, \bibinfo{person}{Arvind Neelakantan}, \bibinfo{person}{Pranav Shyam}, \bibinfo{person}{Girish Sastry}, \bibinfo{person}{Amanda Askell}, \bibinfo{person}{Sandhini Agarwal}, \bibinfo{person}{Ariel Herbert{-}Voss}, \bibinfo{person}{Gretchen Krueger}, \bibinfo{person}{Tom Henighan}, \bibinfo{person}{Rewon Child}, \bibinfo{person}{Aditya Ramesh}, \bibinfo{person}{Daniel~M. Ziegler}, \bibinfo{person}{Jeffrey Wu}, \bibinfo{person}{Clemens Winter}, \bibinfo{person}{Christopher Hesse}, \bibinfo{person}{Mark Chen}, \bibinfo{person}{Eric Sigler}, \bibinfo{person}{Mateusz Litwin}, \bibinfo{person}{Scott Gray}, \bibinfo{person}{Benjamin Chess}, \bibinfo{person}{Jack Clark}, \bibinfo{person}{Christopher Berner}, \bibinfo{person}{Sam McCandlish}, \bibinfo{person}{Alec Radford}, \bibinfo{person}{Ilya Sutskever},
  {and} \bibinfo{person}{Dario Amodei}.} \bibinfo{year}{2020}\natexlab{a}.
\newblock \showarticletitle{Language Models are Few-Shot Learners}. In \bibinfo{booktitle}{\emph{NeurIPS}}.
\newblock


\bibitem[Carranza et~al\mbox{.}(2023)]%
        {carranza2023privacy}
\bibfield{author}{\bibinfo{person}{Aldo~Gael Carranza}, \bibinfo{person}{Rezsa Farahani}, \bibinfo{person}{Natalia Ponomareva}, \bibinfo{person}{Alex Kurakin}, \bibinfo{person}{Matthew Jagielski}, {and} \bibinfo{person}{Milad Nasr}.} \bibinfo{year}{2023}\natexlab{}.
\newblock \showarticletitle{Privacy-Preserving Recommender Systems with Synthetic Query Generation using Differentially Private Large Language Models}.
\newblock \bibinfo{journal}{\emph{arXiv preprint arXiv:2305.05973}} (\bibinfo{year}{2023}).
\newblock


\bibitem[Chen et~al\mbox{.}(2023)]%
        {chen2023bias}
\bibfield{author}{\bibinfo{person}{Jiawei Chen}, \bibinfo{person}{Hande Dong}, \bibinfo{person}{Xiang Wang}, \bibinfo{person}{Fuli Feng}, \bibinfo{person}{Meng Wang}, {and} \bibinfo{person}{Xiangnan He}.} \bibinfo{year}{2023}\natexlab{}.
\newblock \showarticletitle{Bias and debias in recommender system: A survey and future directions}.
\newblock \bibinfo{journal}{\emph{ACM Transactions on Information Systems}} \bibinfo{volume}{41}, \bibinfo{number}{3} (\bibinfo{year}{2023}), \bibinfo{pages}{1--39}.
\newblock


\bibitem[Chen et~al\mbox{.}(2020a)]%
        {chen2020revisiting}
\bibfield{author}{\bibinfo{person}{Lei Chen}, \bibinfo{person}{Le Wu}, \bibinfo{person}{Richang Hong}, \bibinfo{person}{Kun Zhang}, {and} \bibinfo{person}{Meng Wang}.} \bibinfo{year}{2020}\natexlab{a}.
\newblock \showarticletitle{Revisiting graph based collaborative filtering: A linear residual graph convolutional network approach}. In \bibinfo{booktitle}{\emph{Proceedings of the AAAI conference on artificial intelligence}}, Vol.~\bibinfo{volume}{34}. \bibinfo{pages}{27--34}.
\newblock


\bibitem[Chen et~al\mbox{.}(2020b)]%
        {chen2020try}
\bibfield{author}{\bibinfo{person}{Tong Chen}, \bibinfo{person}{Hongzhi Yin}, \bibinfo{person}{Guanhua Ye}, \bibinfo{person}{Zi Huang}, \bibinfo{person}{Yang Wang}, {and} \bibinfo{person}{Meng Wang}.} \bibinfo{year}{2020}\natexlab{b}.
\newblock \showarticletitle{Try this instead: Personalized and interpretable substitute recommendation}. In \bibinfo{booktitle}{\emph{Proceedings of the International ACM SIGIR Conference on Research and Development in Information Retrieval}}. \bibinfo{pages}{891--900}.
\newblock


\bibitem[Chiang et~al\mbox{.}(2023)]%
        {vicuna}
\bibfield{author}{\bibinfo{person}{Wei-Lin Chiang}, \bibinfo{person}{Zhuohan Li}, \bibinfo{person}{Zi Lin}, \bibinfo{person}{Ying Sheng}, \bibinfo{person}{Zhanghao Wu}, \bibinfo{person}{Hao Zhang}, \bibinfo{person}{Lianmin Zheng}, \bibinfo{person}{Siyuan Zhuang}, \bibinfo{person}{Yonghao Zhuang}, \bibinfo{person}{Joseph~E. Gonzalez}, \bibinfo{person}{Ion Stoica}, {and} \bibinfo{person}{Eric~P. Xing}.} \bibinfo{year}{2023}\natexlab{}.
\newblock \bibinfo{title}{Vicuna: An Open-Source Chatbot Impressing GPT-4 with 90\%* ChatGPT Quality}.
\newblock
\newblock
\urldef\tempurl%
\url{https://lmsys.org/blog/2023-03-30-vicuna/}
\showURL{%
\tempurl}


\bibitem[Covington et~al\mbox{.}(2016)]%
        {covington2016deep}
\bibfield{author}{\bibinfo{person}{Paul Covington}, \bibinfo{person}{Jay Adams}, {and} \bibinfo{person}{Emre Sargin}.} \bibinfo{year}{2016}\natexlab{}.
\newblock \showarticletitle{Deep neural networks for youtube recommendations}. In \bibinfo{booktitle}{\emph{Proceedings of the ACM conference on recommender systems}}. \bibinfo{pages}{191--198}.
\newblock


\bibitem[Cui et~al\mbox{.}(2022)]%
        {M6-Rec}
\bibfield{author}{\bibinfo{person}{Zeyu Cui}, \bibinfo{person}{Jianxin Ma}, \bibinfo{person}{Chang Zhou}, \bibinfo{person}{Jingren Zhou}, {and} \bibinfo{person}{Hongxia Yang}.} \bibinfo{year}{2022}\natexlab{}.
\newblock \showarticletitle{M6-Rec: Generative Pretrained Language Models are Open-Ended Recommender Systems}.
\newblock \bibinfo{journal}{\emph{CoRR}}  \bibinfo{volume}{abs/2205.08084} (\bibinfo{year}{2022}).
\newblock


\bibitem[Dai et~al\mbox{.}(2023a)]%
        {uncovering_ChatGPT4Rec}
\bibfield{author}{\bibinfo{person}{Sunhao Dai}, \bibinfo{person}{Ninglu Shao}, \bibinfo{person}{Haiyuan Zhao}, \bibinfo{person}{Weijie Yu}, \bibinfo{person}{Zihua Si}, \bibinfo{person}{Chen Xu}, \bibinfo{person}{Zhongxiang Sun}, \bibinfo{person}{Xiao Zhang}, {and} \bibinfo{person}{Jun Xu}.} \bibinfo{year}{2023}\natexlab{a}.
\newblock \showarticletitle{Uncovering ChatGPT's Capabilities in Recommender Systems}. In \bibinfo{booktitle}{\emph{RecSys}}. \bibinfo{publisher}{{ACM}}, \bibinfo{pages}{1126--1132}.
\newblock


\bibitem[Dai et~al\mbox{.}(2023b)]%
        {dai2023uncovering}
\bibfield{author}{\bibinfo{person}{Sunhao Dai}, \bibinfo{person}{Ninglu Shao}, \bibinfo{person}{Haiyuan Zhao}, \bibinfo{person}{Weijie Yu}, \bibinfo{person}{Zihua Si}, \bibinfo{person}{Chen Xu}, \bibinfo{person}{Zhongxiang Sun}, \bibinfo{person}{Xiao Zhang}, {and} \bibinfo{person}{Jun Xu}.} \bibinfo{year}{2023}\natexlab{b}.
\newblock \showarticletitle{Uncovering ChatGPT's Capabilities in Recommender Systems}.
\newblock \bibinfo{journal}{\emph{arXiv preprint arXiv:2305.02182}} (\bibinfo{year}{2023}).
\newblock


\bibitem[Dai et~al\mbox{.}(2021)]%
        {dai2021adversarial}
\bibfield{author}{\bibinfo{person}{Xinyi Dai}, \bibinfo{person}{Jianghao Lin}, \bibinfo{person}{Weinan Zhang}, \bibinfo{person}{Shuai Li}, \bibinfo{person}{Weiwen Liu}, \bibinfo{person}{Ruiming Tang}, \bibinfo{person}{Xiuqiang He}, \bibinfo{person}{Jianye Hao}, \bibinfo{person}{Jun Wang}, {and} \bibinfo{person}{Yong Yu}.} \bibinfo{year}{2021}\natexlab{}.
\newblock \showarticletitle{An adversarial imitation click model for information retrieval}. In \bibinfo{booktitle}{\emph{Proceedings of the Web Conference}}. \bibinfo{pages}{1809--1820}.
\newblock


\bibitem[Devlin et~al\mbox{.}(2018)]%
        {devlin2018bert}
\bibfield{author}{\bibinfo{person}{Jacob Devlin}, \bibinfo{person}{Ming-Wei Chang}, \bibinfo{person}{Kenton Lee}, {and} \bibinfo{person}{Kristina Toutanova}.} \bibinfo{year}{2018}\natexlab{}.
\newblock \showarticletitle{Bert: Pre-training of deep bidirectional transformers for language understanding}.
\newblock \bibinfo{journal}{\emph{arXiv preprint arXiv:1810.04805}} (\bibinfo{year}{2018}).
\newblock


\bibitem[Gao et~al\mbox{.}(2023)]%
        {gao2023chat}
\bibfield{author}{\bibinfo{person}{Yunfan Gao}, \bibinfo{person}{Tao Sheng}, \bibinfo{person}{Youlin Xiang}, \bibinfo{person}{Yun Xiong}, \bibinfo{person}{Haofen Wang}, {and} \bibinfo{person}{Jiawei Zhang}.} \bibinfo{year}{2023}\natexlab{}.
\newblock \showarticletitle{Chat-rec: Towards interactive and explainable llms-augmented recommender system}.
\newblock \bibinfo{journal}{\emph{arXiv preprint arXiv:2303.14524}} (\bibinfo{year}{2023}).
\newblock


\bibitem[Geng et~al\mbox{.}(2022a)]%
        {P5}
\bibfield{author}{\bibinfo{person}{Shijie Geng}, \bibinfo{person}{Shuchang Liu}, \bibinfo{person}{Zuohui Fu}, \bibinfo{person}{Yingqiang Ge}, {and} \bibinfo{person}{Yongfeng Zhang}.} \bibinfo{year}{2022}\natexlab{a}.
\newblock \showarticletitle{Recommendation as Language Processing {(RLP):} {A} Unified Pretrain, Personalized Prompt {\&} Predict Paradigm {(P5)}}. In \bibinfo{booktitle}{\emph{RecSys}}. \bibinfo{publisher}{{ACM}}, \bibinfo{pages}{299--315}.
\newblock


\bibitem[Geng et~al\mbox{.}(2022b)]%
        {geng2022recommendation}
\bibfield{author}{\bibinfo{person}{Shijie Geng}, \bibinfo{person}{Shuchang Liu}, \bibinfo{person}{Zuohui Fu}, \bibinfo{person}{Yingqiang Ge}, {and} \bibinfo{person}{Yongfeng Zhang}.} \bibinfo{year}{2022}\natexlab{b}.
\newblock \showarticletitle{Recommendation as language processing (rlp): A unified pretrain, personalized prompt \& predict paradigm (p5)}. In \bibinfo{booktitle}{\emph{Proceedings of the 16th ACM Conference on Recommender Systems}}. \bibinfo{pages}{299--315}.
\newblock


\bibitem[Gou et~al\mbox{.}(2023)]%
        {gou2023tora}
\bibfield{author}{\bibinfo{person}{Zhibin Gou}, \bibinfo{person}{Zhihong Shao}, \bibinfo{person}{Yeyun Gong}, \bibinfo{person}{Yujiu Yang}, \bibinfo{person}{Minlie Huang}, \bibinfo{person}{Nan Duan}, \bibinfo{person}{Weizhu Chen}, {et~al\mbox{.}}} \bibinfo{year}{2023}\natexlab{}.
\newblock \showarticletitle{ToRA: A Tool-Integrated Reasoning Agent for Mathematical Problem Solving}.
\newblock \bibinfo{journal}{\emph{arXiv preprint arXiv:2309.17452}} (\bibinfo{year}{2023}).
\newblock


\bibitem[Guo et~al\mbox{.}(2017)]%
        {guo2017deepfm}
\bibfield{author}{\bibinfo{person}{Huifeng Guo}, \bibinfo{person}{Ruiming Tang}, \bibinfo{person}{Yunming Ye}, \bibinfo{person}{Zhenguo Li}, {and} \bibinfo{person}{Xiuqiang He}.} \bibinfo{year}{2017}\natexlab{}.
\newblock \showarticletitle{DeepFM: a factorization-machine based neural network for CTR prediction}.
\newblock \bibinfo{journal}{\emph{arXiv preprint arXiv:1703.04247}} (\bibinfo{year}{2017}).
\newblock


\bibitem[Guo et~al\mbox{.}(2023)]%
        {guo2023gpt4graph}
\bibfield{author}{\bibinfo{person}{Jiayan Guo}, \bibinfo{person}{Lun Du}, \bibinfo{person}{Hengyu Liu}, \bibinfo{person}{Mengyu Zhou}, \bibinfo{person}{Xinyi He}, {and} \bibinfo{person}{Shi Han}.} \bibinfo{year}{2023}\natexlab{}.
\newblock \showarticletitle{Gpt4graph: Can large language models understand graph structured data? an empirical evaluation and benchmarking}.
\newblock \bibinfo{journal}{\emph{arXiv preprint arXiv:2305.15066}} (\bibinfo{year}{2023}).
\newblock


\bibitem[He and McAuley(2016)]%
        {he2016fusing}
\bibfield{author}{\bibinfo{person}{Ruining He} {and} \bibinfo{person}{Julian McAuley}.} \bibinfo{year}{2016}\natexlab{}.
\newblock \showarticletitle{Fusing similarity models with markov chains for sparse sequential recommendation}. In \bibinfo{booktitle}{\emph{IEEE international conference on data mining (ICDM)}}. \bibinfo{pages}{191--200}.
\newblock


\bibitem[He et~al\mbox{.}(2020)]%
        {he2020lightgcn}
\bibfield{author}{\bibinfo{person}{Xiangnan He}, \bibinfo{person}{Kuan Deng}, \bibinfo{person}{Xiang Wang}, \bibinfo{person}{Yan Li}, \bibinfo{person}{Yongdong Zhang}, {and} \bibinfo{person}{Meng Wang}.} \bibinfo{year}{2020}\natexlab{}.
\newblock \showarticletitle{Lightgcn: Simplifying and powering graph convolution network for recommendation}. In \bibinfo{booktitle}{\emph{Proceedings of the International ACM SIGIR Conference on Research and Development in Information Retrieval}}. \bibinfo{pages}{639--648}.
\newblock


\bibitem[He et~al\mbox{.}(2017)]%
        {he2017neural}
\bibfield{author}{\bibinfo{person}{Xiangnan He}, \bibinfo{person}{Lizi Liao}, \bibinfo{person}{Hanwang Zhang}, \bibinfo{person}{Liqiang Nie}, \bibinfo{person}{Xia Hu}, {and} \bibinfo{person}{Tat-Seng Chua}.} \bibinfo{year}{2017}\natexlab{}.
\newblock \showarticletitle{Neural collaborative filtering}. In \bibinfo{booktitle}{\emph{Proceedings of the 26th International Conference on World Wide Web}}. \bibinfo{pages}{173--182}.
\newblock


\bibitem[Hidasi(2015)]%
        {hidasi2015session}
\bibfield{author}{\bibinfo{person}{B Hidasi}.} \bibinfo{year}{2015}\natexlab{}.
\newblock \showarticletitle{Session-based Recommendations with Recurrent Neural Networks}.
\newblock \bibinfo{journal}{\emph{arXiv preprint arXiv:1511.06939}} (\bibinfo{year}{2015}).
\newblock


\bibitem[Hou et~al\mbox{.}(2022)]%
        {hou2022towards}
\bibfield{author}{\bibinfo{person}{Yupeng Hou}, \bibinfo{person}{Shanlei Mu}, \bibinfo{person}{Wayne~Xin Zhao}, \bibinfo{person}{Yaliang Li}, \bibinfo{person}{Bolin Ding}, {and} \bibinfo{person}{Ji-Rong Wen}.} \bibinfo{year}{2022}\natexlab{}.
\newblock \showarticletitle{Towards universal sequence representation learning for recommender systems}. In \bibinfo{booktitle}{\emph{Proceedings of the 28th ACM SIGKDD Conference on Knowledge Discovery and Data Mining}}. \bibinfo{pages}{585--593}.
\newblock


\bibitem[Hou et~al\mbox{.}(2023)]%
        {hou2023large}
\bibfield{author}{\bibinfo{person}{Yupeng Hou}, \bibinfo{person}{Junjie Zhang}, \bibinfo{person}{Zihan Lin}, \bibinfo{person}{Hongyu Lu}, \bibinfo{person}{Ruobing Xie}, \bibinfo{person}{Julian McAuley}, {and} \bibinfo{person}{Wayne~Xin Zhao}.} \bibinfo{year}{2023}\natexlab{}.
\newblock \showarticletitle{Large language models are zero-shot rankers for recommender systems}.
\newblock \bibinfo{journal}{\emph{arXiv preprint arXiv:2305.08845}} (\bibinfo{year}{2023}).
\newblock


\bibitem[Hua et~al\mbox{.}(2023)]%
        {hua2023index}
\bibfield{author}{\bibinfo{person}{Wenyue Hua}, \bibinfo{person}{Shuyuan Xu}, \bibinfo{person}{Yingqiang Ge}, {and} \bibinfo{person}{Yongfeng Zhang}.} \bibinfo{year}{2023}\natexlab{}.
\newblock \showarticletitle{How to index item ids for recommendation foundation models}. In \bibinfo{booktitle}{\emph{Proceedings of the Annual International ACM SIGIR Conference on Research and Development in Information Retrieval in the Asia Pacific Region}}. \bibinfo{pages}{195--204}.
\newblock


\bibitem[Huang and Chang(2022)]%
        {huang2022towards}
\bibfield{author}{\bibinfo{person}{Jie Huang} {and} \bibinfo{person}{Kevin Chen-Chuan Chang}.} \bibinfo{year}{2022}\natexlab{}.
\newblock \showarticletitle{Towards Reasoning in Large Language Models: A Survey}.
\newblock \bibinfo{journal}{\emph{arXiv preprint arXiv:2212.10403}} (\bibinfo{year}{2022}).
\newblock


\bibitem[Kang and McAuley(2018)]%
        {kang2018self}
\bibfield{author}{\bibinfo{person}{Wang-Cheng Kang} {and} \bibinfo{person}{Julian McAuley}.} \bibinfo{year}{2018}\natexlab{}.
\newblock \showarticletitle{Self-attentive sequential recommendation}. In \bibinfo{booktitle}{\emph{2018 IEEE international conference on data mining (ICDM)}}. IEEE, \bibinfo{pages}{197--206}.
\newblock


\bibitem[Koren et~al\mbox{.}(2009)]%
        {koren2009matrix}
\bibfield{author}{\bibinfo{person}{Yehuda Koren}, \bibinfo{person}{Robert Bell}, {and} \bibinfo{person}{Chris Volinsky}.} \bibinfo{year}{2009}\natexlab{}.
\newblock \showarticletitle{Matrix factorization techniques for recommender systems}.
\newblock \bibinfo{journal}{\emph{Computer}} \bibinfo{volume}{42}, \bibinfo{number}{8} (\bibinfo{year}{2009}), \bibinfo{pages}{30--37}.
\newblock


\bibitem[Li et~al\mbox{.}(2023a)]%
        {li2023sailer}
\bibfield{author}{\bibinfo{person}{Haitao Li}, \bibinfo{person}{Qingyao Ai}, \bibinfo{person}{Jia Chen}, \bibinfo{person}{Qian Dong}, \bibinfo{person}{Yueyue Wu}, \bibinfo{person}{Yiqun Liu}, \bibinfo{person}{Chong Chen}, {and} \bibinfo{person}{Qi Tian}.} \bibinfo{year}{2023}\natexlab{a}.
\newblock \showarticletitle{SAILER: Structure-aware Pre-trained Language Model for Legal Case Retrieval}.
\newblock \bibinfo{journal}{\emph{arXiv preprint arXiv:2304.11370}} (\bibinfo{year}{2023}).
\newblock


\bibitem[Li et~al\mbox{.}(2023d)]%
        {RecFormer}
\bibfield{author}{\bibinfo{person}{Jiacheng Li}, \bibinfo{person}{Ming Wang}, \bibinfo{person}{Jin Li}, \bibinfo{person}{Jinmiao Fu}, \bibinfo{person}{Xin Shen}, \bibinfo{person}{Jingbo Shang}, {and} \bibinfo{person}{Julian~J. McAuley}.} \bibinfo{year}{2023}\natexlab{d}.
\newblock \showarticletitle{Text Is All You Need: Learning Language Representations for Sequential Recommendation}. In \bibinfo{booktitle}{\emph{{KDD}}}. \bibinfo{publisher}{{ACM}}, \bibinfo{pages}{1258--1267}.
\newblock


\bibitem[Li et~al\mbox{.}(2020)]%
        {li2020time}
\bibfield{author}{\bibinfo{person}{Jiacheng Li}, \bibinfo{person}{Yujie Wang}, {and} \bibinfo{person}{Julian McAuley}.} \bibinfo{year}{2020}\natexlab{}.
\newblock \showarticletitle{Time interval aware self-attention for sequential recommendation}. In \bibinfo{booktitle}{\emph{Proceedings of the 13th international conference on web search and data mining}}. \bibinfo{pages}{322--330}.
\newblock


\bibitem[Li et~al\mbox{.}(2023e)]%
        {li2023personalized}
\bibfield{author}{\bibinfo{person}{Lei Li}, \bibinfo{person}{Yongfeng Zhang}, {and} \bibinfo{person}{Li Chen}.} \bibinfo{year}{2023}\natexlab{e}.
\newblock \showarticletitle{Personalized prompt learning for explainable recommendation}.
\newblock \bibinfo{journal}{\emph{ACM Transactions on Information Systems}} \bibinfo{volume}{41}, \bibinfo{number}{4} (\bibinfo{year}{2023}), \bibinfo{pages}{1--26}.
\newblock


\bibitem[Li et~al\mbox{.}(2023c)]%
        {li2023exploring}
\bibfield{author}{\bibinfo{person}{Ruyu Li}, \bibinfo{person}{Wenhao Deng}, \bibinfo{person}{Yu Cheng}, \bibinfo{person}{Zheng Yuan}, \bibinfo{person}{Jiaqi Zhang}, {and} \bibinfo{person}{Fajie Yuan}.} \bibinfo{year}{2023}\natexlab{c}.
\newblock \showarticletitle{Exploring the Upper Limits of Text-Based Collaborative Filtering Using Large Language Models: Discoveries and Insights}.
\newblock \bibinfo{journal}{\emph{arXiv preprint arXiv:2305.11700}} (\bibinfo{year}{2023}).
\newblock


\bibitem[Li et~al\mbox{.}(2023b)]%
        {li2023ctrl}
\bibfield{author}{\bibinfo{person}{Xiangyang Li}, \bibinfo{person}{Bo Chen}, \bibinfo{person}{Lu Hou}, {and} \bibinfo{person}{Ruiming Tang}.} \bibinfo{year}{2023}\natexlab{b}.
\newblock \showarticletitle{CTRL: Connect Tabular and Language Model for CTR Prediction}.
\newblock \bibinfo{journal}{\emph{arXiv preprint arXiv:2306.02841}} (\bibinfo{year}{2023}).
\newblock


\bibitem[Lin et~al\mbox{.}(2021)]%
        {lin2021graph}
\bibfield{author}{\bibinfo{person}{Jianghao Lin}, \bibinfo{person}{Weiwen Liu}, \bibinfo{person}{Xinyi Dai}, \bibinfo{person}{Weinan Zhang}, \bibinfo{person}{Shuai Li}, \bibinfo{person}{Ruiming Tang}, \bibinfo{person}{Xiuqiang He}, \bibinfo{person}{Jianye Hao}, {and} \bibinfo{person}{Yong Yu}.} \bibinfo{year}{2021}\natexlab{}.
\newblock \showarticletitle{A Graph-Enhanced Click Model for Web Search}. In \bibinfo{booktitle}{\emph{Proceedings of the International ACM SIGIR Conference on Research and Development in Information Retrieval}}. \bibinfo{pages}{1259--1268}.
\newblock


\bibitem[Lin et~al\mbox{.}(2023)]%
        {lin2023multi}
\bibfield{author}{\bibinfo{person}{Xinyu Lin}, \bibinfo{person}{Wenjie Wang}, \bibinfo{person}{Yongqi Li}, \bibinfo{person}{Fuli Feng}, \bibinfo{person}{See-Kiong Ng}, {and} \bibinfo{person}{Tat-Seng Chua}.} \bibinfo{year}{2023}\natexlab{}.
\newblock \showarticletitle{A multi-facet paradigm to bridge large language model and recommendation}.
\newblock \bibinfo{journal}{\emph{arXiv preprint arXiv:2310.06491}} (\bibinfo{year}{2023}).
\newblock


\bibitem[Lin et~al\mbox{.}(2022)]%
        {lin2022improving}
\bibfield{author}{\bibinfo{person}{Zihan Lin}, \bibinfo{person}{Changxin Tian}, \bibinfo{person}{Yupeng Hou}, {and} \bibinfo{person}{Wayne~Xin Zhao}.} \bibinfo{year}{2022}\natexlab{}.
\newblock \showarticletitle{Improving graph collaborative filtering with neighborhood-enriched contrastive learning}. In \bibinfo{booktitle}{\emph{Proceedings of the ACM Web Conference}}. \bibinfo{pages}{2320--2329}.
\newblock


\bibitem[Linden et~al\mbox{.}(2003)]%
        {linden2003amazon}
\bibfield{author}{\bibinfo{person}{Greg Linden}, \bibinfo{person}{Brent Smith}, {and} \bibinfo{person}{Jeremy York}.} \bibinfo{year}{2003}\natexlab{}.
\newblock \showarticletitle{Amazon. com recommendations: Item-to-item collaborative filtering}.
\newblock \bibinfo{journal}{\emph{IEEE Internet computing}} \bibinfo{volume}{7}, \bibinfo{number}{1} (\bibinfo{year}{2003}), \bibinfo{pages}{76--80}.
\newblock


\bibitem[Liu et~al\mbox{.}(2023b)]%
        {liu2023chatgpt}
\bibfield{author}{\bibinfo{person}{Junling Liu}, \bibinfo{person}{Chao Liu}, \bibinfo{person}{Renjie Lv}, \bibinfo{person}{Kang Zhou}, {and} \bibinfo{person}{Yan Zhang}.} \bibinfo{year}{2023}\natexlab{b}.
\newblock \showarticletitle{Is chatgpt a good recommender? a preliminary study}.
\newblock \bibinfo{journal}{\emph{arXiv preprint arXiv:2304.10149}} (\bibinfo{year}{2023}).
\newblock


\bibitem[Liu et~al\mbox{.}(2023c)]%
        {liu2023pre}
\bibfield{author}{\bibinfo{person}{Peng Liu}, \bibinfo{person}{Lemei Zhang}, {and} \bibinfo{person}{Jon~Atle Gulla}.} \bibinfo{year}{2023}\natexlab{c}.
\newblock \showarticletitle{Pre-train, prompt and recommendation: A comprehensive survey of language modelling paradigm adaptations in recommender systems}.
\newblock \bibinfo{journal}{\emph{arXiv preprint arXiv:2302.03735}} (\bibinfo{year}{2023}).
\newblock


\bibitem[Liu et~al\mbox{.}(2023a)]%
        {liu2023first}
\bibfield{author}{\bibinfo{person}{Qijiong Liu}, \bibinfo{person}{Nuo Chen}, \bibinfo{person}{Tetsuya Sakai}, {and} \bibinfo{person}{Xiao-Ming Wu}.} \bibinfo{year}{2023}\natexlab{a}.
\newblock \showarticletitle{A First Look at LLM-Powered Generative News Recommendation}.
\newblock \bibinfo{journal}{\emph{arXiv preprint arXiv:2305.06566}} (\bibinfo{year}{2023}).
\newblock


\bibitem[Mao et~al\mbox{.}(2023)]%
        {mao2023unitrec}
\bibfield{author}{\bibinfo{person}{Zhiming Mao}, \bibinfo{person}{Huimin Wang}, \bibinfo{person}{Yiming Du}, {and} \bibinfo{person}{Kam-fai Wong}.} \bibinfo{year}{2023}\natexlab{}.
\newblock \showarticletitle{UniTRec: A Unified Text-to-Text Transformer and Joint Contrastive Learning Framework for Text-based Recommendation}.
\newblock \bibinfo{journal}{\emph{The Association for Computational Linguistics}} (\bibinfo{year}{2023}).
\newblock


\bibitem[Minaee et~al\mbox{.}(2024)]%
        {minaee2024large}
\bibfield{author}{\bibinfo{person}{Shervin Minaee}, \bibinfo{person}{Tomas Mikolov}, \bibinfo{person}{Narjes Nikzad}, \bibinfo{person}{Meysam Chenaghlu}, \bibinfo{person}{Richard Socher}, \bibinfo{person}{Xavier Amatriain}, {and} \bibinfo{person}{Jianfeng Gao}.} \bibinfo{year}{2024}\natexlab{}.
\newblock \showarticletitle{Large language models: A survey}.
\newblock \bibinfo{journal}{\emph{arXiv preprint arXiv:2402.06196}} (\bibinfo{year}{2024}).
\newblock


\bibitem[Mysore et~al\mbox{.}(2023)]%
        {mysore2023large}
\bibfield{author}{\bibinfo{person}{Sheshera Mysore}, \bibinfo{person}{Andrew McCallum}, {and} \bibinfo{person}{Hamed Zamani}.} \bibinfo{year}{2023}\natexlab{}.
\newblock \showarticletitle{Large Language Model Augmented Narrative Driven Recommendations}.
\newblock \bibinfo{journal}{\emph{arXiv preprint arXiv:2306.02250}} (\bibinfo{year}{2023}).
\newblock


\bibitem[Qu et~al\mbox{.}(2019)]%
        {qu2019bert}
\bibfield{author}{\bibinfo{person}{Chen Qu}, \bibinfo{person}{Liu Yang}, \bibinfo{person}{Minghui Qiu}, \bibinfo{person}{W~Bruce Croft}, \bibinfo{person}{Yongfeng Zhang}, {and} \bibinfo{person}{Mohit Iyyer}.} \bibinfo{year}{2019}\natexlab{}.
\newblock \showarticletitle{BERT with history answer embedding for conversational question answering}. In \bibinfo{booktitle}{\emph{Proceedings of the International ACM SIGIR Conference on Research and Development in Information Retrieval}}. \bibinfo{pages}{1133--1136}.
\newblock


\bibitem[Raffel et~al\mbox{.}(2020)]%
        {raffel2020exploring}
\bibfield{author}{\bibinfo{person}{Colin Raffel}, \bibinfo{person}{Noam Shazeer}, \bibinfo{person}{Adam Roberts}, \bibinfo{person}{Katherine Lee}, \bibinfo{person}{Sharan Narang}, \bibinfo{person}{Michael Matena}, \bibinfo{person}{Yanqi Zhou}, \bibinfo{person}{Wei Li}, {and} \bibinfo{person}{Peter~J Liu}.} \bibinfo{year}{2020}\natexlab{}.
\newblock \showarticletitle{Exploring the limits of transfer learning with a unified text-to-text transformer}.
\newblock \bibinfo{journal}{\emph{The Journal of Machine Learning Research}} \bibinfo{volume}{21}, \bibinfo{number}{1} (\bibinfo{year}{2020}), \bibinfo{pages}{5485--5551}.
\newblock


\bibitem[Rana et~al\mbox{.}(2023)]%
        {rana2023sayplan}
\bibfield{author}{\bibinfo{person}{Krishan Rana}, \bibinfo{person}{Jesse Haviland}, \bibinfo{person}{Sourav Garg}, \bibinfo{person}{Jad Abou-Chakra}, \bibinfo{person}{Ian Reid}, {and} \bibinfo{person}{Niko Suenderhauf}.} \bibinfo{year}{2023}\natexlab{}.
\newblock \showarticletitle{Sayplan: Grounding large language models using 3d scene graphs for scalable task planning}.
\newblock \bibinfo{journal}{\emph{arXiv preprint arXiv:2307.06135}} (\bibinfo{year}{2023}).
\newblock


\bibitem[Ren et~al\mbox{.}(2024)]%
        {ren2024survey}
\bibfield{author}{\bibinfo{person}{Xubin Ren}, \bibinfo{person}{Jiabin Tang}, \bibinfo{person}{Dawei Yin}, \bibinfo{person}{Nitesh Chawla}, {and} \bibinfo{person}{Chao Huang}.} \bibinfo{year}{2024}\natexlab{}.
\newblock \showarticletitle{A survey of large language models for graphs}. In \bibinfo{booktitle}{\emph{Proceedings of the 30th ACM SIGKDD Conference on Knowledge Discovery and Data Mining}}. \bibinfo{pages}{6616--6626}.
\newblock


\bibitem[Rendle(2010)]%
        {rendle2010factorization}
\bibfield{author}{\bibinfo{person}{Steffen Rendle}.} \bibinfo{year}{2010}\natexlab{}.
\newblock \showarticletitle{Factorization machines}. In \bibinfo{booktitle}{\emph{IEEE International Conference on Data Mining}}. \bibinfo{pages}{995--1000}.
\newblock


\bibitem[Rendle et~al\mbox{.}(2009)]%
        {rendle2009bpr}
\bibfield{author}{\bibinfo{person}{Steffen Rendle}, \bibinfo{person}{Christoph Freudenthaler}, \bibinfo{person}{Zeno Gantner}, {and} \bibinfo{person}{Lars Schmidt-Thieme}.} \bibinfo{year}{2009}\natexlab{}.
\newblock \showarticletitle{BPR: Bayesian personalized ranking from implicit feedback}. In \bibinfo{booktitle}{\emph{Proceedings of the Conference on Uncertainty in Artificial Intelligence}}. \bibinfo{pages}{452--461}.
\newblock


\bibitem[Rendle et~al\mbox{.}(2012)]%
        {rendle2012bpr}
\bibfield{author}{\bibinfo{person}{Steffen Rendle}, \bibinfo{person}{Christoph Freudenthaler}, \bibinfo{person}{Zeno Gantner}, {and} \bibinfo{person}{Lars Schmidt-Thieme}.} \bibinfo{year}{2012}\natexlab{}.
\newblock \showarticletitle{BPR: Bayesian personalized ranking from implicit feedback}.
\newblock \bibinfo{journal}{\emph{arXiv preprint arXiv:1205.2618}} (\bibinfo{year}{2012}).
\newblock


\bibitem[Rendle et~al\mbox{.}(2010)]%
        {rendle2010factorizing}
\bibfield{author}{\bibinfo{person}{Steffen Rendle}, \bibinfo{person}{Christoph Freudenthaler}, {and} \bibinfo{person}{Lars Schmidt-Thieme}.} \bibinfo{year}{2010}\natexlab{}.
\newblock \showarticletitle{Factorizing personalized markov chains for next-basket recommendation}. In \bibinfo{booktitle}{\emph{Proceedings of the international conference on World wide web}}. \bibinfo{pages}{811--820}.
\newblock


\bibitem[Sakata et~al\mbox{.}(2019)]%
        {sakata2019faq}
\bibfield{author}{\bibinfo{person}{Wataru Sakata}, \bibinfo{person}{Tomohide Shibata}, \bibinfo{person}{Ribeka Tanaka}, {and} \bibinfo{person}{Sadao Kurohashi}.} \bibinfo{year}{2019}\natexlab{}.
\newblock \showarticletitle{FAQ retrieval using query-question similarity and BERT-based query-answer relevance}. In \bibinfo{booktitle}{\emph{Proceedings of the International ACM SIGIR Conference on Research and Development in Information Retrieval}}. \bibinfo{pages}{1113--1116}.
\newblock


\bibitem[Schick et~al\mbox{.}(2023)]%
        {schick2023toolformer}
\bibfield{author}{\bibinfo{person}{Timo Schick}, \bibinfo{person}{Jane Dwivedi-Yu}, \bibinfo{person}{Roberto Dess{\`\i}}, \bibinfo{person}{Roberta Raileanu}, \bibinfo{person}{Maria Lomeli}, \bibinfo{person}{Luke Zettlemoyer}, \bibinfo{person}{Nicola Cancedda}, {and} \bibinfo{person}{Thomas Scialom}.} \bibinfo{year}{2023}\natexlab{}.
\newblock \showarticletitle{Toolformer: Language models can teach themselves to use tools}.
\newblock \bibinfo{journal}{\emph{arXiv preprint arXiv:2302.04761}} (\bibinfo{year}{2023}).
\newblock


\bibitem[Sun et~al\mbox{.}(2019)]%
        {sun2019bert4rec}
\bibfield{author}{\bibinfo{person}{Fei Sun}, \bibinfo{person}{Jun Liu}, \bibinfo{person}{Jian Wu}, \bibinfo{person}{Changhua Pei}, \bibinfo{person}{Xiao Lin}, \bibinfo{person}{Wenwu Ou}, {and} \bibinfo{person}{Peng Jiang}.} \bibinfo{year}{2019}\natexlab{}.
\newblock \showarticletitle{BERT4Rec: Sequential recommendation with bidirectional encoder representations from transformer}. In \bibinfo{booktitle}{\emph{Proceedings of the 28th ACM international conference on information and knowledge management}}. \bibinfo{pages}{1441--1450}.
\newblock


\bibitem[Sun et~al\mbox{.}(2024)]%
        {sun2024determlr}
\bibfield{author}{\bibinfo{person}{Hongda Sun}, \bibinfo{person}{Weikai Xu}, \bibinfo{person}{Wei Liu}, \bibinfo{person}{Jian Luan}, \bibinfo{person}{Bin Wang}, \bibinfo{person}{Shuo Shang}, \bibinfo{person}{Ji-Rong Wen}, {and} \bibinfo{person}{Rui Yan}.} \bibinfo{year}{2024}\natexlab{}.
\newblock \showarticletitle{Determlr: Augmenting llm-based logical reasoning from indeterminacy to determinacy}. In \bibinfo{booktitle}{\emph{Proceedings of the 62nd Annual Meeting of the Association for Computational Linguistics (Volume 1: Long Papers)}}. \bibinfo{pages}{9828--9862}.
\newblock


\bibitem[Sun et~al\mbox{.}(2023)]%
        {sun2023chatgpt}
\bibfield{author}{\bibinfo{person}{Weiwei Sun}, \bibinfo{person}{Lingyong Yan}, \bibinfo{person}{Xinyu Ma}, \bibinfo{person}{Pengjie Ren}, \bibinfo{person}{Dawei Yin}, {and} \bibinfo{person}{Zhaochun Ren}.} \bibinfo{year}{2023}\natexlab{}.
\newblock \showarticletitle{Is ChatGPT Good at Search? Investigating Large Language Models as Re-Ranking Agent}.
\newblock \bibinfo{journal}{\emph{arXiv preprint arXiv:2304.09542}} (\bibinfo{year}{2023}).
\newblock


\bibitem[Touvron et~al\mbox{.}(2023)]%
        {llama}
\bibfield{author}{\bibinfo{person}{Hugo Touvron}, \bibinfo{person}{Thibaut Lavril}, \bibinfo{person}{Gautier Izacard}, \bibinfo{person}{Xavier Martinet}, \bibinfo{person}{Marie{-}Anne Lachaux}, \bibinfo{person}{Timoth{\'{e}}e Lacroix}, \bibinfo{person}{Baptiste Rozi{\`{e}}re}, \bibinfo{person}{Naman Goyal}, \bibinfo{person}{Eric Hambro}, \bibinfo{person}{Faisal Azhar}, \bibinfo{person}{Aur{\'{e}}lien Rodriguez}, \bibinfo{person}{Armand Joulin}, \bibinfo{person}{Edouard Grave}, {and} \bibinfo{person}{Guillaume Lample}.} \bibinfo{year}{2023}\natexlab{}.
\newblock \showarticletitle{LLaMA: Open and Efficient Foundation Language Models}.
\newblock \bibinfo{journal}{\emph{CoRR}}  \bibinfo{volume}{abs/2302.13971} (\bibinfo{year}{2023}).
\newblock


\bibitem[Wang et~al\mbox{.}(2024)]%
        {wang2024llms}
\bibfield{author}{\bibinfo{person}{Duo Wang}, \bibinfo{person}{Yuan Zuo}, \bibinfo{person}{Fengzhi Li}, {and} \bibinfo{person}{Junjie Wu}.} \bibinfo{year}{2024}\natexlab{}.
\newblock \showarticletitle{LLMs as Zero-shot Graph Learners: Alignment of GNN Representations with LLM Token Embeddings}.
\newblock \bibinfo{journal}{\emph{arXiv preprint arXiv:2408.14512}} (\bibinfo{year}{2024}).
\newblock


\bibitem[Wang et~al\mbox{.}(2023)]%
        {wang2023generative}
\bibfield{author}{\bibinfo{person}{Wenjie Wang}, \bibinfo{person}{Xinyu Lin}, \bibinfo{person}{Fuli Feng}, \bibinfo{person}{Xiangnan He}, {and} \bibinfo{person}{Tat-Seng Chua}.} \bibinfo{year}{2023}\natexlab{}.
\newblock \showarticletitle{Generative recommendation: Towards next-generation recommender paradigm}.
\newblock \bibinfo{journal}{\emph{arXiv preprint arXiv:2304.03516}} (\bibinfo{year}{2023}).
\newblock


\bibitem[Wang et~al\mbox{.}(2019)]%
        {wang2019neural}
\bibfield{author}{\bibinfo{person}{Xiang Wang}, \bibinfo{person}{Xiangnan He}, \bibinfo{person}{Meng Wang}, \bibinfo{person}{Fuli Feng}, {and} \bibinfo{person}{Tat-Seng Chua}.} \bibinfo{year}{2019}\natexlab{}.
\newblock \showarticletitle{Neural graph collaborative filtering}. In \bibinfo{booktitle}{\emph{Proceedings of the 42nd international ACM SIGIR conference on Research and development in Information Retrieval}}. \bibinfo{pages}{165--174}.
\newblock


\bibitem[Wang et~al\mbox{.}(2022)]%
        {wang2022towards}
\bibfield{author}{\bibinfo{person}{Xiaolei Wang}, \bibinfo{person}{Kun Zhou}, \bibinfo{person}{Ji-Rong Wen}, {and} \bibinfo{person}{Wayne~Xin Zhao}.} \bibinfo{year}{2022}\natexlab{}.
\newblock \showarticletitle{Towards unified conversational recommender systems via knowledge-enhanced prompt learning}. In \bibinfo{booktitle}{\emph{Proceedings of the ACM SIGKDD Conference on Knowledge Discovery and Data Mining}}. \bibinfo{pages}{1929--1937}.
\newblock


\bibitem[Wei et~al\mbox{.}(2023)]%
        {wei2023llmrec}
\bibfield{author}{\bibinfo{person}{Wei Wei}, \bibinfo{person}{Xubin Ren}, \bibinfo{person}{Jiabin Tang}, \bibinfo{person}{Qinyong Wang}, \bibinfo{person}{Lixin Su}, \bibinfo{person}{Suqi Cheng}, \bibinfo{person}{Junfeng Wang}, \bibinfo{person}{Dawei Yin}, {and} \bibinfo{person}{Chao Huang}.} \bibinfo{year}{2023}\natexlab{}.
\newblock \showarticletitle{LLMRec: Large Language Models with Graph Augmentation for Recommendation}.
\newblock \bibinfo{journal}{\emph{The ACM International Conference on Web Search and Data Mining}} (\bibinfo{year}{2023}).
\newblock


\bibitem[Wu et~al\mbox{.}(2024)]%
        {wu2024survey}
\bibfield{author}{\bibinfo{person}{Likang Wu}, \bibinfo{person}{Zhi Zheng}, \bibinfo{person}{Zhaopeng Qiu}, \bibinfo{person}{Hao Wang}, \bibinfo{person}{Hongchao Gu}, \bibinfo{person}{Tingjia Shen}, \bibinfo{person}{Chuan Qin}, \bibinfo{person}{Chen Zhu}, \bibinfo{person}{Hengshu Zhu}, \bibinfo{person}{Qi Liu}, {et~al\mbox{.}}} \bibinfo{year}{2024}\natexlab{}.
\newblock \showarticletitle{A survey on large language models for recommendation}.
\newblock \bibinfo{journal}{\emph{World Wide Web}} \bibinfo{volume}{27}, \bibinfo{number}{5} (\bibinfo{year}{2024}), \bibinfo{pages}{60}.
\newblock


\bibitem[Xi et~al\mbox{.}(2023)]%
        {xi2023bird}
\bibfield{author}{\bibinfo{person}{Yunjia Xi}, \bibinfo{person}{Jianghao Lin}, \bibinfo{person}{Weiwen Liu}, \bibinfo{person}{Xinyi Dai}, \bibinfo{person}{Weinan Zhang}, \bibinfo{person}{Rui Zhang}, \bibinfo{person}{Ruiming Tang}, {and} \bibinfo{person}{Yong Yu}.} \bibinfo{year}{2023}\natexlab{}.
\newblock \showarticletitle{A Bird's-eye View of Reranking: from List Level to Page Level}. In \bibinfo{booktitle}{\emph{Proceedings of the ACM International Conference on Web Search and Data Mining}}. \bibinfo{pages}{1075--1083}.
\newblock


\bibitem[Xie et~al\mbox{.}(2022)]%
        {xie2022contrastive}
\bibfield{author}{\bibinfo{person}{Xu Xie}, \bibinfo{person}{Fei Sun}, \bibinfo{person}{Zhaoyang Liu}, \bibinfo{person}{Shiwen Wu}, \bibinfo{person}{Jinyang Gao}, \bibinfo{person}{Jiandong Zhang}, \bibinfo{person}{Bolin Ding}, {and} \bibinfo{person}{Bin Cui}.} \bibinfo{year}{2022}\natexlab{}.
\newblock \showarticletitle{Contrastive learning for sequential recommendation}. In \bibinfo{booktitle}{\emph{IEEE International Conference on Data Engineering}}. \bibinfo{pages}{1259--1273}.
\newblock


\bibitem[Yang et~al\mbox{.}(2019a)]%
        {yang2019end}
\bibfield{author}{\bibinfo{person}{Wei Yang}, \bibinfo{person}{Yuqing Xie}, \bibinfo{person}{Aileen Lin}, \bibinfo{person}{Xingyu Li}, \bibinfo{person}{Luchen Tan}, \bibinfo{person}{Kun Xiong}, \bibinfo{person}{Ming Li}, {and} \bibinfo{person}{Jimmy Lin}.} \bibinfo{year}{2019}\natexlab{a}.
\newblock \showarticletitle{End-to-end open-domain question answering with bertserini}.
\newblock \bibinfo{journal}{\emph{arXiv preprint arXiv:1902.01718}} (\bibinfo{year}{2019}).
\newblock


\bibitem[Yang et~al\mbox{.}(2019b)]%
        {yang2019simple}
\bibfield{author}{\bibinfo{person}{Wei Yang}, \bibinfo{person}{Haotian Zhang}, {and} \bibinfo{person}{Jimmy Lin}.} \bibinfo{year}{2019}\natexlab{b}.
\newblock \showarticletitle{Simple applications of BERT for ad hoc document retrieval}.
\newblock \bibinfo{journal}{\emph{arXiv preprint arXiv:1903.10972}} (\bibinfo{year}{2019}).
\newblock


\bibitem[Yang et~al\mbox{.}(2023)]%
        {liu2013once}
\bibfield{author}{\bibinfo{person}{Wei Yang}, \bibinfo{person}{Haotian Zhang}, {and} \bibinfo{person}{Jimmy Lin}.} \bibinfo{year}{2023}\natexlab{}.
\newblock \showarticletitle{ONCE: Boosting Content-based Recommendation with Both Open- and Closed-source Large Language Models}.
\newblock \bibinfo{journal}{\emph{arXiv preprint arXiv:2305.06566}} (\bibinfo{year}{2023}).
\newblock


\bibitem[Ye et~al\mbox{.}(2024)]%
        {ye2024language}
\bibfield{author}{\bibinfo{person}{Ruosong Ye}, \bibinfo{person}{Caiqi Zhang}, \bibinfo{person}{Runhui Wang}, \bibinfo{person}{Shuyuan Xu}, {and} \bibinfo{person}{Yongfeng Zhang}.} \bibinfo{year}{2024}\natexlab{}.
\newblock \showarticletitle{Language is all a graph needs}. In \bibinfo{booktitle}{\emph{Findings of the Association for Computational Linguistics: EACL 2024}}. \bibinfo{pages}{1955--1973}.
\newblock


\bibitem[Yu et~al\mbox{.}(2023a)]%
        {yu2023xsimgcl}
\bibfield{author}{\bibinfo{person}{Junliang Yu}, \bibinfo{person}{Xin Xia}, \bibinfo{person}{Tong Chen}, \bibinfo{person}{Lizhen Cui}, \bibinfo{person}{Nguyen Quoc~Viet Hung}, {and} \bibinfo{person}{Hongzhi Yin}.} \bibinfo{year}{2023}\natexlab{a}.
\newblock \showarticletitle{XSimGCL: Towards extremely simple graph contrastive learning for recommendation}.
\newblock \bibinfo{journal}{\emph{IEEE Transactions on Knowledge and Data Engineering}} (\bibinfo{year}{2023}).
\newblock


\bibitem[Yu et~al\mbox{.}(2022)]%
        {yu2022graph}
\bibfield{author}{\bibinfo{person}{Junliang Yu}, \bibinfo{person}{Hongzhi Yin}, \bibinfo{person}{Xin Xia}, \bibinfo{person}{Tong Chen}, \bibinfo{person}{Lizhen Cui}, {and} \bibinfo{person}{Quoc Viet~Hung Nguyen}.} \bibinfo{year}{2022}\natexlab{}.
\newblock \showarticletitle{Are graph augmentations necessary? simple graph contrastive learning for recommendation}. In \bibinfo{booktitle}{\emph{Proceedings of the 45th international ACM SIGIR conference on research and development in information retrieval}}. \bibinfo{pages}{1294--1303}.
\newblock


\bibitem[Yu et~al\mbox{.}(2023b)]%
        {yu2023self}
\bibfield{author}{\bibinfo{person}{Junliang Yu}, \bibinfo{person}{Hongzhi Yin}, \bibinfo{person}{Xin Xia}, \bibinfo{person}{Tong Chen}, \bibinfo{person}{Jundong Li}, {and} \bibinfo{person}{Zi Huang}.} \bibinfo{year}{2023}\natexlab{b}.
\newblock \showarticletitle{Self-supervised learning for recommender systems: A survey}.
\newblock \bibinfo{journal}{\emph{IEEE Transactions on Knowledge and Data Engineering}} (\bibinfo{year}{2023}).
\newblock


\bibitem[Yuan et~al\mbox{.}(2022)]%
        {yuan2022tenrec}
\bibfield{author}{\bibinfo{person}{Guanghu Yuan}, \bibinfo{person}{Fajie Yuan}, \bibinfo{person}{Yudong Li}, \bibinfo{person}{Beibei Kong}, \bibinfo{person}{Shujie Li}, \bibinfo{person}{Lei Chen}, \bibinfo{person}{Min Yang}, \bibinfo{person}{Chenyun Yu}, \bibinfo{person}{Bo Hu}, \bibinfo{person}{Zang Li}, {et~al\mbox{.}}} \bibinfo{year}{2022}\natexlab{}.
\newblock \showarticletitle{Tenrec: A Large-scale Multipurpose Benchmark Dataset for Recommender Systems}.
\newblock \bibinfo{journal}{\emph{Advances in Neural Information Processing Systems}}  \bibinfo{volume}{35} (\bibinfo{year}{2022}), \bibinfo{pages}{11480--11493}.
\newblock


\bibitem[Yuan et~al\mbox{.}(2023)]%
        {yuan2023go}
\bibfield{author}{\bibinfo{person}{Zheng Yuan}, \bibinfo{person}{Fajie Yuan}, \bibinfo{person}{Yu Song}, \bibinfo{person}{Youhua Li}, \bibinfo{person}{Junchen Fu}, \bibinfo{person}{Fei Yang}, \bibinfo{person}{Yunzhu Pan}, {and} \bibinfo{person}{Yongxin Ni}.} \bibinfo{year}{2023}\natexlab{}.
\newblock \showarticletitle{Where to go next for recommender systems? id-vs. modality-based recommender models revisited}.
\newblock \bibinfo{journal}{\emph{arXiv preprint arXiv:2303.13835}} (\bibinfo{year}{2023}).
\newblock


\bibitem[Zeng et~al\mbox{.}(2021)]%
        {zeng2021knowledge}
\bibfield{author}{\bibinfo{person}{Zheni Zeng}, \bibinfo{person}{Chaojun Xiao}, \bibinfo{person}{Yuan Yao}, \bibinfo{person}{Ruobing Xie}, \bibinfo{person}{Zhiyuan Liu}, \bibinfo{person}{Fen Lin}, \bibinfo{person}{Leyu Lin}, {and} \bibinfo{person}{Maosong Sun}.} \bibinfo{year}{2021}\natexlab{}.
\newblock \showarticletitle{Knowledge transfer via pre-training for recommendation: A review and prospect}.
\newblock \bibinfo{journal}{\emph{Frontiers in Big Data}}  \bibinfo{volume}{4} (\bibinfo{year}{2021}), \bibinfo{pages}{602071}.
\newblock


\bibitem[Zhang et~al\mbox{.}(2024a)]%
        {agent4rec}
\bibfield{author}{\bibinfo{person}{An Zhang}, \bibinfo{person}{Yuxin Chen}, \bibinfo{person}{Leheng Sheng}, \bibinfo{person}{Xiang Wang}, {and} \bibinfo{person}{Tat-Seng Chua}.} \bibinfo{year}{2024}\natexlab{a}.
\newblock \showarticletitle{On Generative Agents in Recommendation}. In \bibinfo{booktitle}{\emph{SIGIR}}.
\newblock


\bibitem[Zhang et~al\mbox{.}(2024b)]%
        {zhang2024llm4dyg}
\bibfield{author}{\bibinfo{person}{Zeyang Zhang}, \bibinfo{person}{Xin Wang}, \bibinfo{person}{Ziwei Zhang}, \bibinfo{person}{Haoyang Li}, \bibinfo{person}{Yijian Qin}, {and} \bibinfo{person}{Wenwu Zhu}.} \bibinfo{year}{2024}\natexlab{b}.
\newblock \showarticletitle{LLM4DyG: Can Large Language Models Solve Spatial-Temporal Problems on Dynamic Graphs?}. In \bibinfo{booktitle}{\emph{Proceedings of the 30th ACM SIGKDD Conference on Knowledge Discovery and Data Mining}}. \bibinfo{pages}{4350--4361}.
\newblock


\bibitem[Zhao et~al\mbox{.}(2021)]%
        {zhao2021variational}
\bibfield{author}{\bibinfo{person}{Jing Zhao}, \bibinfo{person}{Pengpeng Zhao}, \bibinfo{person}{Lei Zhao}, \bibinfo{person}{Yanchi Liu}, \bibinfo{person}{Victor~S Sheng}, {and} \bibinfo{person}{Xiaofang Zhou}.} \bibinfo{year}{2021}\natexlab{}.
\newblock \showarticletitle{Variational self-attention network for sequential recommendation}. In \bibinfo{booktitle}{\emph{2021 IEEE 37th International Conference on Data Engineering (ICDE)}}. IEEE, \bibinfo{pages}{1559--1570}.
\newblock


\bibitem[Zhao et~al\mbox{.}(2023)]%
        {zhao2023survey}
\bibfield{author}{\bibinfo{person}{Wayne~Xin Zhao}, \bibinfo{person}{Kun Zhou}, \bibinfo{person}{Junyi Li}, \bibinfo{person}{Tianyi Tang}, \bibinfo{person}{Xiaolei Wang}, \bibinfo{person}{Yupeng Hou}, \bibinfo{person}{Yingqian Min}, \bibinfo{person}{Beichen Zhang}, \bibinfo{person}{Junjie Zhang}, \bibinfo{person}{Zican Dong}, {et~al\mbox{.}}} \bibinfo{year}{2023}\natexlab{}.
\newblock \showarticletitle{A survey of large language models}.
\newblock \bibinfo{journal}{\emph{arXiv preprint arXiv:2303.18223}} (\bibinfo{year}{2023}).
\newblock


\bibitem[Zhao et~al\mbox{.}(2024)]%
        {ToolRec}
\bibfield{author}{\bibinfo{person}{Yuyue Zhao}, \bibinfo{person}{Jiancan Wu}, \bibinfo{person}{Xiang Wang}, \bibinfo{person}{Wei Tang}, \bibinfo{person}{Dingxian Wang}, {and} \bibinfo{person}{Maarten de Rijke}.} \bibinfo{year}{2024}\natexlab{}.
\newblock \showarticletitle{Let Me Do It For You: Towards LLM Empowered Recommendation via Tool Learning}. In \bibinfo{booktitle}{\emph{SIGIR}}.
\newblock


\bibitem[Zhou et~al\mbox{.}(2018)]%
        {zhou2018deep}
\bibfield{author}{\bibinfo{person}{Guorui Zhou}, \bibinfo{person}{Xiaoqiang Zhu}, \bibinfo{person}{Chenru Song}, \bibinfo{person}{Ying Fan}, \bibinfo{person}{Han Zhu}, \bibinfo{person}{Xiao Ma}, \bibinfo{person}{Yanghui Yan}, \bibinfo{person}{Junqi Jin}, \bibinfo{person}{Han Li}, {and} \bibinfo{person}{Kun Gai}.} \bibinfo{year}{2018}\natexlab{}.
\newblock \showarticletitle{Deep interest network for click-through rate prediction}. In \bibinfo{booktitle}{\emph{Proceedings of the ACM SIGKDD international conference on knowledge discovery \& data mining}}. \bibinfo{pages}{1059--1068}.
\newblock


\end{thebibliography}

% \appendix

\end{document}